\documentclass[usenatbib,usegraphicx,useAMS,usenatbib]{mn2e}

\newcommand{\uchii}{{UC H{\scriptsize II} }}
\newcommand{\hii}{{H{\scriptsize II} }}

\newcommand{\water}{H$_2$O}
\newcommand{\nh}{NH$_3$}

\newcommand{\kms}{km\,s$^{-1}$}

\title[HOPS I: Techniques and \water~masers]{The \water~southern Galactic Plane Survey
(HOPS): I. Techniques and \water~maser data}
\author[Walsh et al.]{A. J. Walsh$^{1}$\thanks{E-mail:
andrew.walsh@jcu.edu.au}, S. L. Breen $^{2}$, T. Britton$^2$, K. J. Brooks$^2$, M. G. Burton$^{3}$, \newauthor
M. R. Cunningham$^3$, J. A. Green$^2$, L. Harvey-Smith$^2$, L. Hindson$^{2,4}$, M. G. Hoare$^5$, \newauthor
B. Indermuehle$^2$, P. A. Jones$^{2,6}$, N. Lo$^{6,7}$, S. N. Longmore$^8$, V. Lowe$^{2,3}$, C. J. Phillips$^2$, \newauthor
C. R. Purcell$^5$, M. A. Thompson $^4$, J. S. Urquhart$^2$, M. A. Voronkov$^2$ \newauthor
G. L. White$^1$ and M. T. Whiting$^2$ \\
$^{1}$School of Engineering and Physical Sciences, James Cook University, Townsville, QLD 4811, Australia; \\
$^{2}$CSIRO Astronomy and Space Science, PO BOX 76, Epping, NSW 1710, Australia; \\
$^{3}$School of Physics, University of New South Wales, Sydney, NSW 2052, Australia; \\
$^{4}$Centre for Astrophysics Research, Science and Technology Research Institute, University of Hertfordshire, AL10 9AB, UK; \\
$^{5}$School of Physics and Astronomy, University of Leeds, Leeds, LS2 9JT, UK; \\
$^{6}$Departamento de Astronom\'ia, Universidad de Chile, Casilla 36-D, Santiago, Chile\\
$^{7}$Laboratoire AIM Paris-Saclay, CEA/Irfu - Uni. Paris Did\'erot - CNRS/INSU, 91191 Gif-sur-Yvette, France\\
$^{8}$Harvard-Smithsonian Center for Astrophysics, 60 Garden Street, Cambridge, MA 02138, USA}
\begin{document}



\maketitle

\label{firstpage}

\begin{abstract}
We present first results of the {\bf H}$_2${\bf O} Southern Galactic
{\bf P}lane {\bf S}urvey (HOPS),
using the Mopra radiotelescope with a broad band backend and a beam size of
about 2\arcmin. We have observed 100 square degrees of the southern Galactic
plane at 12\,mm (19.5 to 27.5\,GHz), including spectral line emission from
\water~masers, multiple metastable transitions of ammonia, cyanoacetylene,
methanol and radio recombination lines. In this paper, we report on the
characteristics of the survey and \water~maser emission. We find 540
\water~masers, of which 334 are new detections. The strongest maser is
3933\,Jy and the weakest is 0.7\,Jy, with 62 masers over 100\,Jy. In 14 maser
sites, the spread in velocity of the \water~maser emission exceeds
100\,km\,s$^{-1}$. In one region, the \water~maser velocities are separated
by 351.3\,km\,s$^{-1}$. The rms noise levels are typically between 1-2\,Jy,
with 95\% of the survey under 2\,Jy. We estimate completeness limits of
98\% at around 8.4\,Jy and 50\% at around 5.5\,Jy. We estimate
that there are between 800 and 1500 \water~masers in the Galaxy that are
detectable in a survey with similar completeness limits to HOPS. We report possible
masers in NH$_3$ (11,9) and (8,6) emission towards G19.61-0.23 and in the
NH$_3$ (3,3) line towards G23.33-0.30.

\end{abstract}


\section{Introduction}
Understanding how the interstellar medium (ISM) is linked to stellar
birth and death in our Galaxy is a major problem in astrophysics today.
In order to understand the processes involved, it is often useful to conduct
large-scale Galactic plane surveys. Such surveys as the International
Galactic Plane Survey \citep{taylor03,mcclure05,stil06} in H{\sc I}, as well as
surveys of $^{12}$CO \citep{dame01} and $^{13}$CO \citep{jackson06}
have been widely used in tracing out Galactic structure and motions. But
these popular tracers are typically insensitive to high density gas. In order
to understand the processes at the beginning of star formation, where gas
and dust have accumulated to high density regions, it is necessary
to choose a tracer of such high densities, as well as other signposts of star
formation activity.

We have completed a survey of 100 square degrees of the southern Galaxy in
multiple spectral lines, using the Mopra radiotelescope in the 12\,mm band (HOPS -
The {\bf H}$_2${\bf O} southern Galactic {\bf P}lane {\bf S}urvey). Our main
target lines are the \water~(6$_{1,6}$--5$_{2,3}$) maser line, \nh~(1,1),
(2,2) and (3,3) inversion transitions, radio recombination lines H62$\alpha$
and H69$\alpha$ and HCCCN (3--2).

\water~masers are an important signpost of unusual astrophysical conditions
such as outflows and shocked gas.
They are known to occur in both high and low-mass star forming regions
(eg. \citealt{forster99,claussen96}),
late M-type stars \citep{dickinson76}, planetary nebulae \citep{miranda01},
Mira variables \citep{hinkle79}, Asymptotic Giant Branch stars
\citep{barlow96} and the centres of active galaxies \citep{claussen84}.
The majority of currently known \water~masers are found towards regions of
star formation within our Galaxy. However, an untargeted survey is required
to determine the relative occurrence of bright \water~masers with other types
of astrophysical objects. There is some evidence that within star forming regions,
\water~masers may be observable at very early stages (eg. \citealt{forster00})
of evolution. There is evidence that methanol masers may also be visible at
these early stages (eg. \citealt{walsh98}). The relative occurrence of these
two masers can be assessed through the two untargeted surveys: the Methanol
Multibeam Survey \citep{green09} and HOPS, described here.
Recent research has focused on determining parallax distances to \water~masers
(eg. \citealt{imai07}). A long term goal is to accurately establish the
dimensions of our Galaxy using bright maser sources throughout the Galaxy
(eg. \citealt{reid09}).
It is hoped that the bright \water~masers discovered in HOPS may
be used for such distance determinations.

Thermal \nh~emission in our Galaxy typically traces high density ($\sim
10^4 {\rm cm}^{-3}$; \citealt{evans99}) gas. Properties of \nh~spectral
line emission can be used to understand the physical conditions of
the gas. Lower order (J,K) inversion transitions of \nh~commonly display
hyperfine structure (eg. \citealt{ho83}) which can be used to calculate the
optical depth of the emission, to reduce confusing factors of optically thick
emission, when trying to measure the total column density of gas.
Multiple inversion transitions of \nh~are
also commonly seen, which can be used to estimate the temperature of the
excited gas. These inversion transitions occur within a few GHz of each other,
making it possible to observe them simultaneously with the same telescope
and similar setup. This greatly eliminates sources of uncertainty when comparing
multiple transitions, which in turn makes interpretation more reliable.

In cold, dense regions most molecules and ions are known
to freeze-out onto dust grains, with CO, CS and HCO$^+$ being classic examples
\citep{bergin97,willacy98,caselli99,tafalla02,bergin02}.
Thus, such species are unreliable tracers of the densest and coldest regions
in our Galaxy. CO also has a low effective critical density $\sim10^2$\,cm$^{-3}$
\citep{evans99} making it a better tracer of diffuse gas and
susceptible to becoming optically thick in dense regions. CO, CS and HCO$^+$
are also known to be good tracers of outflow activity in star forming regions.
Therefore a Galactic map of emission in these tracers can be difficult
to interpret in terms of identifying quiescent gas. On the other hand,
\nh~appears to be more robust than CO, CS or HCO$^+$ against freeze-out onto
dust grains (eg. \citealt{aikawa01}). This means that \nh~can be used to
probe the colder denser regions.
In addition to this, N$_2$H$^+$, which is found under conditions very similar
to \nh, is known to avoid regions of outflow (eg. \citealt{walsh07a}), tracing
only the quiescent gas on large scales. This makes \nh~a very reliable tracer
of cool, dense gas.  Combined with the information derived from \nh~hyperfine
structure and multiple inversion transitions, \nh~can be used to reliably
characterise the dense, quiescent gas component of the ISM.

In this paper, we concentrate on describing the survey as well as presenting
global properties of the \water~maser data. In Paper II (Purcell et al. 2011,
{\em in preparation})
we will present the NH$_3$ (1,1) and (2,2) data and cloud catalogue as well
as describing the emission finding algorithm. Paper III (Longmore at al.
2011, {\em in preparation}) will detail the thermal line fitting
routines used on the NH$_3$ data and will present the physical properties
of the unresolved clouds in the catalogue. Results from other spectral
lines will be reported in later papers.
In follow-up work, we will accurately measure the positions of the \water~masers
using the Australia Telescope Compact Array (ATCA).

\section{Mopra Characterisation}

The Australia Telescope National Facility Mopra telescope is a 22\,m antenna
located 26\,km outside the town of Coonabarrabran in New South Wales, Australia.
It is at an elevation of 850 metres above sea level and at a latitude of
31$^\circ$ south.

The receivers use Indium Phosphide High Electron Mobility Transistor (HEMT)
Monolithic Microwave Integrated Circuits (MMICs) as the amplifying elements.
The receiver systems require no tuning and include noise diodes for system
temperature determination. The nominal operating range for the 12\,mm receiver is
from 16 to 27\,GHz, however we have found the receiver still performs well
at frequencies as high as 27.5\,GHz. Further general information on Mopra can
be found in \citet{urquhart10}.

\subsection{The Mopra Spectrometer}

The Mopra Spectrometer (MOPS) is the digital filterband backend used for the
observations. It comprises an 8.3\,GHz total bandwidth, split into four
overlapping intermediate frequencies (IFs), each with a width of 2.2\,GHz.
MOPS can be configured in two modes,
either broadband mode, where the full 8.3\,GHz is available, or in zoom
mode, which was used for HOPS. In zoom mode, each of the four IFs contains
four zoom bands of 137.5\,MHz each, available to take spectra. The positioning
of the four 137.5\,MHz zoom bands within each IF is highly flexible, which
allows the user to observe virtually any of up to four spectral windows within
each 2.2\,GHz IF. Thus, it is possible to simultaneously observe up to
16 spectral windows throughout the full 8.3\,GHz bandwidth. Each spectral
window consists of 4096 channels, which is equivalent to a bandwidth and
velocity resolution of 2114\,\kms~and 0.52\,\kms~at 19.5\,GHz, or
1499\,\kms~and 0.37\,\kms~at 27.5\,GHz, respectively. Table \ref{tab1} gives
the details of centre frequencies for each band we observed, as well as the
stronger spectral lines found within the bands.

\begin{table*}
  \caption{MOPS zoom band settings and strong lines found in each band. The first
column lists the MOPS band number, with the total frequency range covered given in
the second column. The third column lists the spectral lines that we might detect
in HOPS. The spectral line frequency is given in column 5. Column 6 indicates
whether the spectral line is a thermal line, a maser line or both. The last
column give the reference for the spectral line frequency. Spectral lines
that were detected in HOPS are shown in bold.}
\label{tab1}
  \begin{tabular}{ccllcl}
  \hline
Band & Frequency Range & Spectral                                & Frequency &  Maser or   & Frequency\\
     &  covered (MHz)  & Line                                    &   (MHz)   & Thermal$^1$ & Reference\\
\hline
 1 & 19518 - 19655 & {\bf H69$\alpha$}                           & 19591.110 & thermal    & \citet{lilley68}\\
 2 & 19932 - 20069 & CH$_3$OH (2$_{1,1}$--3$_{0,3}$)E            & 19967.396 & maser (II) & \citet{mehrotra85}\\
 3 & 20691 - 20828 & {\bf NH$_3$(8,6)}                           & 20719.221 & both       & \citet{poynter75}\\
 3 & 20691 - 20828 & NH$_3$(9,7)                                 & 20735.452 & thermal    & \citet{poynter75}\\
 3 & 20691 - 20828 & NH$_3$(7,5)                                 & 20804.830 & thermal    & \citet{poynter75}\\
 4 & 21036 - 21173 & {\bf NH$_3$(11,9)}                          & 21070.739 & both       & \citet{poynter75}\\
 4 & 21036 - 21173 & NH$_3$(4,1)                                 & 21134.311 & thermal    & \citet{poynter75}\\
 5 & 22145 - 22277 & {\bf H$_2$O (6$_{1,6}$--5$_{2,3}$)}         & 22235.080 & maser      & \citet{kukolich69}\\
 6 & 22278 - 22415 & C$_2$S (2$_1$--1$_0$)                       & 22344.030 & thermal    & \citet{kaifu87}\\
 7 & 23037 - 23174 & CH$_3$OH (9$_{2,7}$--10$_{1,10}$)A$^+$      & 23121.024 & maser (II) & \citet{mehrotra85}\\
 7 & 23037 - 23174 & NH$_3$(2,1)                                 & 23098.819 & thermal    & \citet{kukolich70}\\
 8 & 23382 - 23519 & {\bf H65$\alpha$}                           & 23404.280 & thermal    & \citet{lilley68}\\
 8 & 23382 - 23519 & {\bf CH$_3$OH (10$_{1,9}$--9$_{2,8}$)A$^-$} & 23444.778 & maser (I)  & \citet{mehrotra85}\\
 9 & 23658 - 23795 & {\bf NH$_3$(1,1)}                           & 23694.471 & thermal    & \citet{kukolich67}\\
 9 & 23658 - 23795 & {\bf NH$_3$(2,2)}                           & 23722.634 & thermal    & \citet{kukolich67}\\
10 & 23796 - 23933 & {\bf NH$_3$(3,3)}                           & 23870.130 & both       & \citet{kukolich67}\\
11 & 24900 - 25037 & {\bf CH$_3$OH (3$_{2,1}$--3$_{1,2}$)E}      & 24928.707 & maser (I)  & \citet{mueller04}\\
11 & 24900 - 25037 & {\bf CH$_3$OH (4$_{2,2}$--4$_{1,3}$)E}      & 24933.468 & maser (I)  & \citet{gaines74}\\
11 & 24900 - 25037 & {\bf CH$_3$OH (2$_{2,0}$--2$_{1,1}$)E}      & 24934.382 & maser (I)  & \citet{gaines74}\\
11 & 24900 - 25037 & {\bf CH$_3$OH (5$_{2,3}$--5$_{1,4}$)E}      & 24959.079 & maser (I)  & \citet{mehrotra85}\\
11 & 24900 - 25037 & {\bf CH$_3$OH (6$_{2,4}$--6$_{1,5}$)E}      & 25018.123 & maser (I)  & \citet{mehrotra85}\\
12 & 25038 - 25175 & {\bf CH$_3$OH (7$_{2,5}$--7$_{1,6}$)E}      & 25124.872 & maser (I)  & \citet{mehrotra85}\\
12 & 25038 - 25175 & {\bf NH$_3$(6,6)}                           & 25056.025 & both       & \citet{kukolich70}\\
13 & 26556 - 26693 & {\bf HC$_5$N (10--9)}                       & 26626.533 & thermal    & \citet{jennings82}\\
14 & 26832 - 26969 & {\bf H62$\alpha$}                           & 26939.170 & thermal    & \citet{lilley68}\\
14 & 26832 - 26969 & CH$_3$OH (12$_{2,10}$--12$_{1,11}$)E        & 26847.205 & maser (I)  & \citet{mueller04}\\
15 & 27246 - 27383 & {\bf HCCCN (3--2)}                        & 27294.078 & thermal    & \citet{lafferty78}\\
16 & 27384 - 27521 & {\bf NH$_3$(9,9)}                           & 27477.943 & thermal    & \citet{poynter75}\\
16 & 27384 - 27521 & CH$_3$OH (13$_{2,11}$--13$_{1,12}$)E        & 27472.501 & maser (I)  & \citet{mueller04}\\
\hline
\end{tabular}
\medskip\\
$^1$CH$_3$OH masers are identified as either Class I or II.
\end{table*}

For the \water~maser observations, two contiguous bands (5 and 6) can be
used together to search for masers in the velocity range of $-$2424 to
+1216\,\kms. At the \water~maser frequency, each channel corresponds to
0.45\,\kms.

\subsection{Mopra beam shape}
\begin{figure}
\includegraphics[width=0.48\textwidth]{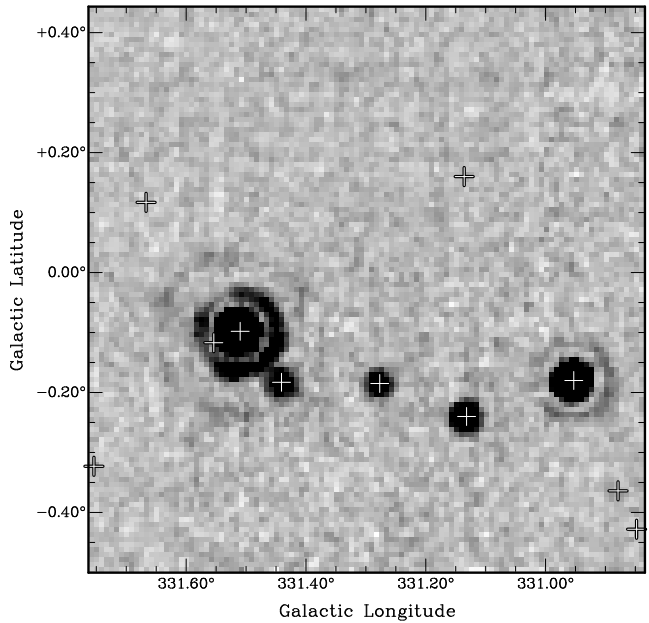}
\caption{Water maser emission in one spectral channel at $-$89.02\,km\,s$^{-1}$.
Two strong masers can be seen - G331.51-0.09 (left) and G330.96-0.18 (right) - which both
show diffraction rings. Plus symbols indicate the positions of identified water
masers. Since the image is only a single channel of the data cube, not all
masers are seen in the image, as they do not all emit at this velocity.}
\label{beam_image}
\end{figure}

Bright water masers can be used to assess the shape of the Mopra beam
at 22\,GHz. However, due to their highly variable nature, water masers are
not used to determine the efficiency of the telescope.
The bright water maser G331.51-0.09 was mapped as part of HOPS. The strongest
emission is found in the spectral channel at velocity $-89.02$\,km\,s$^{-1}$
and is shown in Figure \ref{beam_image}. This Figure
shows two diffraction rings around the maser peak. We note that a secondary
water maser site (G331.44-0.19) coincides with the outer ring from
G331.51-0.09 and can be seen in Figure \ref{beam_image}. There is also
a maser (G331.56-0.12) that is located close to the inner ring, but the
maser emission is not significant at this radial velocity and is thus not
seen in the Figure. Due to the confusion with G331.44-0.19, we do not fit
the rings around G331.51-0.09. Instead, the maser G330.96-0.18, shown in Figure
\ref{beam_image} is used, which also shows the inner beam ring and does not
appear to have any other confusing masers nearby.

\begin{figure}
\includegraphics[width=0.48\textwidth]{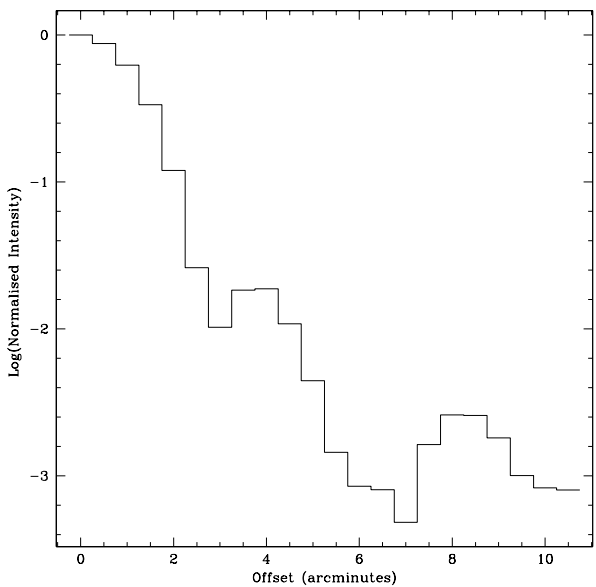}
\caption{Azimuthally averaged beam profile around the water maser G330.96-0.18.
The intensity scale has been normalised to the brightest pixel at the centre
and is shown with a log scale on the y-axis.}
\label{beam_profile}
\end{figure}

In Figure \ref{beam_profile}, we show an azimuthally averaged histogram for
G330.96-0.09. The histogram shows both the inner and outer beam rings.
The inner ring occurs
at a radius of about 3.8\arcmin~and contains approximately 14\% of the flux
of the main peak. The outer ring occurs at about 8.3\arcmin~and contains
approximately 3.5\% of the flux of the main peak. We estimate the FWHM of
the beam from the radial profile to be 2.2\arcmin.

Recent observations by \citet{urquhart10} of the water masers in Orion-KL
have been used to characterise the Mopra radiotelescope. We find that the
positions and intensities of the rings, as well as the size of the beam that
we derive above agree well with their results.

\subsection{Mopra efficiency and stability}
We adopt values from \citet{urquhart10} for
the main beam size (2.2\arcmin~at 22.2\,GHz), efficiency (0.54 at 22.2\,GHz)
and Jy\,K$^{-1}$ conversion factor (12.5 at 22.2\,GHz).

During the course of the HOPS observations, we regularly performed observations
of a number of well known line sources in the sky, including Orion. These
observations were typically conducted once each night and consisted of a
position-switch with 2 minutes on-source integration. The standard HOPS
zoom configuration was used (Table \ref{tab1}). The purpose of these observations
is to assess the stability of the telescope system over time and through different
observing conditions. Whilst the zoom configuration includes strong maser lines,
especially the \water~maser line, we did not use this line for our calibration
as the intensity is known to vary with time \citep{felli07}. Instead, we found the
strong radio recombination emission lines (H69$\alpha$ and H62$\alpha$) well suited
for calibration, as they are not expected to significantly vary in intensity over
the timescale of our observations.

We measure the integrated intensity of the lines by fitting the spectrum with a
single gaussian curve and determining the area under that curve, using standard
routines in the ASAP software package\footnote{The ATNF Spectral Analysis Package;
http://svn.atnf.csiro.au/trac/asap}.

\begin{figure*}
\includegraphics[width=0.9\textwidth]{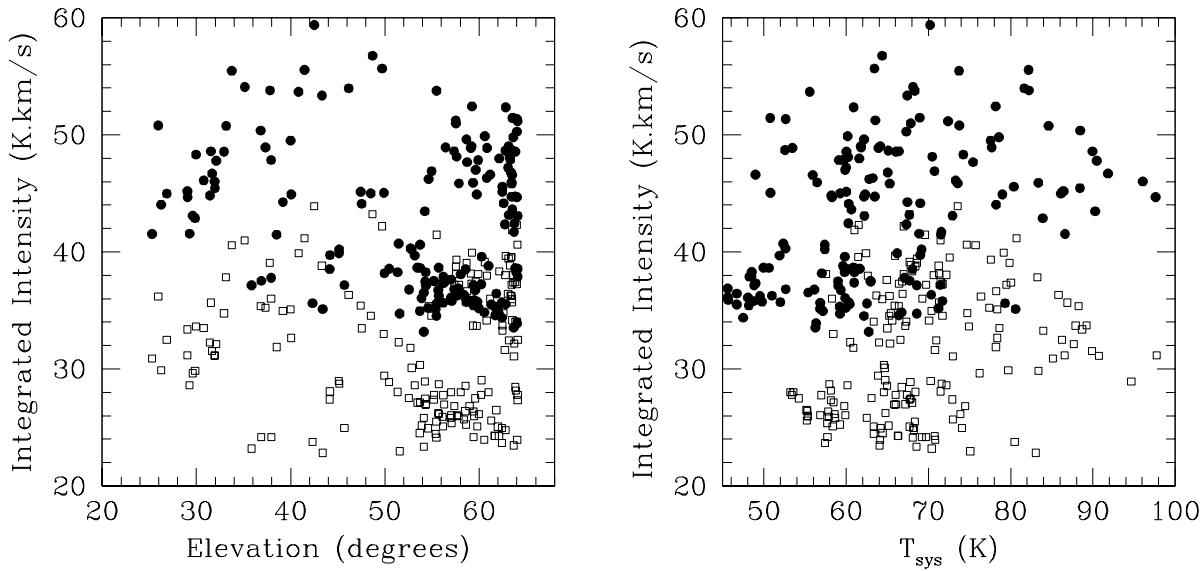}
\caption{Distribution of integrated intensities for the H69$\alpha$ line (filled circles)
and H62$\alpha$ line (open squares) in Orion, plotted against Elevation (left) and system
temperature (right). The distributions show there is a spread of integrated
intensities, implying that there is an inherent uncertainty in the HOPS absolute
flux scale of about 30 per cent.}
\label{rrlplot}
\end{figure*}

Figure \ref{rrlplot} shows the distribution of measured integrated intensities
for both radio recombination lines throughout the observations. Filled circles
indicate integrated intensities for the H69$\alpha$ line and open squares indicate
integrated intensities for the H62$\alpha$ line. We might expect that if there is
significant atmospheric attenuation, then the integrated intesities might be
correlated with elevation and/or system temperature. However, Figure \ref{rrlplot}
shows that the distribution of integrated intensities for each line does not
appear to correlate with either elevation or system temperature. But Figure
\ref{rrlplot} does show a scatter of integrated intensities. We calculate the standard
deviation of integrated intensities is 15 per cent for H69$\alpha$ and 16 per cent
for H62$\alpha$. Thus, about 95 per cent (two standard deviations) of all integrated
intensities are within 30 per cent of the mean integrated intesity. We assign
30 per cent as the uncertainty in the intensities of the HOPS data. Since the
radio recombination line emission in Orion is extended, it is possible that the actual
uncertainty is less then this as some of the scatter in Figure \ref{rrlplot} may
be due to small pointing errors of the telescope. Thus we consider 30 per cent
as an upper limit.

\begin{figure}
\includegraphics[width=0.48\textwidth]{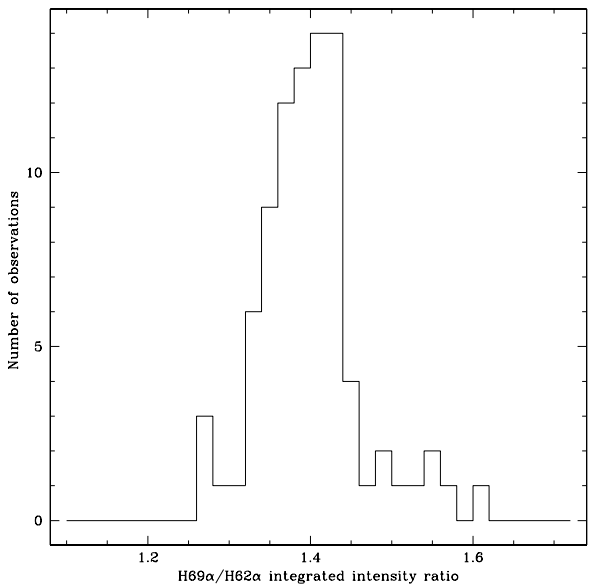}
\caption{Distribution of the ratio of H69$\alpha$/H62$\alpha$ integrated intensities
in Orion. The ratio has a mean of 1.37 and standard deviation of 0.07, showing that the ratio
of intensities is well defined.}
\label{rrlratio}
\end{figure}

Figure \ref{rrlplot} does show a significant difference in the integrated intensities
of the two radio recombination lines, with the H62$\alpha$ line weaker than the H69$\alpha$
line, on average. Figure \ref{rrlratio} shows the distribution of the ratio of
H69$\alpha$/H62$\alpha$ integrated intensities for simultaneous observations.
The mean of this distribution is 1.37, with a standard deviation of 0.07. This shows that
the ratio of intensities appears stable over time. Therefore, whilst the absolute
intensity scale may be uncertain by a factor of 30 per cent, the relative intensity
scale of different spectral lines measured using simultaneous observations is likely
to be no worse than 5 per cent. We note that the higher intensity of the H69$\alpha$ line,
compared to the H62$\alpha$ line is most likely because the beam at 19.6\,GHz (the
H69$\alpha$ line frequency) is larger than at 26.9\,GHz (the H62$\alpha$ line frequency)
and encompassess more extended radio recombination line emission.

\section{Survey Design}

Previous observations of the Galactic plane in continuum \citep{schuller09},
thermal line \citep{jackson06,dame01} and methanol maser emission \citep{walsh98,green09} indicate that
molecular material is concentrated in a region within about 30$^\circ$ in
Galactic longitude of the
Galactic centre. The survey region for HOPS was tailored to cover the bulk
of this emission, where the nominal survey region covered Galactic longitudes
290$^\circ < l < 360^\circ$ and 0$^\circ < l < 30^\circ$. The bulk of methanol masers \citep{walsh98,
green09} appear to be confined to within half a degree of the Galactic
Plane and \water~and methanol masers are commonly found in the same star
forming regions, so the survey covered nominal Galactic latitudes from
$-0.5^\circ < b < +0.5^\circ$. Our choice of survey region was also based on the
amount of observing time that would be reasonably allocated in order
to complete the project.

Based on our pilot observations \citep{walsh08}, we found the most suitable
method was to divide the survey region into blocks of
0.5$^\circ$ by 0.5$^\circ$.
Each block was observed twice in the on-the-fly (OTF) mapping mode, scanning
once in Galactic longitude and once in Galactic latitude.
Adjacent scans were separated by 51\arcsec, to give Nyquist
sampling of the beam at the highest observing frequency (27.5\,GHz). At lower frequencies,
the observations consequently oversample the mapped
region. The scanning rate was 15\arcsec\,s$^{-1}$ and spectra
were stored every 2 seconds, giving a 30\arcsec~spacing
between spectra in each row.
Each block requires about 2 hours of continuous
observations to complete. Thus, each square degree of sky requires
approximately 8 hours to be fully observed. We calculate that each position
was observed for effectively one minute on-source, using this method.

The positions of reference observations for the 0.5$^\circ$ by 0.5$^\circ$ blocks
were chosen to typically lie at b=+0.6$^\circ$ or b=-0.6$^\circ$, close to one
corner of each block. Each reference position was initially checked for emission
with a position switch observation, with 2 minutes on-source integration time.
Any potential reference position showing emission was discarded and a new reference
position was chosen and checked in the same manner.

In order to minimise variations in sensitivity across the survey region,
observations were generally limited to elevations greater than 30$^\circ$,
although observations were occasionally made at slightly lower elevations
during exceptionally good
weather conditions. During good weather and at high elevation, typical
system temperatures were 50\,K at 19.5\,GHz and 65\,K at 27.5\,GHz. We discarded
all data with a system temperature of over 120\,K. This
upper limit was usually reached when observing through average weather
conditions close to an elevation of 30$^\circ$ or during poor weather
conditions. We also discarded data where the system temperature varied quickly
due to clouds passing through the telescope beam. Such variations were
typically 20\,K or greater and had the effect of producing striped features
in the reduced data.

Pointing observations were conducted at the start of the
survey at a variety of positions across the sky. We found that the global
pointing model of Mopra was never more than 15 arcseconds from the true
position. We estimate that for the majority of HOPS observations the pointing
was better than 10 arcseconds, which is less than 10\% of the beam width.
Consequently, regular pointing updates were not used, in favour of the global
pointing model.

\section{Data reduction}
Data were initially reduced using {\sc livedata} and {\sc gridzilla}\footnote{Developed by
Mark Calabretta: http://www.atnf.csiro.au/people/mcalabre/livedata.html},
which are both AIPS++ packages written for the Parkes radiotelescope and
adapted for Mopra.  {\sc Livedata} performs a bandpass calibration for
each row along the scanning direction, using the preceding reference scan. A first order polynomial
(ie. a straight line) is fit to the baseline and then subtracted. In order
to minimise the effects of noisy channels at the start and end of each
zoom window, the first and last 150 channels were masked before performing
the baseline fitting.

{\sc Gridzilla} regrids and combines the data from multiple scanning directions
onto a data cube. We chose to regrid onto pixels of 30 $\times$ 30 arcseconds.
Data were automatically weighted according to their system temperature and data
with system temperatures above 120\,K were discarded.
Data cubes were then processed in {\sc miriad} to smooth and/or bin the data,
before the creation of peak temperature maps, which were used to find sources
of emission. Peak temperature maps are created in {\sc miriad}, using the
``moment=-2'' option in the task {\sc moment}. Peak temperature maps are
two-dimensional maps similar to moment zero, (integrated intensity) maps
more commonly used
in radio astronomy, but each pixel in the two-dimensional peak temperature
map corresponds to the highest intensity pixel of the spectrum from the
full data cube, at each position on the sky. We found that the peak temperature
maps were more sensitive than moment zero maps in detecting weak emission,
especially with imperfect baselines and wide bandwidths that were features
of the HOPS data.

\section{Source finding}
We employed a variety of methods to detect \water~maser emission.
These same methods were used to detect
other rarer masers (see \S\ref{othermasers}). Initially, we
smoothed the data with a 90 $\times$ 90 arcsecond two-dimensional Gaussian
kernel. We found that this size effectively reduced the noise by about 15\%,
compared to the maser emission. This kernel is slightly smaller than the size
of the beam (132\arcsec), however, we found that a smoothing kernel the
size of the beam produced maps that were not as sensitive to identifying
weak masers close to the noise level.

After smoothing, the data were converted to peak temperature maps. These maps
were visually inspected for maser candidates. Masers were initially identified
by inspecting the spectrum at the position of a maser candidate in the peak
temperature map and looking for a peak that contained at least two adjacent
channels where at least one channel was greater than 3 $\times$ the rms
noise level and where the sum of all emission channels was
greater than 5 $\times$ the rms noise level. These values were chosen as
they appear to give the best efficiency for detection of real features, with
little contamination. Any weak maser candidates were then followed
up with a position-switch observation, with 2 minutes on-source integration
time, which typically had rms noise levels a factor of 1.4 better than the
original map. Any maser candidate that was not seen in the followup spectrum was
considered spurious and discarded.

After identifying masers with the above method, the spatially-smoothed data
cube was then spectrally binned (along the velocity axis), with three channels
combined into each bin. A new peak temperature map was made from this 
spatially-smoothed and spectrally-binned map then
visually inspected for other masers. This method was used because some masers
show line widths significantly greater than the channel width
(0.45\,km\,s$^{-1}$). Binning the data in this way allows us to pick out weak
masers with line widths approximately 1.4\,km\,s$^{-1}$. We also binned the
data with eight channels per bin, but did not find any new detections over the
unbinned and 3-channel binned data.

We note that our method is well suited to finding weak \water~masers that
typically have large enough line widths to appear in more than one channel.
However, the method may miss some weak methanol masers, which are known to
have very narrow line widths, since the methanol maser may appear in a single
channel and be flagged as a noise spike.

\subsection{Positional Accuracy}
\label{positional}
\begin{figure}
\includegraphics[width=0.48\textwidth]{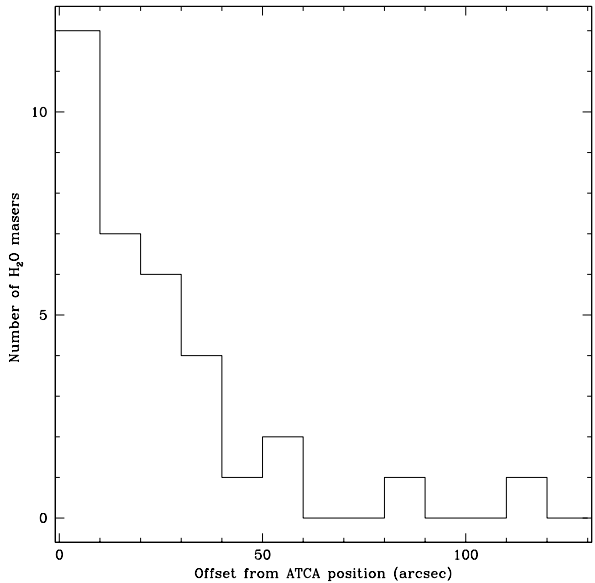}
\caption{Distribution of Mopra \water~maser offsets, compared to ATCA positions,
using \citet{breen10}. Note that there are two sites offset by more than 60\arcsec.
These are G000.53+0.18 and G337.64+0.15. The emission features in our spectra
of these sites lie close to, but not coincident with, emission features from
\citet{breen10}. Thus it is possible that both sites detected in HOPS may be
separate to those found by \citet{breen10}.}
\label{positions}
\end{figure}

The reported positions of \water~masers (see \S\ref{results}) are based
on the position of the brightest emission pixel in the data cube. Whilst
the Mopra beam is 2.2\arcmin~at 22.2\,GHz, the positions of bright masers should
be more accurate than this. Some of the detected \water~masers have been
previously observed with the ATCA \citep{breen10} to higher positional accuracy
than our Mopra observations. We can therefore compare the positions of those
masers that appear in both surveys. Figure \ref{positions} shows the distribution
of offsets for the masers we have detected with Mopra, compared to the positions
given by \citet{breen10}. We find that nearly all of our \water~maser positions
are no more than one arcminute offset from the corresponding ATCA position. We
also find that the median offset is 20\arcsec, indicating that most of the reported
positions of masers are probably no more than 20\arcsec~off the real positions.

We also note that of the 34 maser sites that overlap with \citep{breen10}, 9
(ie. 26 per cent) are identified as multiple sites by \citep{breen10} but are
not resolved by the Mopra beam in our observations. We expect that many more
maser spectra identified in this work may consist of multiple maser sites.

\section{Results}
\label{results}
Here we report results on the \water~maser emission detected in HOPS. Occasional
detections of rare masers are
reported in \S\ref{othermasers} and results from other thermal spectral lines
such as NH$_3$, radio recombination lines and HCCCN (3--2) will be reported in later
papers.

\water~maser emission was found toward 540 distinct sites in the region observed in
HOPS. The
properties of these sites are summarised in Table \ref{330masers}. The source name
(column 1) is derived from the Galactic coordinates. Columns 2 and 3 list
the coordinates of the site. Peak
flux density, velocity and FWHM of the strongest maser spot in the spectrum are
listed in columns 4, 5 and 6, respectively. The minimum
and maximum velocities over which emission has been detected are given in columns 7 and
8, respectively. The last column lists
whether a maser is a new detection or previously reported. Reference numbers to previously
reported masers in the last
column are the first detection of the maser and are given in Table \ref{refs}. Spectra of
the masers are presented in Figure \ref{spectra}. The strongest maser
(G27.18$-$0.08) has a peak flux density of 3933\,Jy and the weakest maser
(G305.56$+$0.01) has a peak flux density of 0.7\,Jy. Note that the weakest maser was
detected because it shows a broad line FWHM (5.0\,km\,s$^{-1}$). The median peak flux
density is 11.4\,Jy. We detect 62 masers (9\%) with peak flux densities over
100\,Jy. 334 of the detected \water~maser sites are new detections (62\%).
Note that since the Mopra beam for these observations is 2\arcmin~it is likely
that many of the maser spectra are composed of multiple maser sites, so that 540
maser sites is almost certainly a lower limit.

\begin{table*}
  \caption{Properties of detected water masers. Column 1 lists the source name,
derived from the Galactic coordinates. Right ascension and declination coordinates are
given in columns 2 and 3, respectively. Properties
of the strongest maser spot are given in columns 4, 5 and 6: column 4 lists
the peak flux density, column 5 lists the line centre velocity and column 6 lists
the line FWHM of the brightest maser spot. For maser sites with multiple maser spots,
the minimum and maximum velocity of the maser spots are given in columns 7 and 8, respectively.
Otherwise, for maser sites with only a single maser spot, no velocity range is given.
Column 9 lists whether the maser site is a
new detection (``NEW'') or a previous detection. Previous detections are listed as
numbers which refer to Table \ref{refs}. If a previous detection also identifies
the maser site with an evolved star, this is also noted. If a previously known
evolved star is associated with a newly discovered \water~maser, then `NEW' is
written, as well as the reference number for the evolved star association.}
\label{330masers}
\begin{tabular}{ccccccccl}
  \hline
   Source    &     RA      &      Dec          & \multicolumn{3}{c}{Strongest Maser Spot} & \multicolumn{2}{c}{Velocity Range} & Comments$^1$\\
    Name     &   (J2000)   &    (J2000)        & Peak Flux    & Velocity     & FWHM       &  Min  &   Max  & \\
             &   (h m s)   &($^\circ$~\arcmin~\arcsec)& Density (Jy) &(km\,s$^{-1}$)&(km\,s$^{-1}$)& \multicolumn{2}{c}{(km\,s$^{-1}$)}& \\
  \hline
G0.05$-$0.22 &  17 46 35.9 &    -29 00 15      &    7.4       &    12.4      &    0.8     &   11.7 & 14.7  & 1\\
G0.31$-$0.18 &  17 47 01.6 &    -28 46 00      &   39.2       &    10.6      &    0.8     &   -6.3 & 21.4  & 2\\
G0.31$-$0.21 &  17 47 10.3 &    -28 46 48      &    6.6       &    21.5      &    0.9     &   10.3 & 22.4  & 2\\
G0.34$+$0.09 &  17 46 04.0 &    -28 36 04      &   36.6       &     7.1      &    0.6     &          &     & 1\\
G0.37$-$0.17 &  17 47 09.7 &    -28 42 22      &   22.4       &    -9.1      &    0.9     &   -9.7 & -5.2  & 1\\
\hline
\end{tabular}
\begin{flushleft}
$^1$``NEW'' indicates a newly detected maser, otherwise a number (refer to Table \ref{refs}) is given
for a previous detection. The majority of masers are presumed to be associated with star formation.
Previously known associations with evolved stars are referenced, but we await higher precision positions to
associate the majority of masers.
\end{flushleft}
\end{table*}

\begin{table*}
  \contcaption{...}
  \begin{tabular}{ccccccccl}
  \hline
   Source     &     RA      &      Dec          & \multicolumn{3}{c}{Strongest Maser Spot} & \multicolumn{2}{c}{Velocity Range} & Comments$^1$\\
    Name      &   (J2000)   &    (J2000)        & Peak Flux    & Velocity     & FWHM       &  Min  &   Max  & \\
              &   (h m s)   &($^\circ$~\arcmin~\arcsec)& Density (Jy) &(km\,s$^{-1}$)&(km\,s$^{-1}$)& \multicolumn{2}{c}{(km\,s$^{-1}$)}& \\
  \hline
G8.67$-$0.36  &  18 06 19.7 &    -21 37 30      &   29.8       &    24.7      &    0.7     &   -5.5 & 44.8  & 12\\
G8.70$-$0.43  &  18 06 38.5 &    -21 37 57      &   4.9        &    32.8      &    1.3     &        &       & NEW\\
G8.73$-$0.39  &  18 06 34.2 &    -21 35 40      &   6.0        &    46.7      &    0.7     &   30.4 & 47.4  & NEW\\
G8.83$-$0.03  &  18 05 25.7 &    -21 19 24      &   38.9       &    -9.5      &    1.7     &  -13.3 & -7.2  & 7\\
G8.93$-$0.02  &  18 05 35.6 &    -21 13 44      &   88.9       &     7.0      &    0.9     &    5.7 & 49.2  & 13 - M supergiant\\
\hline
\end{tabular}
\medskip\\
\end{table*}

\begin{table*}
  \contcaption{...}
  \begin{tabular}{ccccccccl}
  \hline
   Source    &     RA      &      Dec          & \multicolumn{3}{c}{Strongest Maser Spot} & \multicolumn{2}{c}{Velocity Range} &Comments$^1$\\
    Name     &   (J2000)   &    (J2000)        & Peak Flux    & Velocity     & FWHM       &  Min  &   Max  & \\
             &   (h m s)   &($^\circ$~\arcmin~\arcsec)& Density (Jy) &(km\,s$^{-1}$)&(km\,s$^{-1}$)& \multicolumn{2}{c}{(km\,s$^{-1}$)}& \\
  \hline
G18.55$+$0.04 & 18 24 38.0 &    -12 45 21      &   10.7       &    32.2      &    0.5     &   31.3 & 35.2  & NEW\\
G18.61$-$0.08 & 18 25 10.3 &    -12 45 30      &    6.0       &    41.3      &    1.2     &   40.0 & 52.4  & NEW\\
G18.73$-$0.22 & 18 25 55.3 &    -12 42 49      &   28.5       &    36.7      &    1.3     &   32.2 & 46.3  & NEW\\
G18.88$-$0.47 & 18 27 07.0 &    -12 41 55      &    7.7       &    65.1      &    1.5     &        &       & NEW\\
G19.00$-$0.23 & 18 26 27.8 &    -12 28 57      &   21.8       &   -15.9      &    7.2     &        &       & NEW\\
\hline
\end{tabular}
\medskip\\
\end{table*}

\begin{table*}
  \contcaption{...}
  \begin{tabular}{ccccccccl}
  \hline
   Source    &     RA      &      Dec          & \multicolumn{3}{c}{Strongest Maser Spot} & \multicolumn{2}{c}{Velocity Range} & Comments$^1$\\
    Name     &   (J2000)   &    (J2000)        & Peak Flux    & Velocity     & FWHM       &  Min  &   Max  & \\
             &   (h m s)   &($^\circ$~\arcmin~\arcsec)& Density (Jy) &(km\,s$^{-1}$)&(km\,s$^{-1}$)& \multicolumn{2}{c}{(km\,s$^{-1}$)} & \\
  \hline
G27.33$+$0.18  & 18 40 33.7 &    -04 54 30     &   20.0       &    33.9      &    0.9     &   33.1 & 38.4  & NEW, 29 - Carbon Star\\
G27.37$-$0.16  & 18 41 49.6 &    -05 01 24     &   25.6       &    88.2      &    2.2     &   86.1 & 98.9  & 31\\
G27.41$-$0.05  & 18 41 31.1 &    -04 56 12     &    5.0       &   108.7      &    2.0     &  100.9 & 109.9 & NEW\\
G27.50$-$0.28  & 18 42 31.1 &    -04 57 44     &    2.9       &   -15.4      &    1.3     &  -39.6 & -11.2 & NEW\\
G27.60$+$0.08  & 18 41 23.7 &    -04 42 21     &    4.1       &    35.4      &    0.9     &   23.4 & 41.6  & NEW\\
\hline
\end{tabular}
\medskip\\
\end{table*}

\begin{table*}
  \contcaption{...}
  \begin{tabular}{ccccccccl}
  \hline
   Source    &     RA      &      Dec          & \multicolumn{3}{c}{Strongest Maser Spot} & \multicolumn{2}{c}{Velocity Range} & Comments$^1$\\
    Name     &   (J2000)   &    (J2000)        & Peak Flux    & Velocity     & FWHM       &  Min  &   Max  & \\
             &   (h m s)   &($^\circ$~\arcmin~\arcsec)& Density (Jy) &(km\,s$^{-1}$)&(km\,s$^{-1}$)& \multicolumn{2}{c}{(km\,s$^{-1}$)}& \\
  \hline
G305.36$+$0.15 & 13 12 34.3 &    -62 37 13     &  457.6       &   -35.6      &    3.0     &  -46.3 & -29.7 & 42\\
G305.36$+$0.21 & 13 12 33.8 &    -62 33 33     &  170.9       &   -33.6      &    1.2     &  -90.5 & 52.7  & 43\\
G305.56$+$0.01 & 13 14 24.4 &    -62 44 48     &    0.7       &   -37.7      &    5.0     &        &       & 40\\
G305.66$-$0.10 & 13 15 24.1 &    -62 50 52     &    6.7       &   -68.9      &    1.8     &        &       & 44\\
G305.73$+$0.08 & 13 15 49.4 &    -62 39 19     &    4.1       &   -43.9      &    3.8     &        &       & 40\\
\hline
\end{tabular}
\medskip\\
\end{table*}

\begin{table*}
  \contcaption{...}
  \begin{tabular}{ccccccccl}
  \hline
   Source    &     RA      &      Dec          & \multicolumn{3}{c}{Strongest Maser Spot} & \multicolumn{2}{c}{Velocity Range} & Comments$^1$\\
    Name     &   (J2000)   &    (J2000)        & Peak Flux    & Velocity     & FWHM       &  Min  &   Max  & \\
             &   (h m s)   &($^\circ$~\arcmin~\arcsec)& Density (Jy) &(km\,s$^{-1}$)&(km\,s$^{-1}$)& \multicolumn{2}{c}{(km\,s$^{-1}$)}& \\
  \hline
G319.83$-$0.19 & 15 06 52.2 &    -58 32 47     &   16.0       &   -12.1      &    2.9     &  -15.6 & -0.5  & 38\\
G320.23$-$0.28 & 15 09 52.8 &    -58 25 31     &    5.7       &   -61.7      &    1.3     &  -67.5 & -58.9 & 38\\
G320.24$-$0.08 & 15 09 09.8 &    -58 14 55     &    3.7       &    -9.9      &    1.1     &        &       & NEW\\
G320.26$-$0.30 & 15 10 06.6 &    -58 25 38     &    7.0       &  -116.3      &    2.5     & -127.9 & -108.4& 38\\
G320.34$-$0.20 & 15 10 16.2 &    -58 17 57     &   23.8       &  -107.7      &    1.3     & -118.3 & -102.3& NEW\\
\hline
\end{tabular}
\medskip\\
\end{table*}

\begin{table*}
  \contcaption{...}
  \begin{tabular}{ccccccccl}
  \hline
   Source    &     RA      &      Dec          & \multicolumn{3}{c}{Strongest Maser Spot} & \multicolumn{2}{c}{Velocity Range} & Comments$^1$\\
    Name     &   (J2000)   &    (J2000)        & Peak Flux    & Velocity     & FWHM       &  Min  &   Max  & \\
             &   (h m s)   &($^\circ$~\arcmin~\arcsec)& Density (Jy) &(km\,s$^{-1}$)&(km\,s$^{-1}$)& \multicolumn{2}{c}{(km\,s$^{-1}$)}& \\
  \hline
G332.61$-$0.15 & 16 17 24.7 &    -50 45 18     &    4.7       &   -26.7      &    0.9     &  -29.8 & -24.1 & 2\\
G333.00$-$0.43 & 16 20 24.9 &    -50 40 37     &   11.7       &  -110.4      &    2.0     & -147.0 & -93.8 & 40\\
G333.03$-$0.45 & 16 20 37.2 &    -50 40 27     &    5.0       &   -60.1      &    1.8     &  -61.6 & -51.0 & 40\\
G333.06$-$0.48 & 16 20 55.9 &    -50 40 38     &    4.4       &     1.3      &    0.8     &  -10.9 &  2.8  & 47\\
G333.06$-$0.10 & 16 19 15.1 &    -50 24 20     &    8.8       &   -37.5      &    1.3     &  -49.4 & -32.2 & NEW\\
\hline
\end{tabular}
\medskip\\
\end{table*}

\begin{table*}
  \contcaption{...}
  \begin{tabular}{ccccccccl}
  \hline
   Source    &     RA      &      Dec          & \multicolumn{3}{c}{Strongest Maser Spot} & \multicolumn{2}{c}{Velocity Range} & Comments$^1$\\
    Name     &   (J2000)   &    (J2000)        & Peak Flux    & Velocity     & FWHM       &  Min  &   Max & \\
             &   (h m s)   &($^\circ$~\arcmin~\arcsec)& Density (Jy) &(km\,s$^{-1}$)&(km\,s$^{-1}$)& \multicolumn{2}{c}{(km\,s$^{-1}$)}& \\
  \hline
G339.16$-$0.29 & 16 45 06.1 &    -46 04 09     &    3.1       &   -19.7      &    1.1     &        &       & NEW\\
G339.20$-$0.02 & 16 44 05.8 &    -45 51 46     &   10.5       &   -14.6      &    0.7     &        &       & NEW\\
G339.48$+$0.18 & 16 44 15.7 &    -45 31 26     &    2.9       &   -90.9      &    0.8     &        &       & NEW\\
G339.52$+$0.27 & 16 44 01.2 &    -45 25 55     &    2.8       &   -14.5      &    0.7     &  -30.7 & -13.7 & NEW\\ 
G339.54$-$0.42 & 16 47 05.4 &    -45 52 11     &    1.2       &   -48.4      &    5.7     &        &       & NEW\\
\hline
\end{tabular}
\medskip\\
\end{table*}

\begin{table*}
  \contcaption{...}
  \begin{tabular}{ccccccccl}
  \hline
   Source    &     RA      &      Dec          & \multicolumn{3}{c}{Strongest Maser Spot} & \multicolumn{2}{c}{Velocity Range} & Comments$^1$\\
    Name     &   (J2000)   &    (J2000)        & Peak Flux    & Velocity     & FWHM       &  Min  &   Max  & \\
             &   (h m s)   &($^\circ$~\arcmin~\arcsec)& Density (Jy) &(km\,s$^{-1}$)&(km\,s$^{-1}$)& \multicolumn{2}{c}{(km\,s$^{-1}$)}& \\
  \hline
G346.53$-$0.12 & 17 09 33.8 &    -40 12 14     &   11.0       &   -47.4      &    0.6     &        &       & NEW\\
G347.06$+$0.00 & 17 10 43.6 &    -39 42 17     &    3.7       &    -9.7      &    1.9     &        &       & NEW\\
G347.23$+$0.02 & 17 11 10.0 &    -39 33 39     &    6.8       &   -77.8      &    1.2     &  -89.5 & -73.5 & NEW\\
G347.30$+$0.13 & 17 10 54.7 &    -39 25 54     &   42.3       &   -82.6      &    2.4     &  -98.4 & -75.9 & 46\\
G347.32$+$0.02 & 17 11 26.9 &    -39 29 12     &    3.2       &    -8.5      &    1.1     &        &       & NEW\\
\hline
\end{tabular}
\medskip\\
\end{table*}

\begin{table*}
  \contcaption{...}
  \begin{tabular}{ccccccccl}
  \hline
   Source    &     RA      &      Dec          & \multicolumn{3}{c}{Strongest Maser Spot} & \multicolumn{2}{c}{Velocity Range} & Comments$^1$\\
    Name     &   (J2000)   &    (J2000)        & Peak Flux    & Velocity     & FWHM       &  Min  &   Max  & \\
             &   (h m s)   &($^\circ$~\arcmin~\arcsec)& Density (Jy) &(km\,s$^{-1}$)&(km\,s$^{-1}$)& \multicolumn{2}{c}{(km\,s$^{-1}$)}& \\
  \hline
G359.44$-$0.10 & 17 44 41.1 &    -29 28 08     &   18.1       &   -58.4      &    2.2     &        &       & 3\\
G359.51$-$0.15 & 17 45 02.4 &    -29 25 59     &    5.6       &     7.2      &    0.8     &        &       & 53\\
G359.62$-$0.25 & 17 45 40.4 &    -29 23 35     &  106.4       &   -15.9      &    1.5     &  -25.5 & 11.6  & 3\\
G359.73$-$0.38 & 17 46 28.9 &    -29 21 56     &   42.2       &    -4.8      &    0.6     &  -14.8 & -1.6  & 1\\
G359.94$-$0.15 & 17 46 03.4 &    -29 03 57     &   10.6       &   -18.7      &    1.7     &  -20.9 & 24.7  & 55\\
G359.97$-$0.46 & 17 47 21.6 &    -29 12 08     &   26.3       &    26.8      &    0.7     &    3.9 & 27.5  & 3\\
\hline
\end{tabular}
\medskip\\
\end{table*}

\begin{table}
  \caption{References for Table \ref{330masers}.}
\label{refs}
  \begin{tabular}{cl}
  \hline
Comment in            &  Reference \\
Table \ref{330masers} &            \\
  \hline
         1            & \citet{caswell11}\\
         2            & \citet{breen10}\\
         3            & \citet{caswell83}\\
         4            & \citet{genzel77}\\
         5            & \citet{mehringer98}\\
         6            & \citet{benson90}\\
\hline
\end{tabular}
\end{table}
\clearpage

\begin{figure*}
\includegraphics[width=\textwidth]{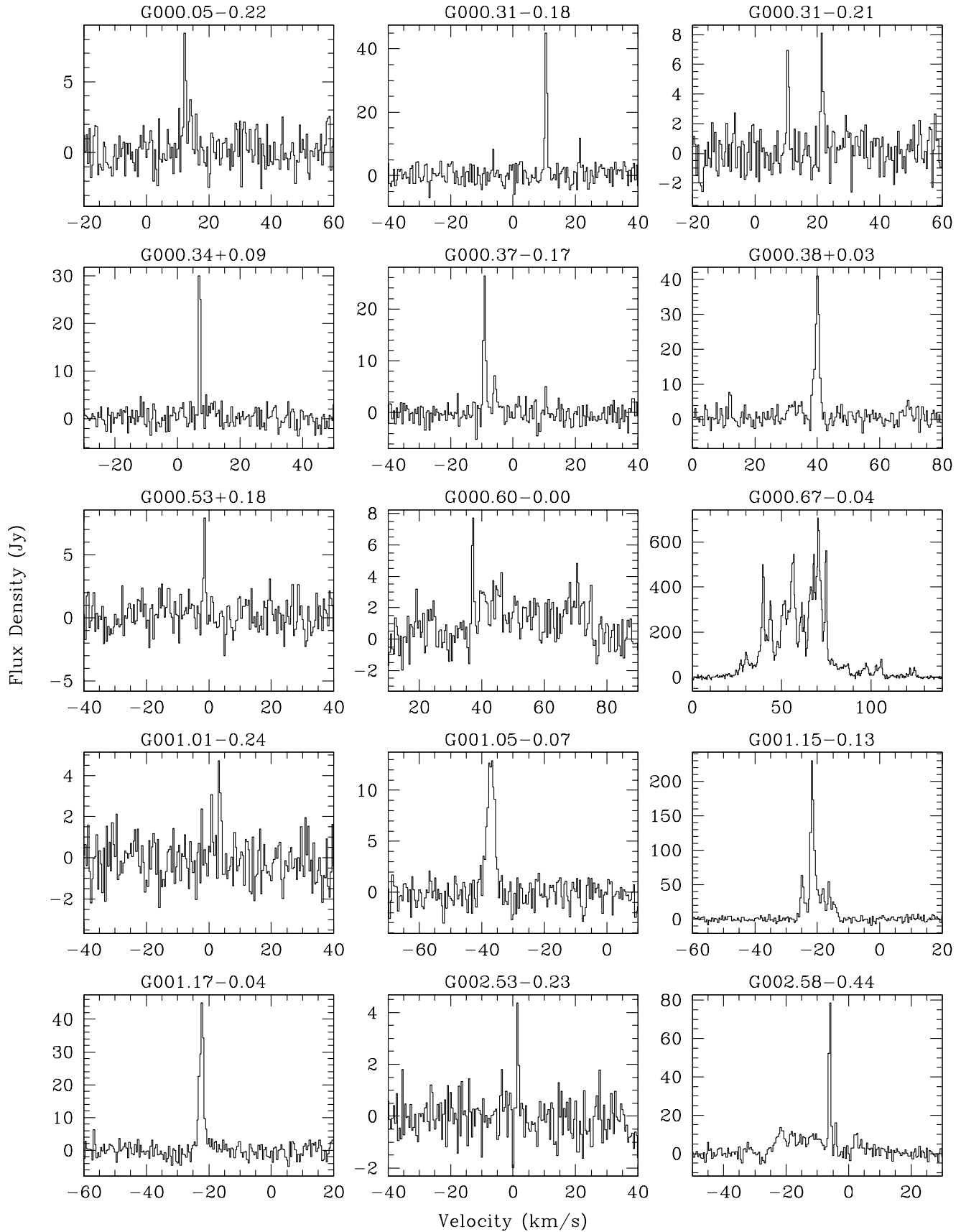}
\caption{Spectra of water masers. The source name is given
at the top of each panel.}
\label{spectra}
\end{figure*}

\begin{figure*}
\includegraphics[width=\textwidth]{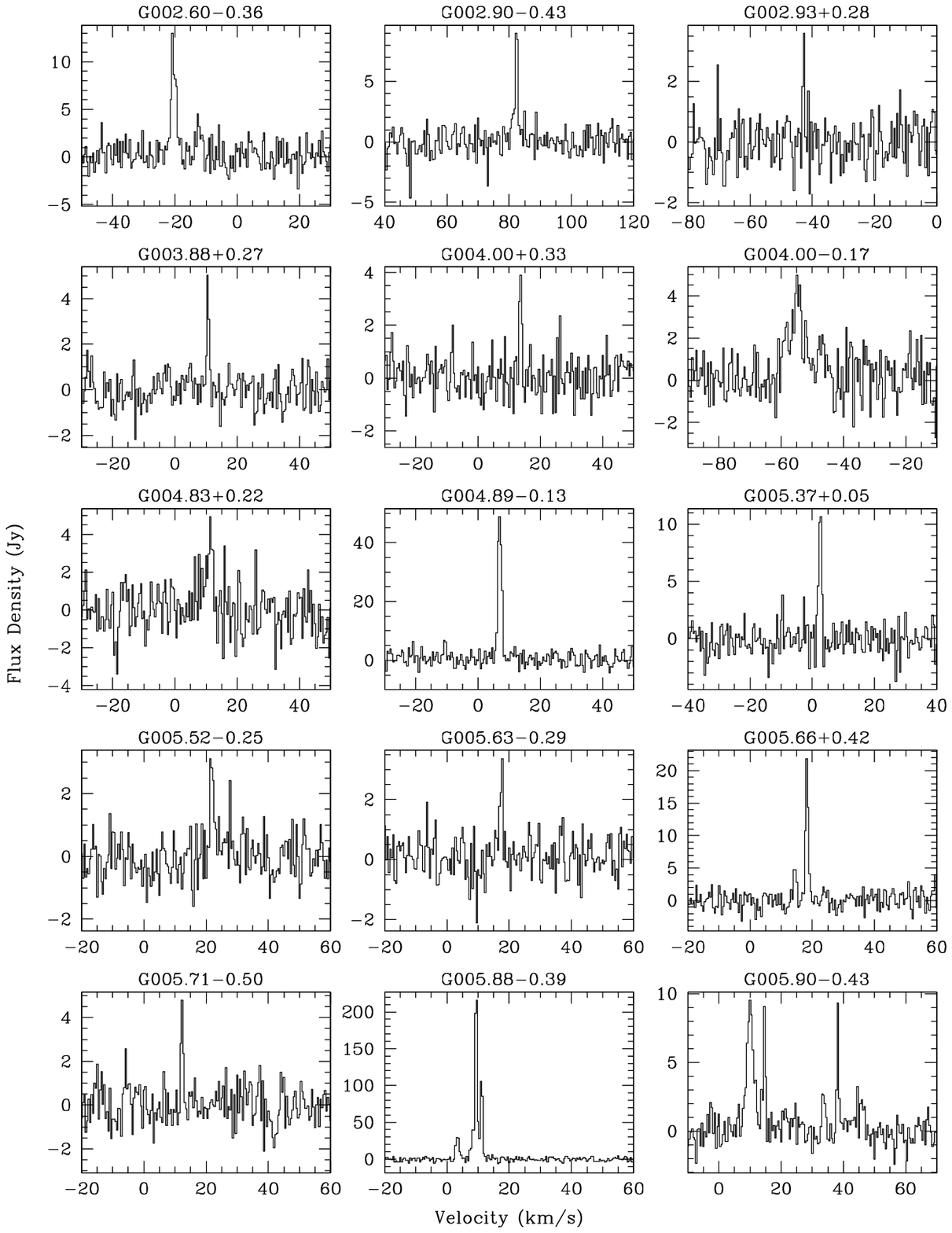}
\contcaption{...}
\end{figure*}

\begin{figure*}
\includegraphics[width=\textwidth]{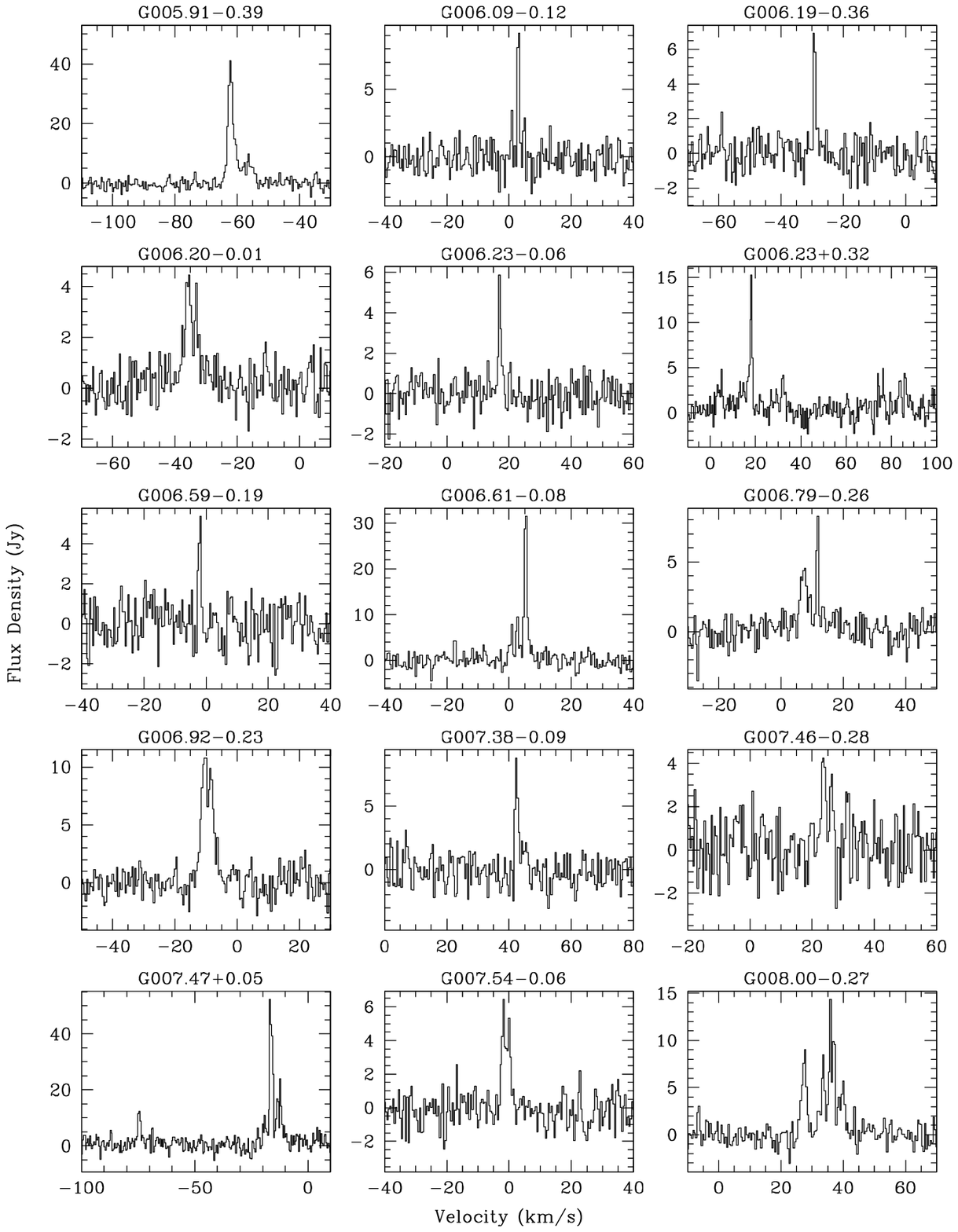}
\contcaption{...}
\end{figure*}

\begin{figure*}
\includegraphics[width=\textwidth]{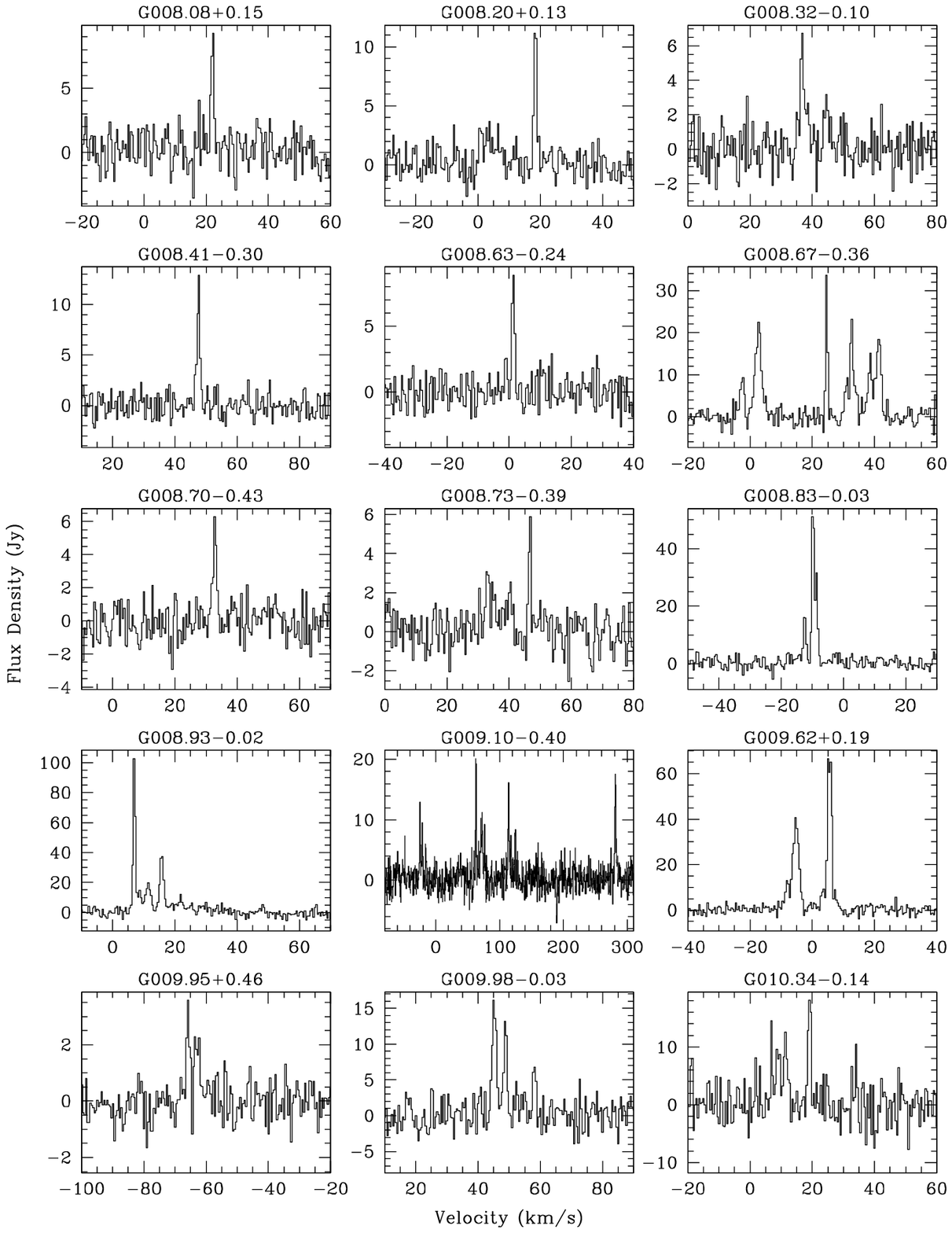}
\contcaption{...}
\end{figure*}

\begin{figure*}
\includegraphics[width=\textwidth]{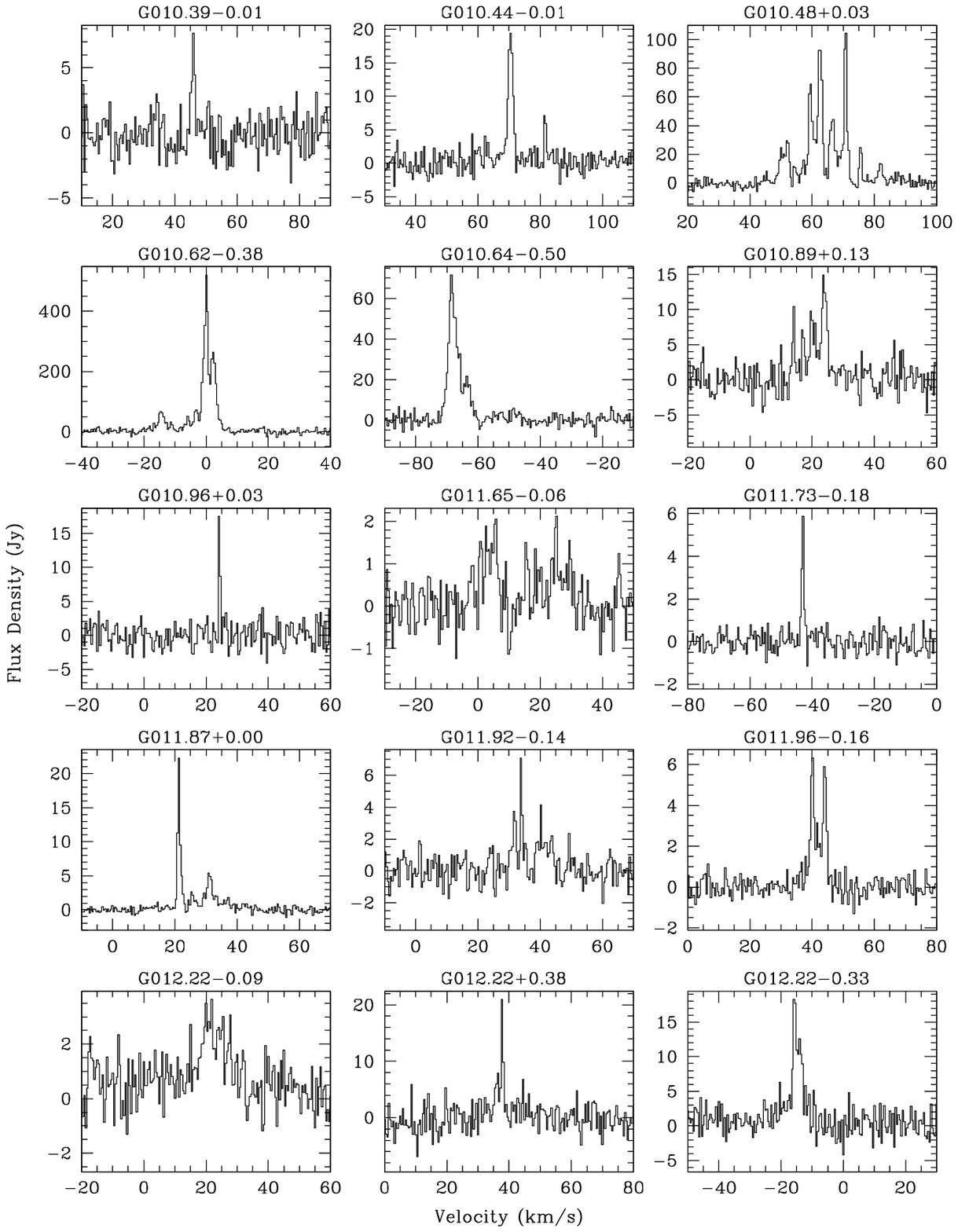}
\contcaption{...}
\end{figure*}

\begin{figure*}
\includegraphics[width=\textwidth]{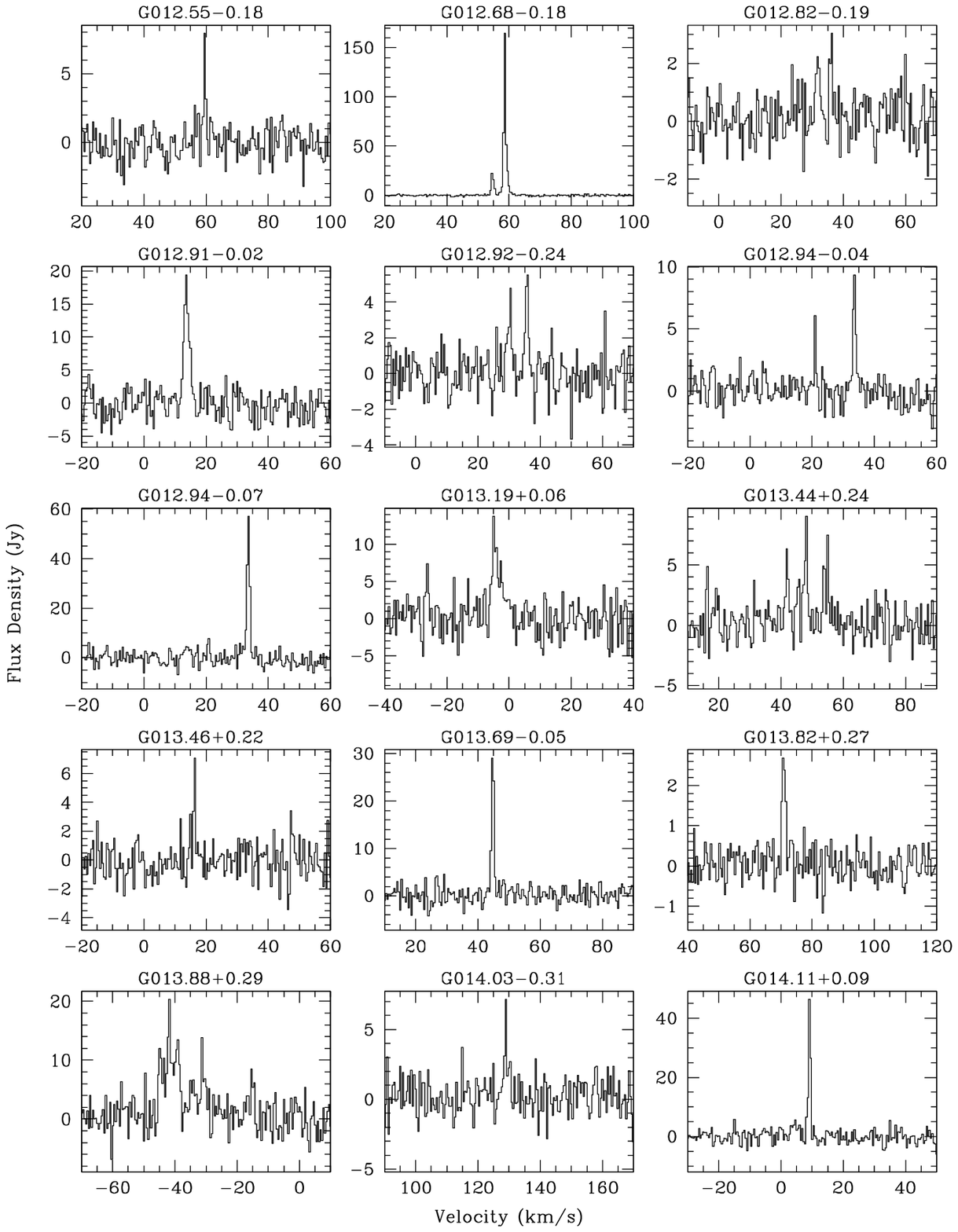}
\contcaption{...}
\end{figure*}
\clearpage

\begin{figure*}
\includegraphics[width=\textwidth]{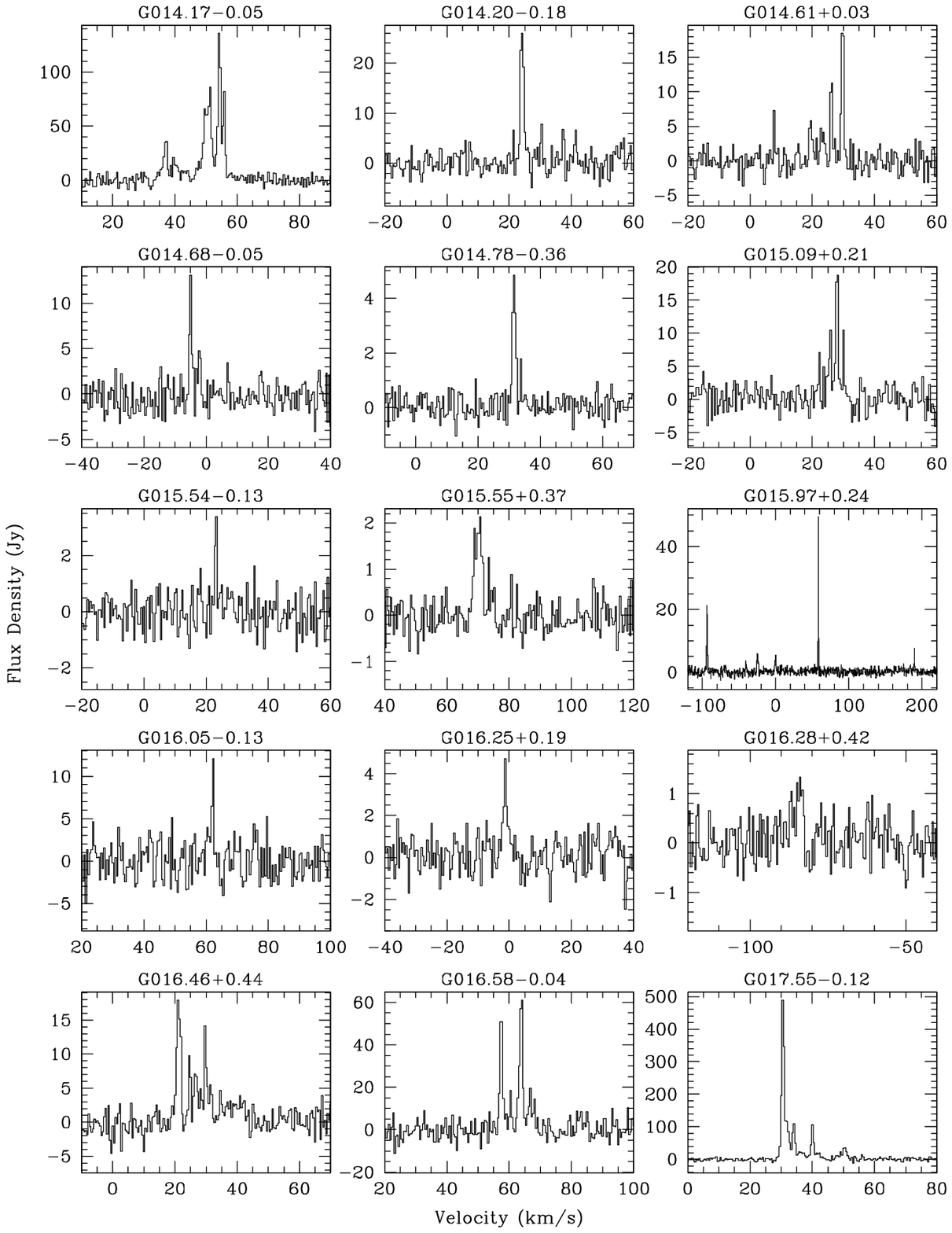}
\contcaption{...}
\end{figure*}

\begin{figure*}
\includegraphics[width=\textwidth]{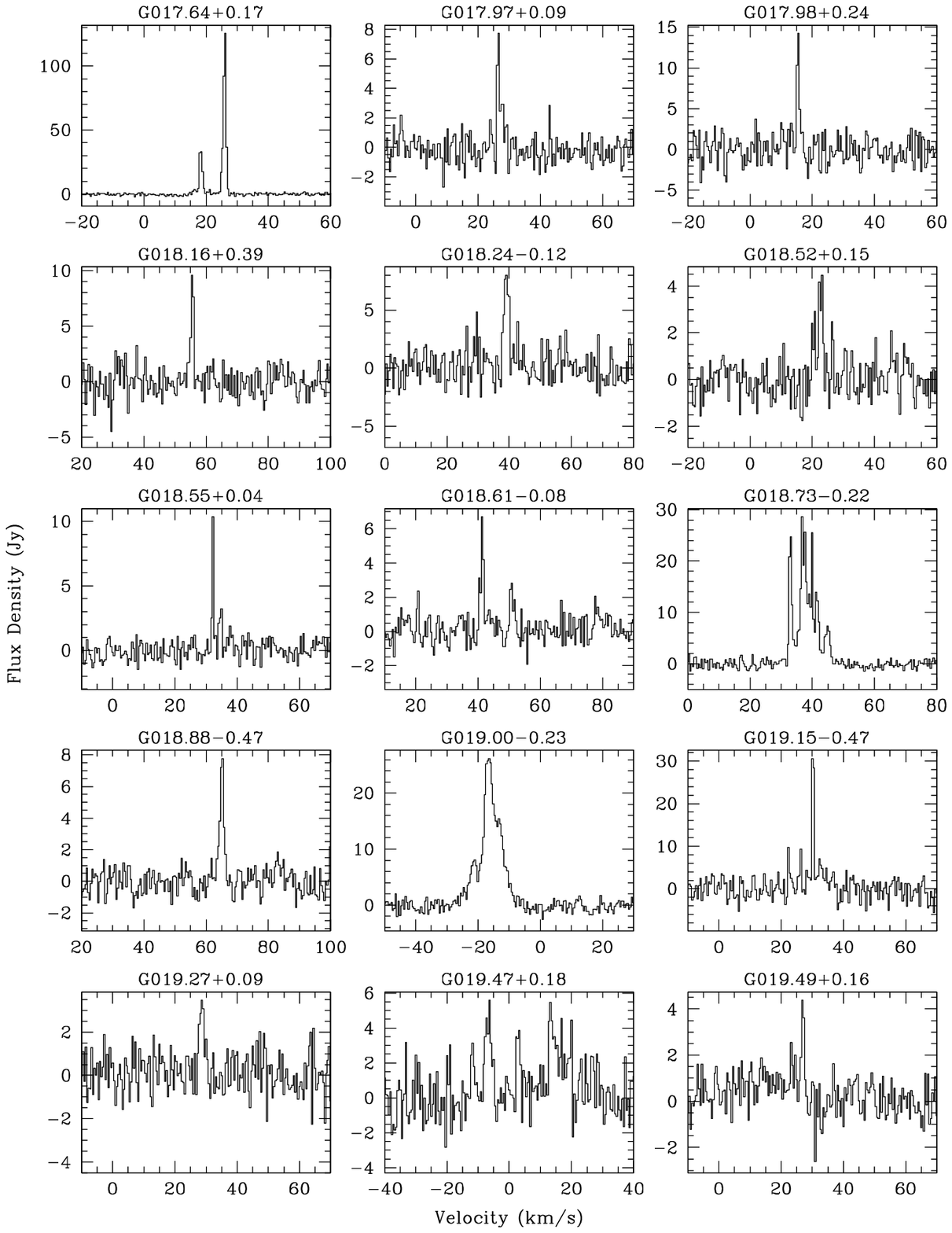}
\contcaption{...}
\end{figure*}

\begin{figure*}
\includegraphics[width=\textwidth]{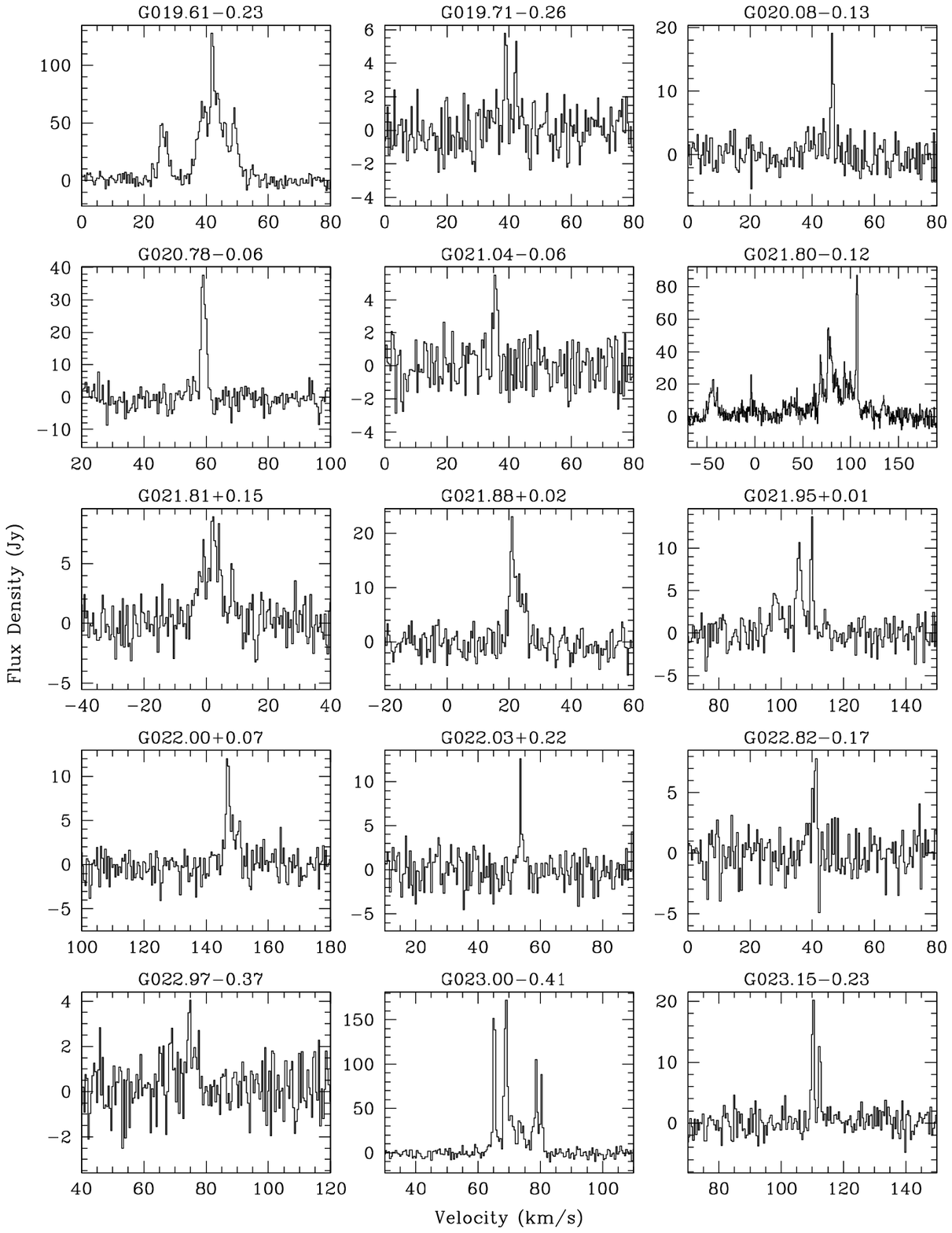}
\contcaption{...}
\end{figure*}

\clearpage
\begin{figure*}
\includegraphics[width=\textwidth]{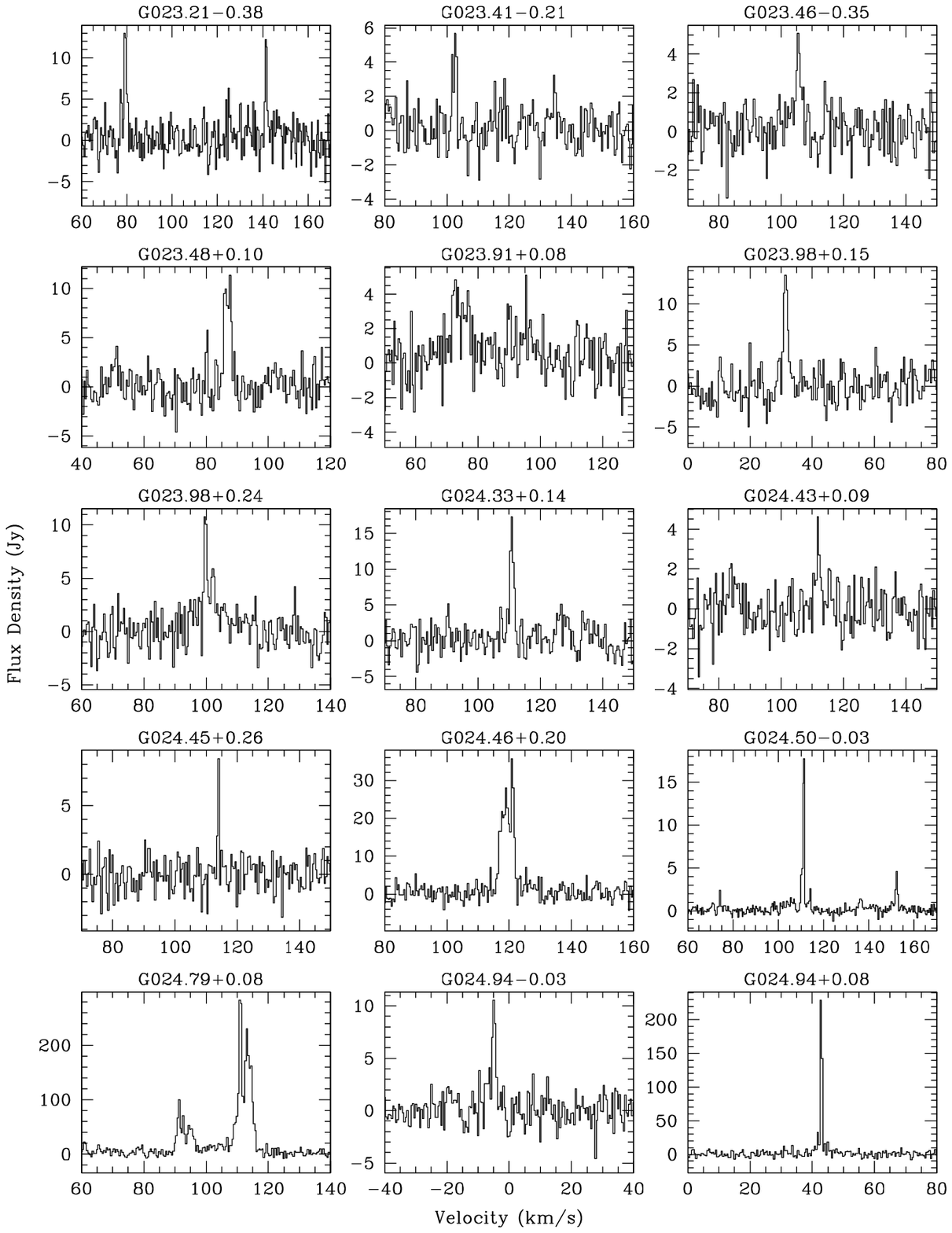}
\contcaption{...}
\end{figure*}

\begin{figure*}
\includegraphics[width=\textwidth]{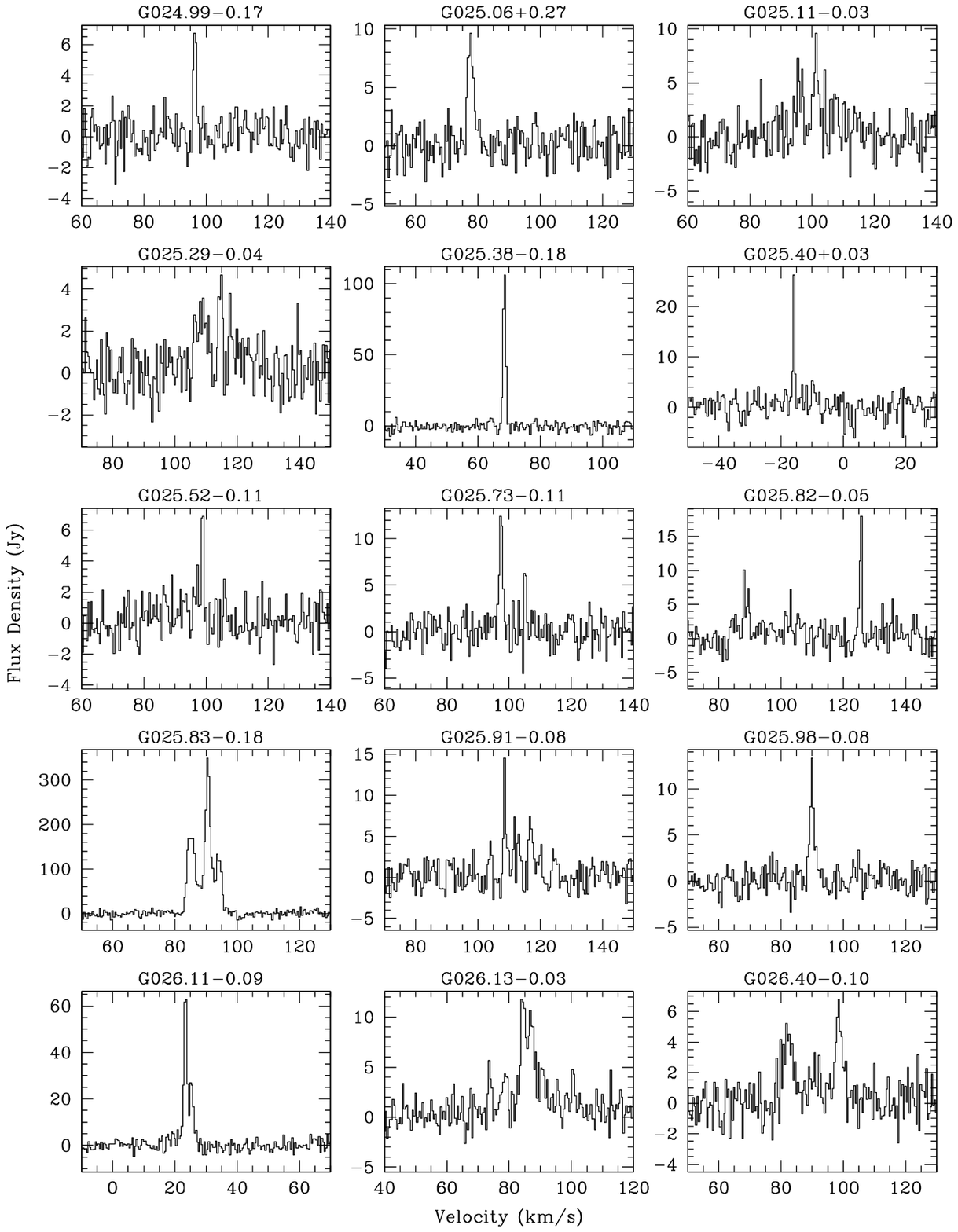}
\contcaption{...}
\end{figure*}

\begin{figure*}
\includegraphics[width=\textwidth]{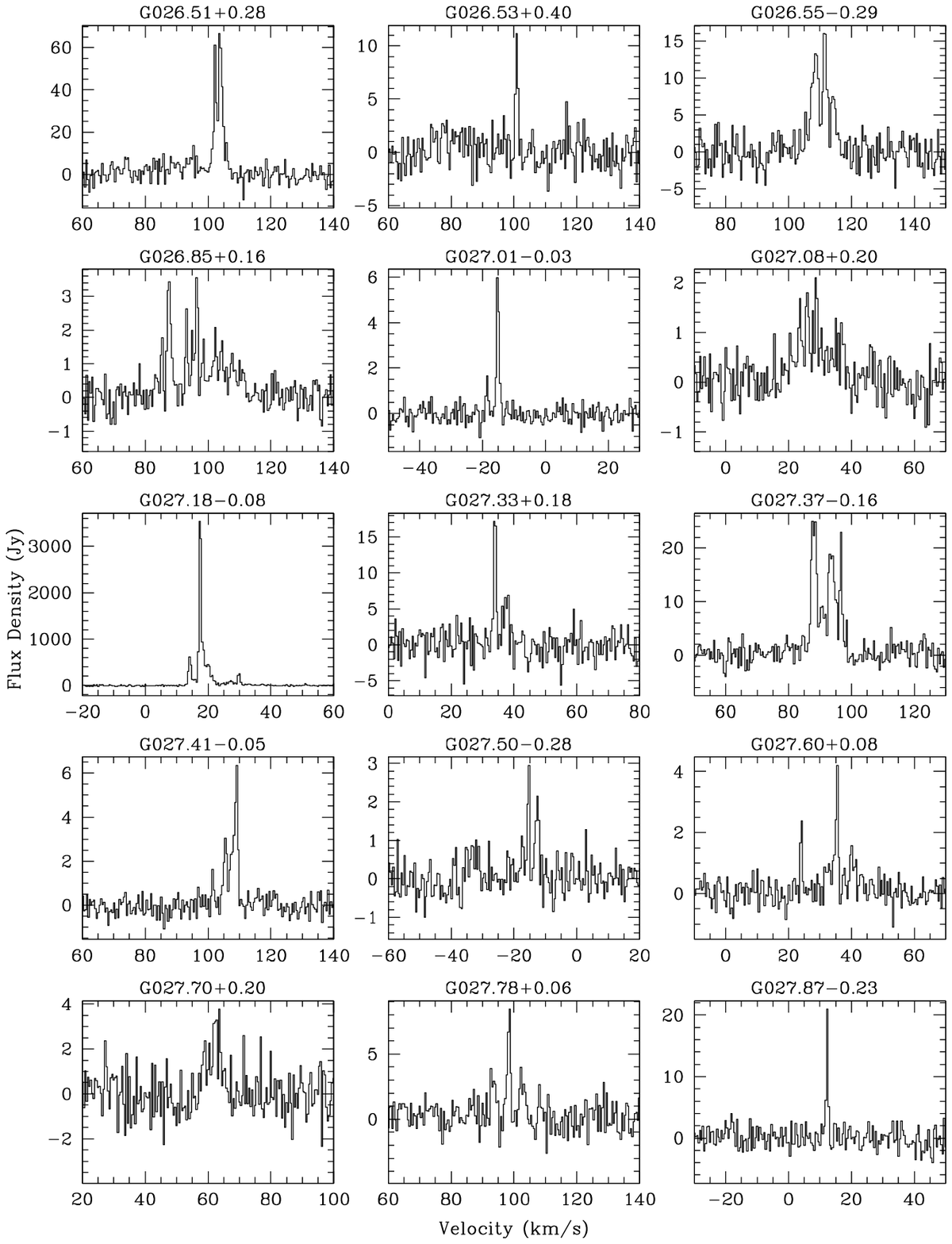}
\contcaption{...}
\end{figure*}

\clearpage
\begin{figure*}
\includegraphics[width=\textwidth]{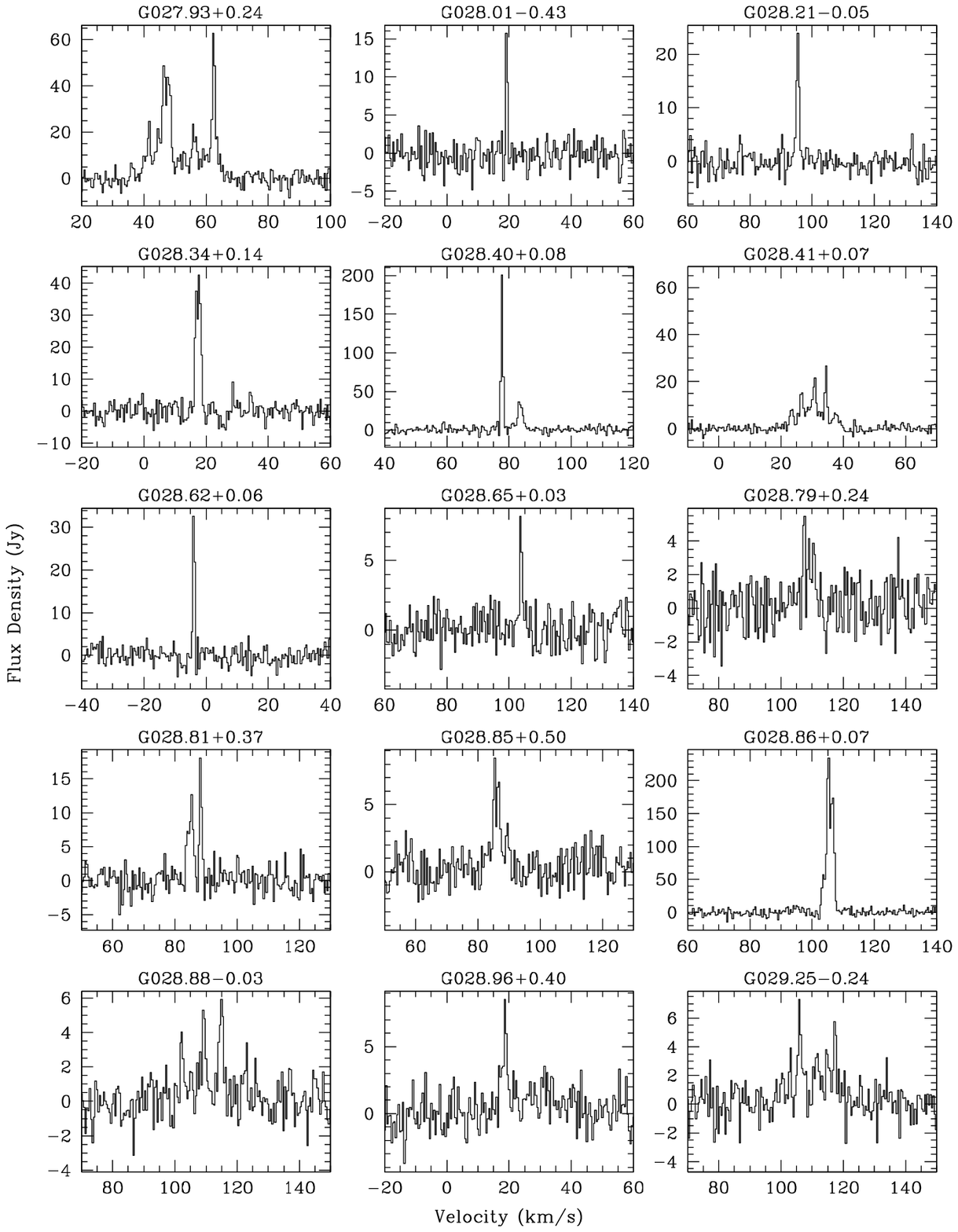}
\contcaption{...}
\end{figure*}

\begin{figure*}
\includegraphics[width=\textwidth]{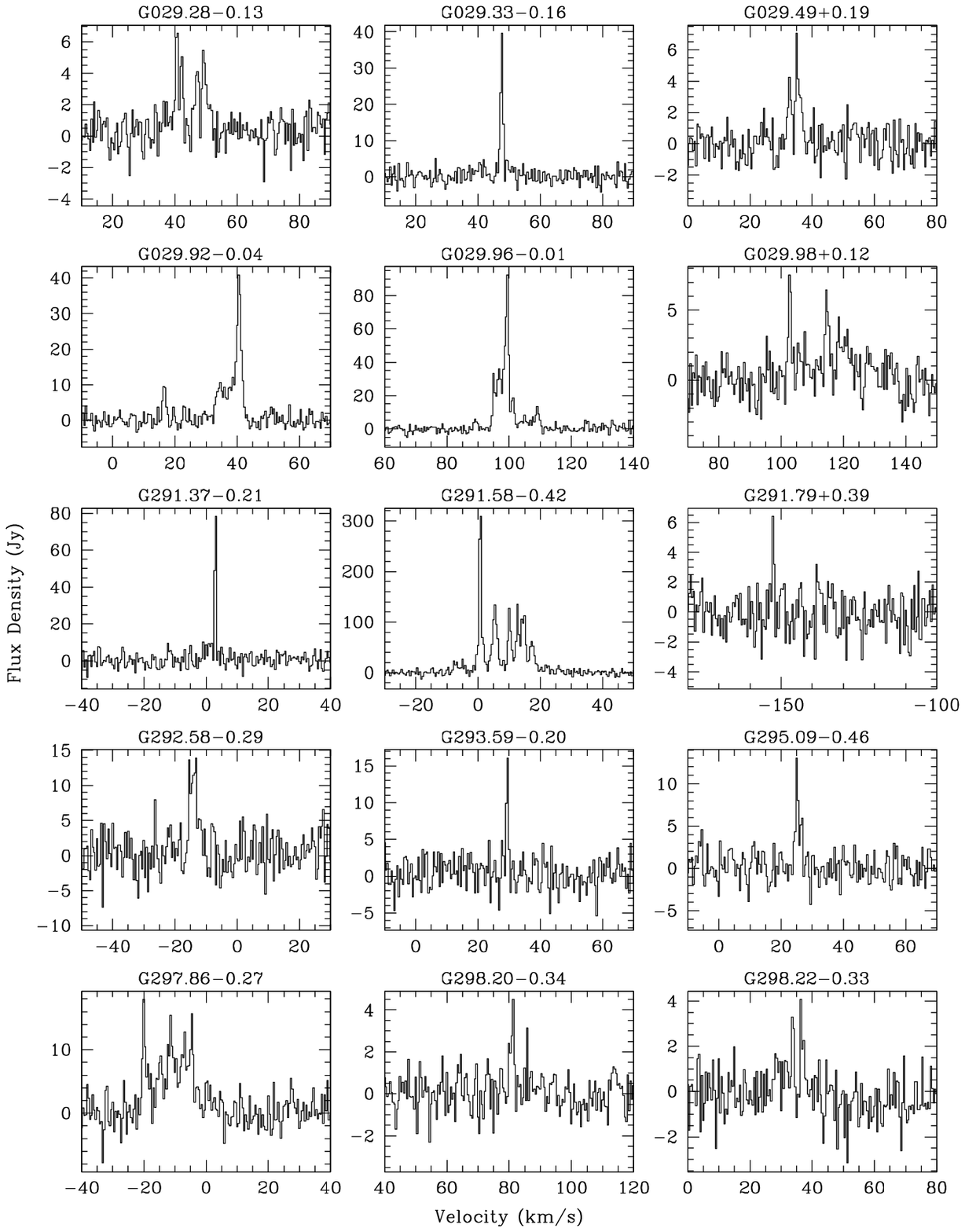}
\contcaption{...}
\end{figure*}

\begin{figure*}
\includegraphics[width=\textwidth]{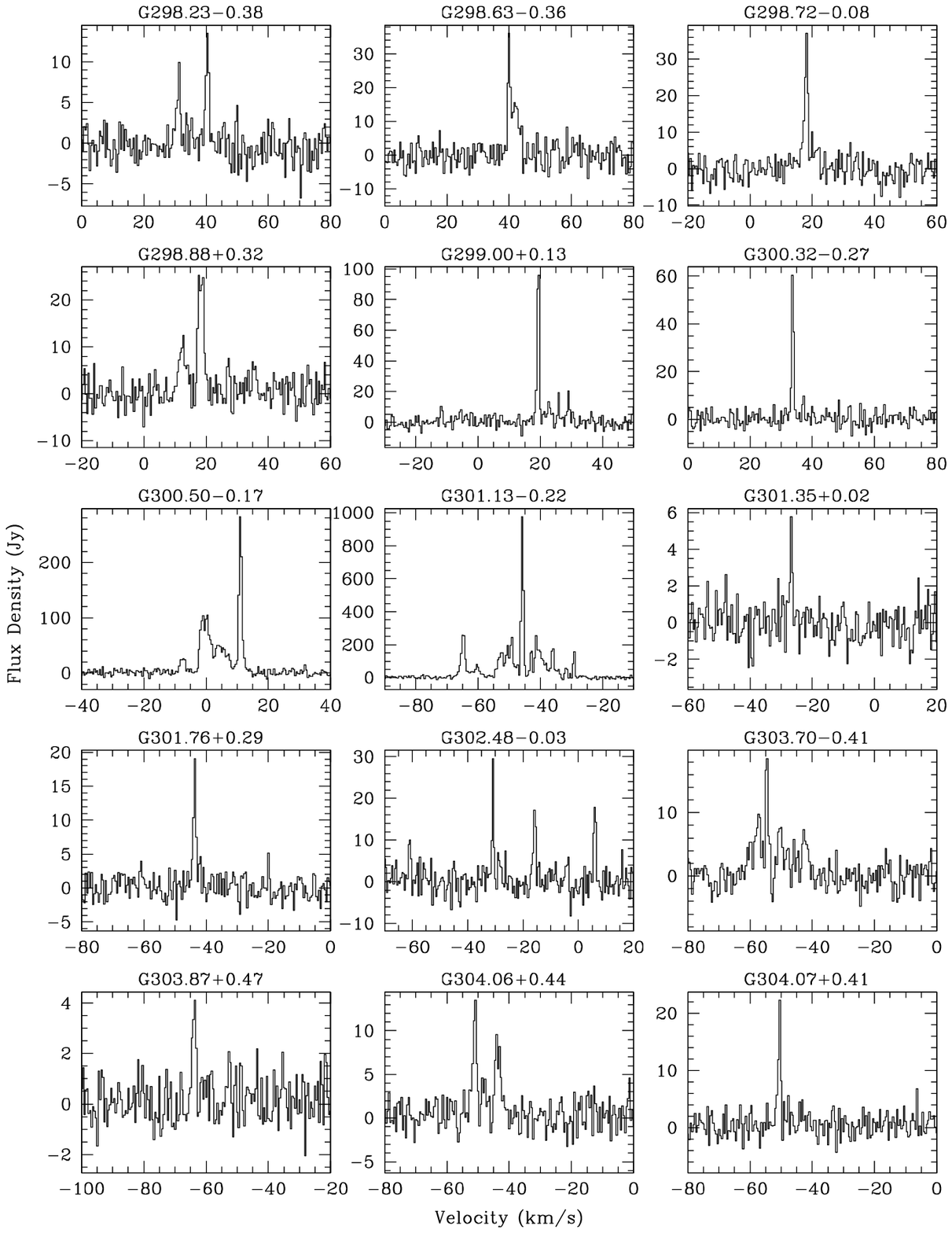}
\contcaption{...}
\end{figure*}

\clearpage
\begin{figure*}
\includegraphics[width=\textwidth]{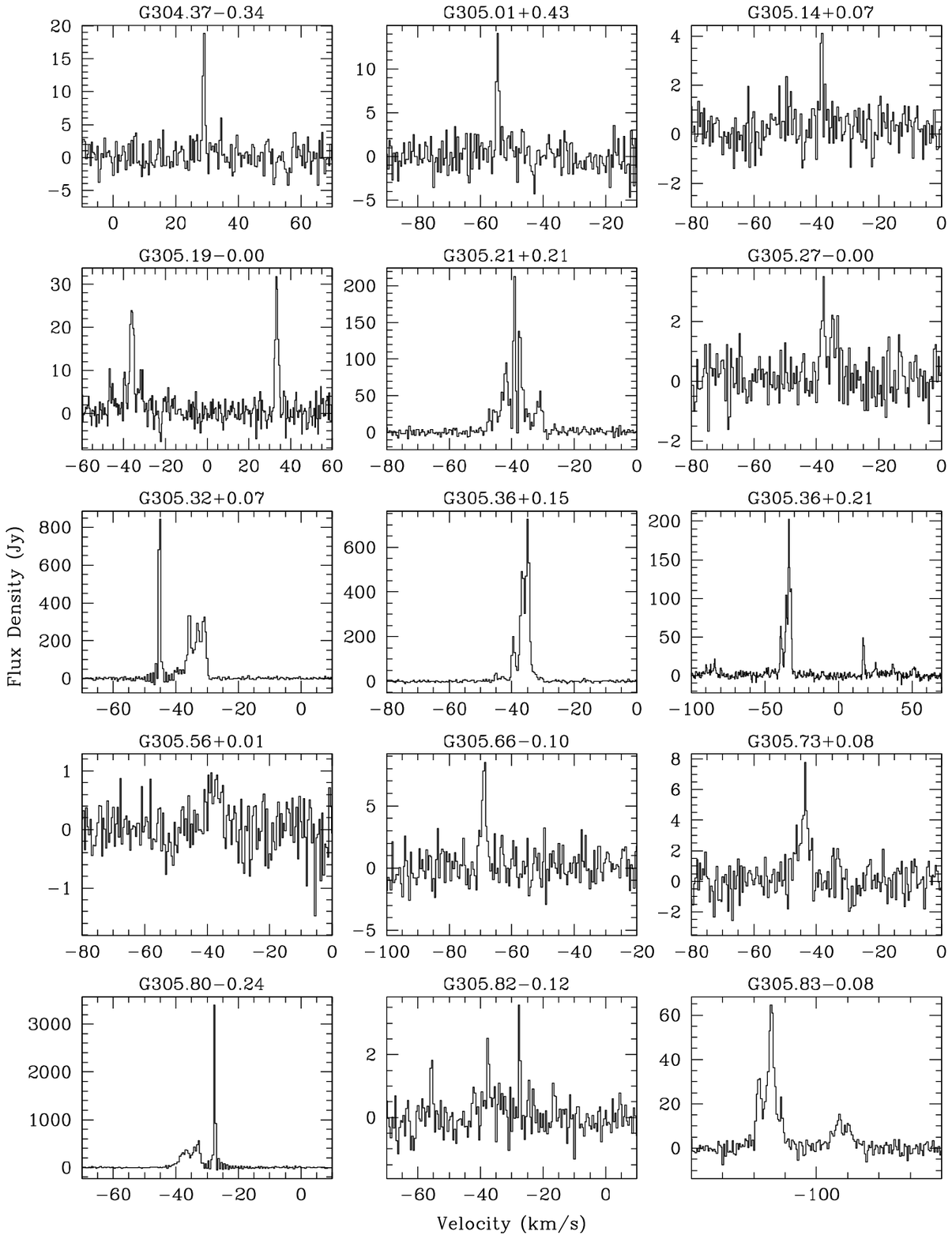}
\contcaption{...}
\end{figure*}

\begin{figure*}
\includegraphics[width=\textwidth]{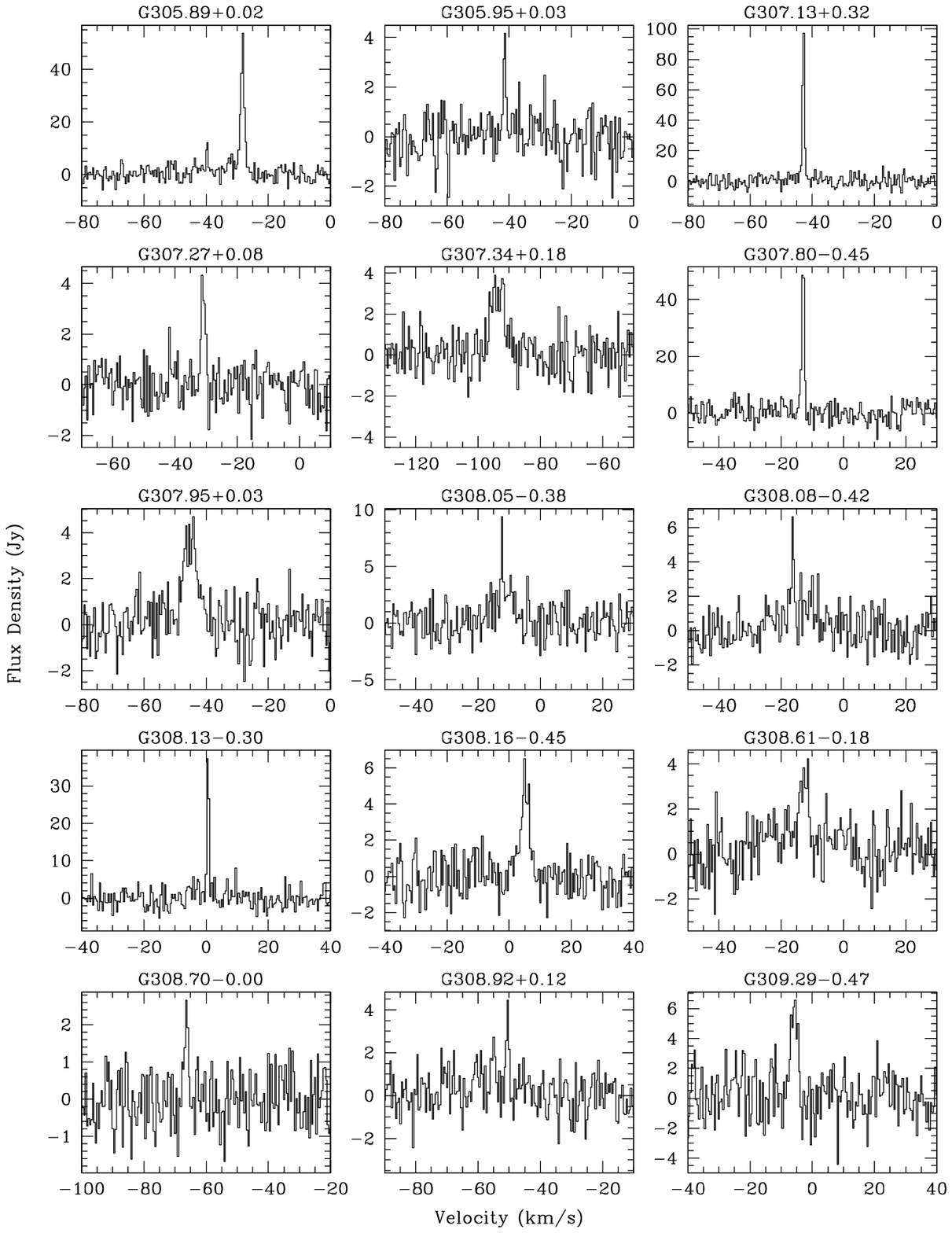}
\contcaption{...}
\end{figure*}

\begin{figure*}
\includegraphics[width=\textwidth]{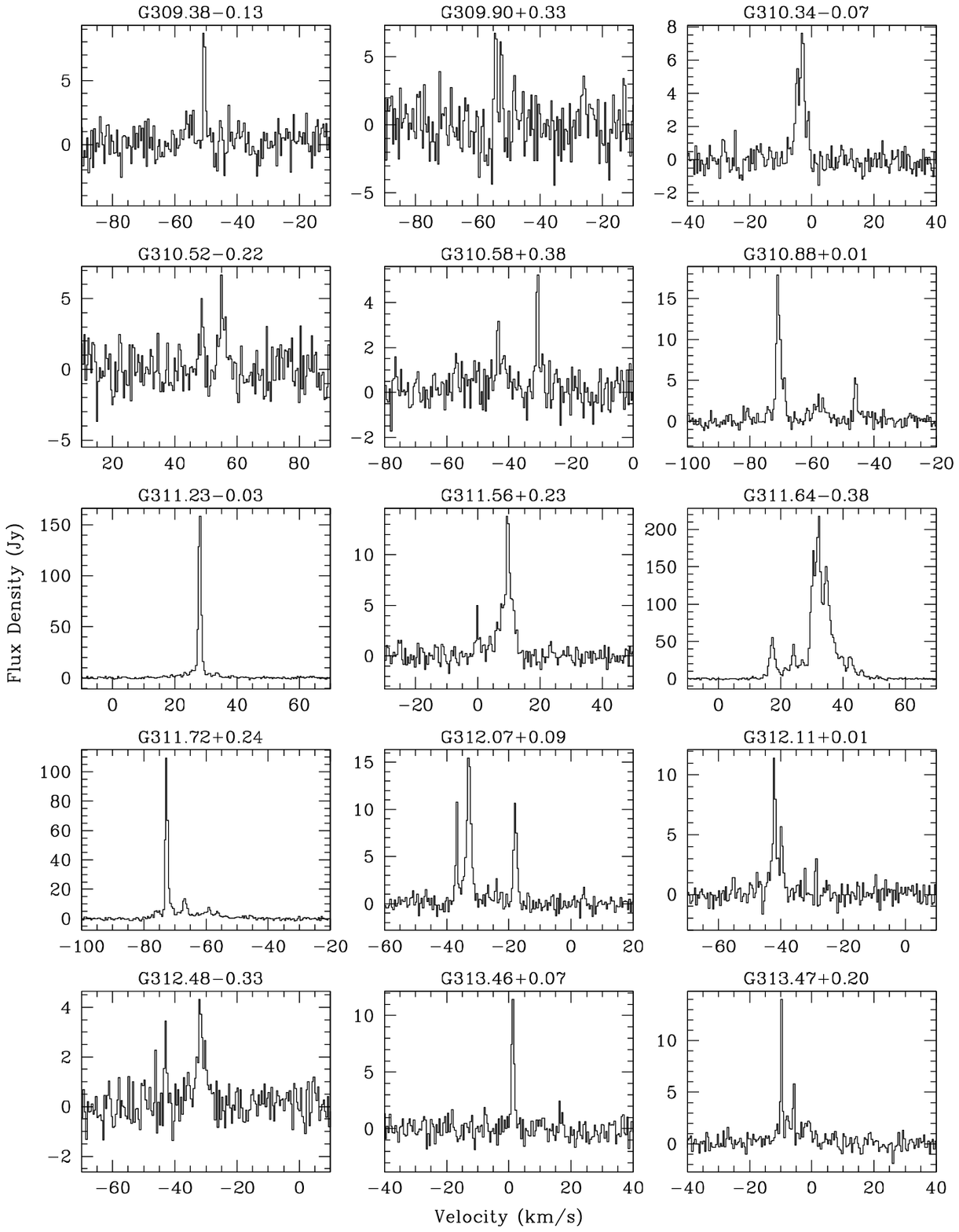}
\contcaption{...}
\end{figure*}

\clearpage
\begin{figure*}
\includegraphics[width=\textwidth]{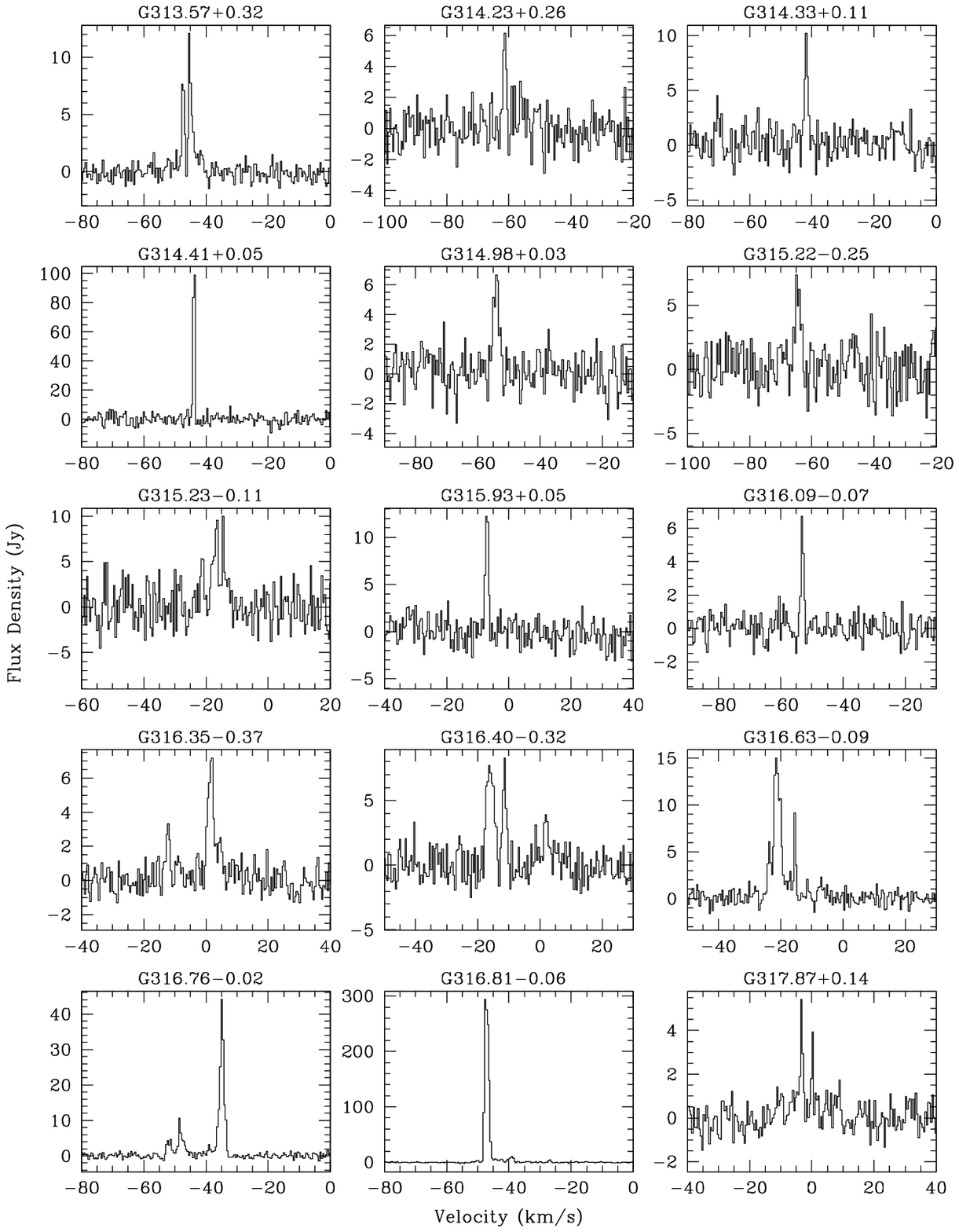}
\contcaption{...}
\end{figure*}

\begin{figure*}
\includegraphics[width=\textwidth]{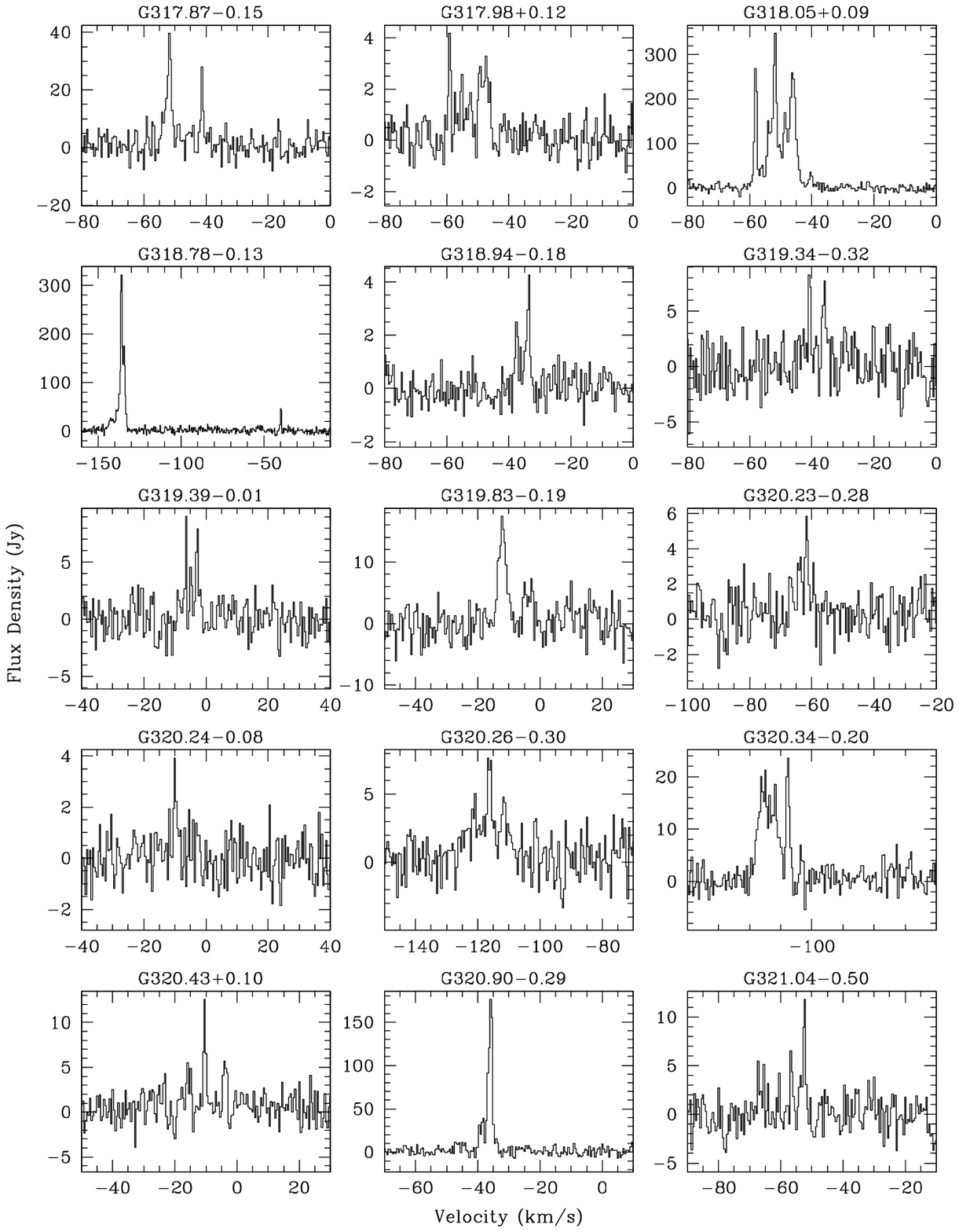}
\contcaption{...}
\end{figure*}

\begin{figure*}
\includegraphics[width=\textwidth]{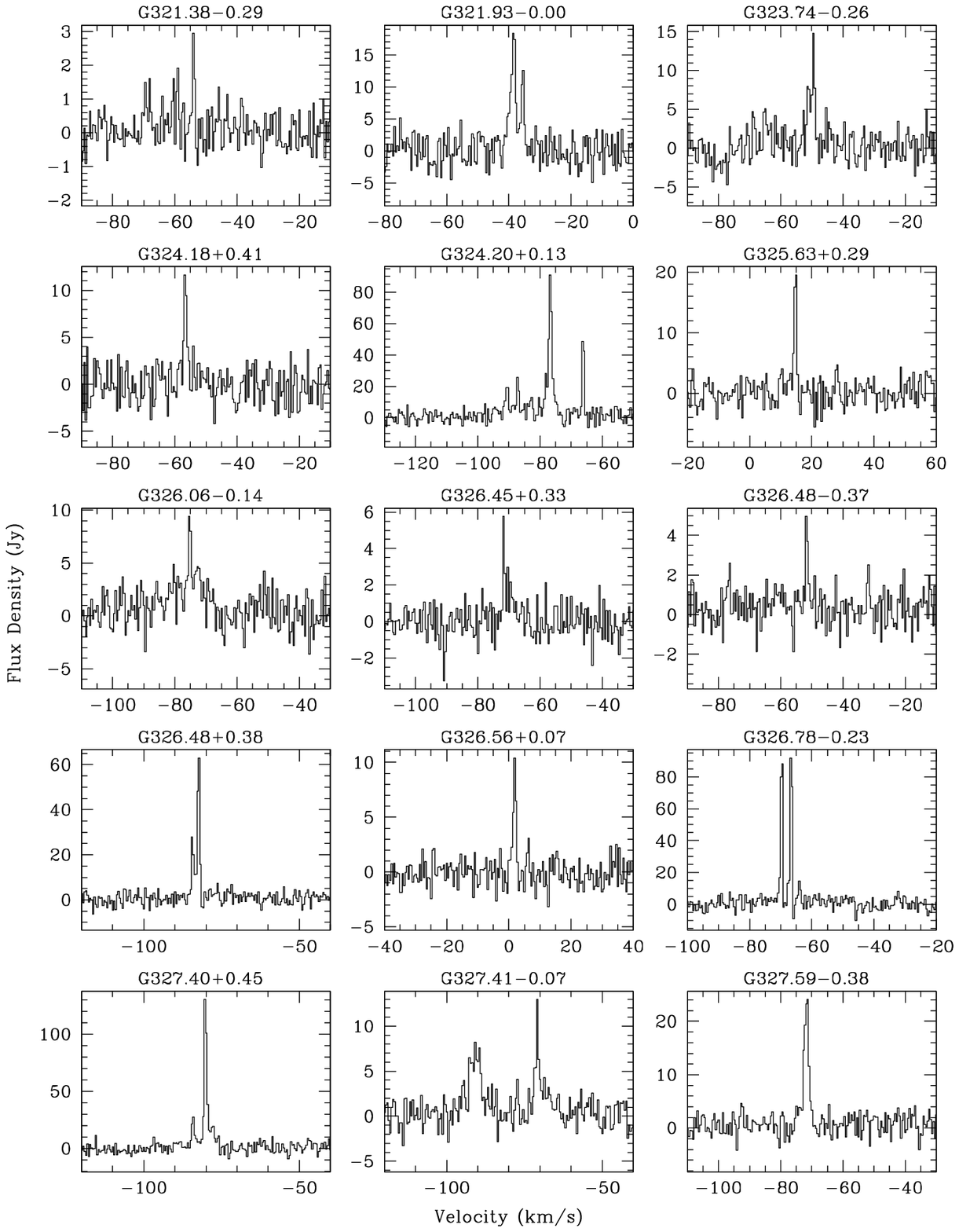}
\contcaption{...}
\end{figure*}

\clearpage
\begin{figure*}
\includegraphics[width=\textwidth]{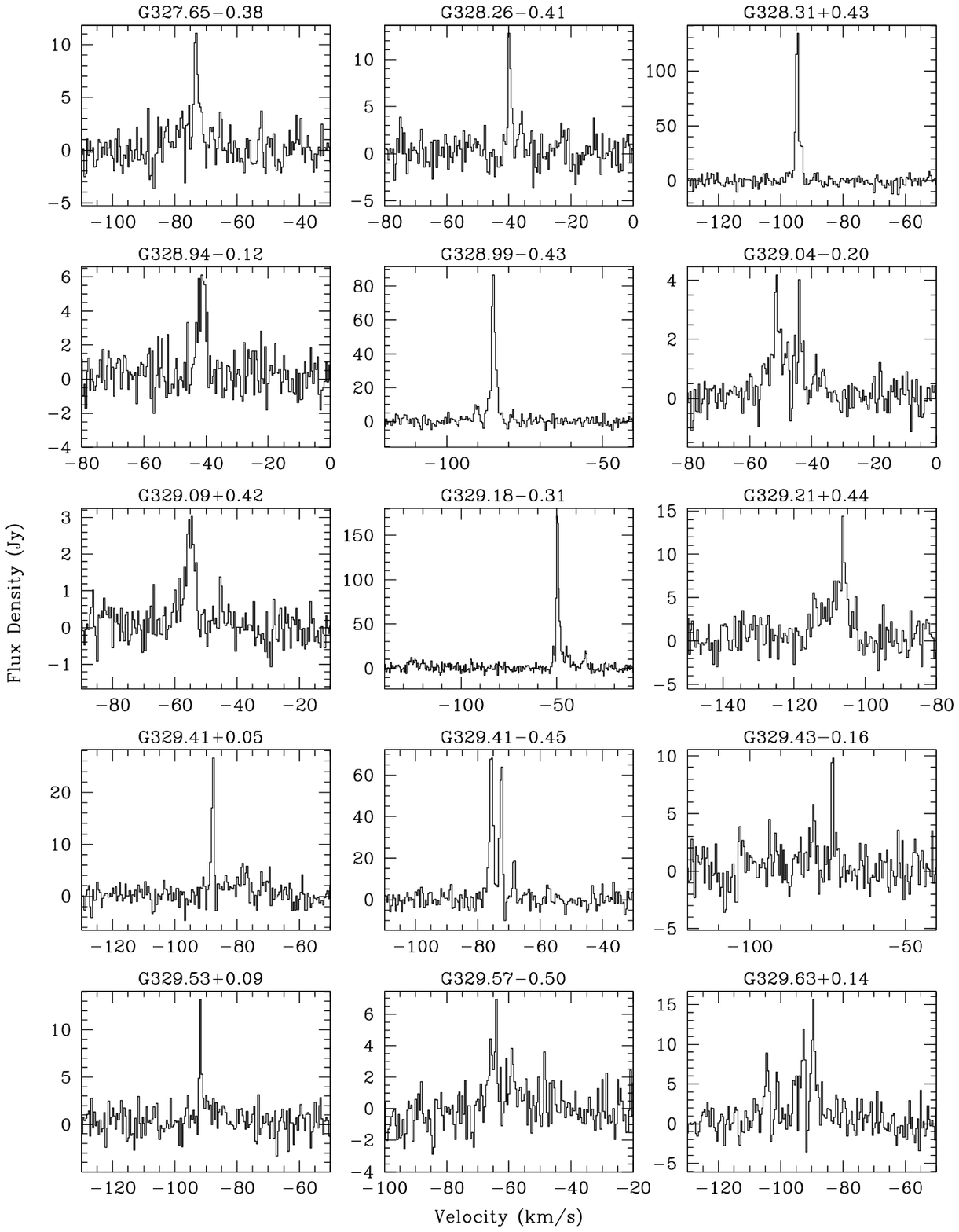}
\contcaption{...}
\end{figure*}

\begin{figure*}
\includegraphics[width=\textwidth]{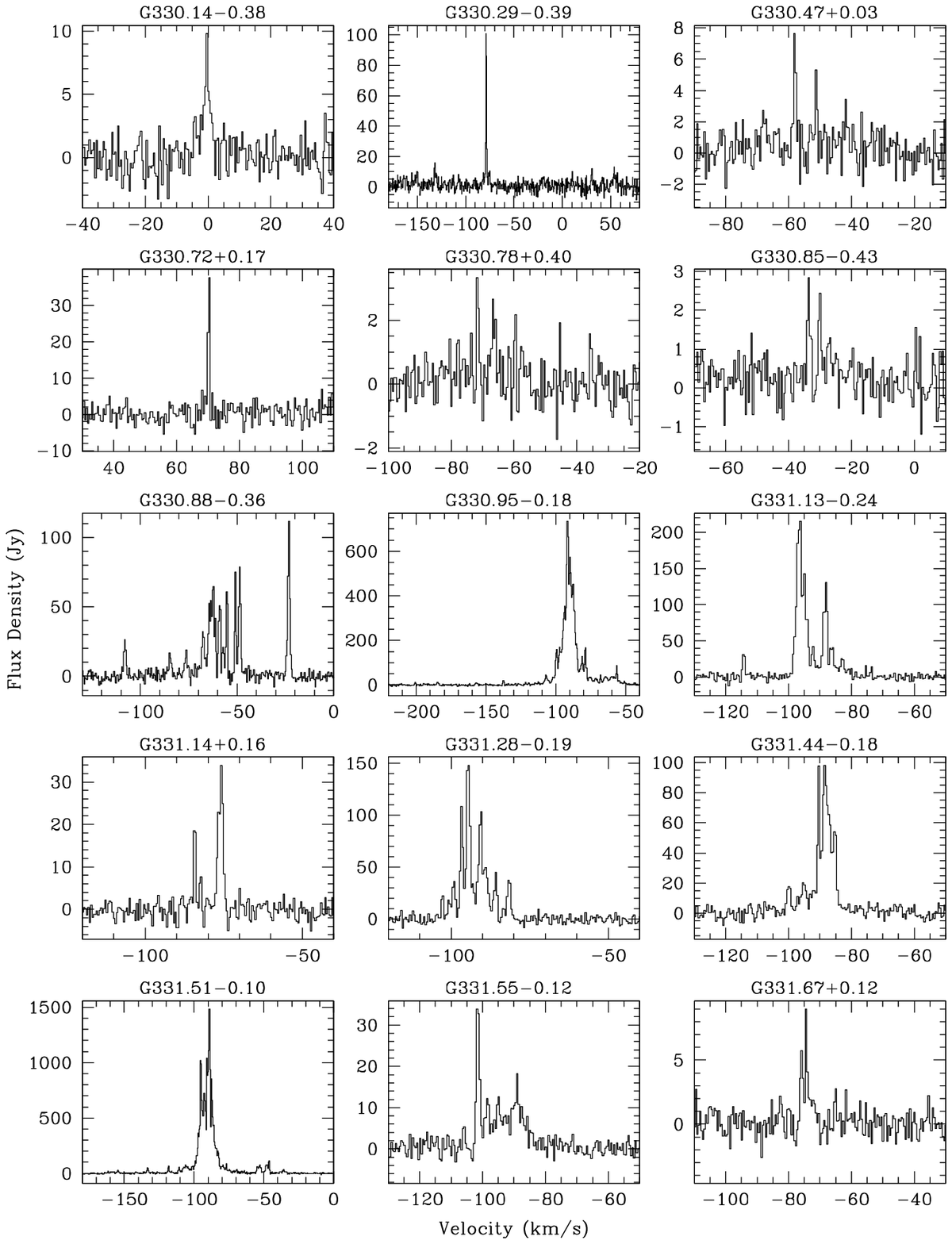}
\contcaption{...}
\end{figure*}

\begin{figure*}
\includegraphics[width=\textwidth]{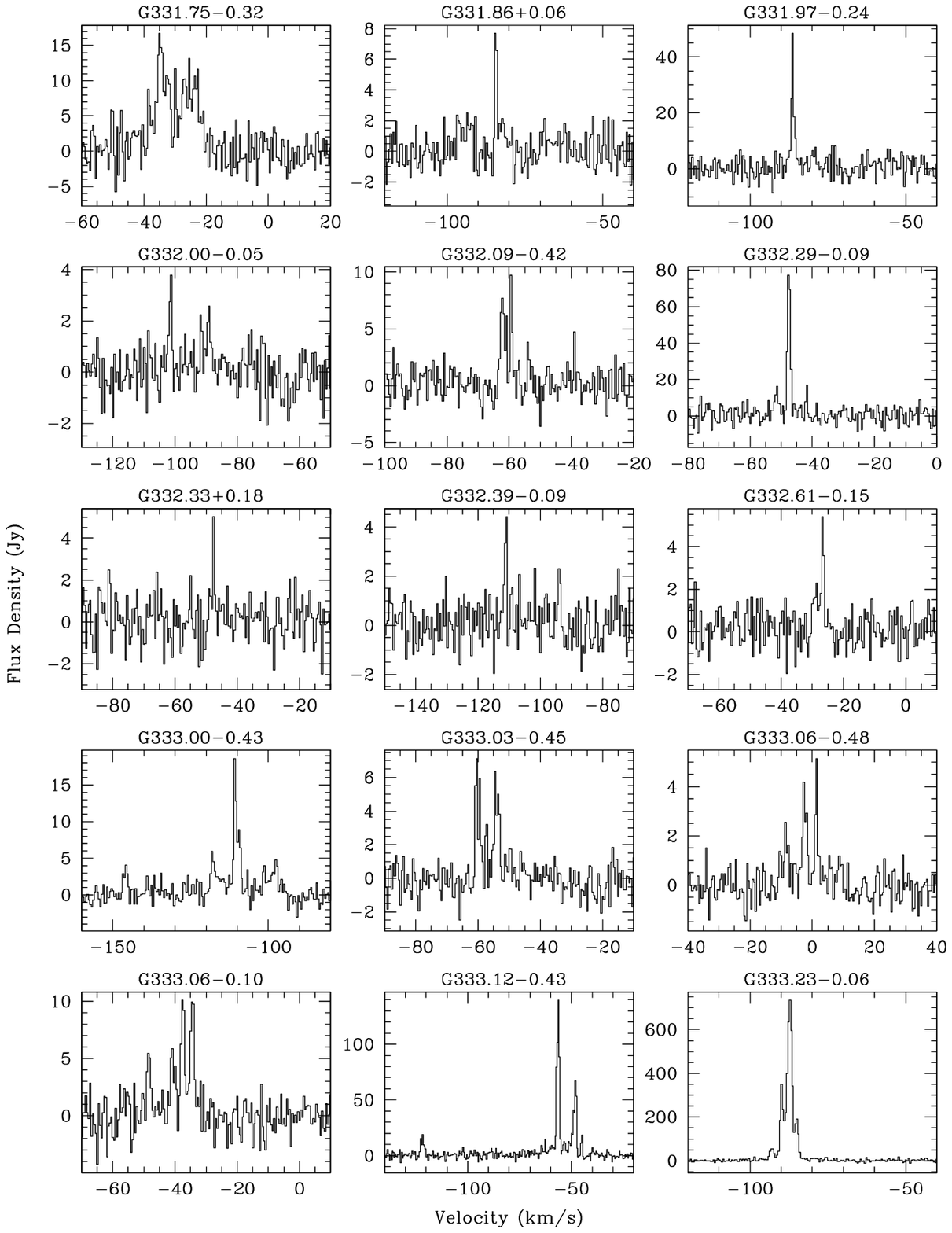}
\contcaption{...}
\end{figure*}

\clearpage
\begin{figure*}
\includegraphics[width=\textwidth]{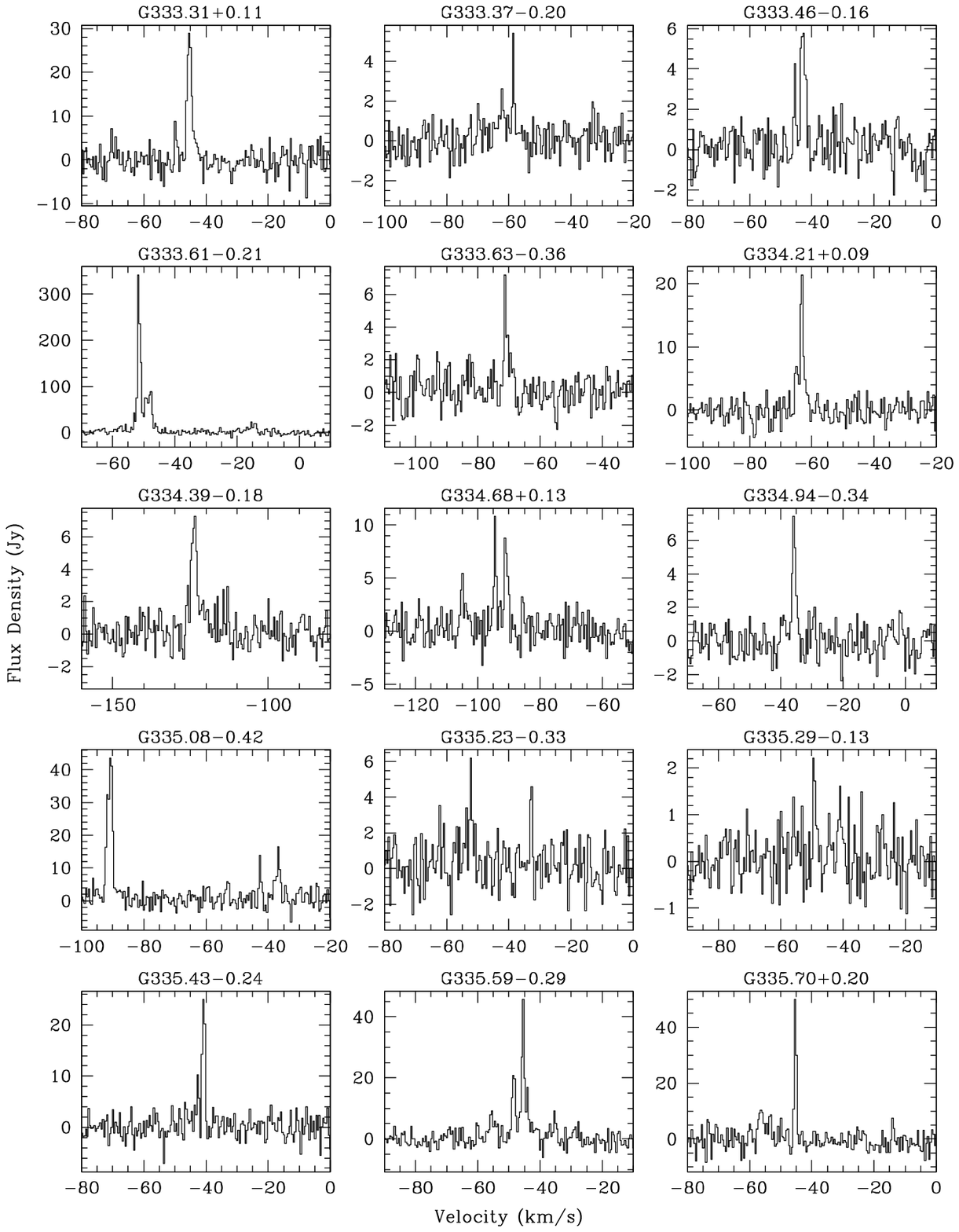}
\contcaption{...}
\end{figure*}

\begin{figure*}
\includegraphics[width=\textwidth]{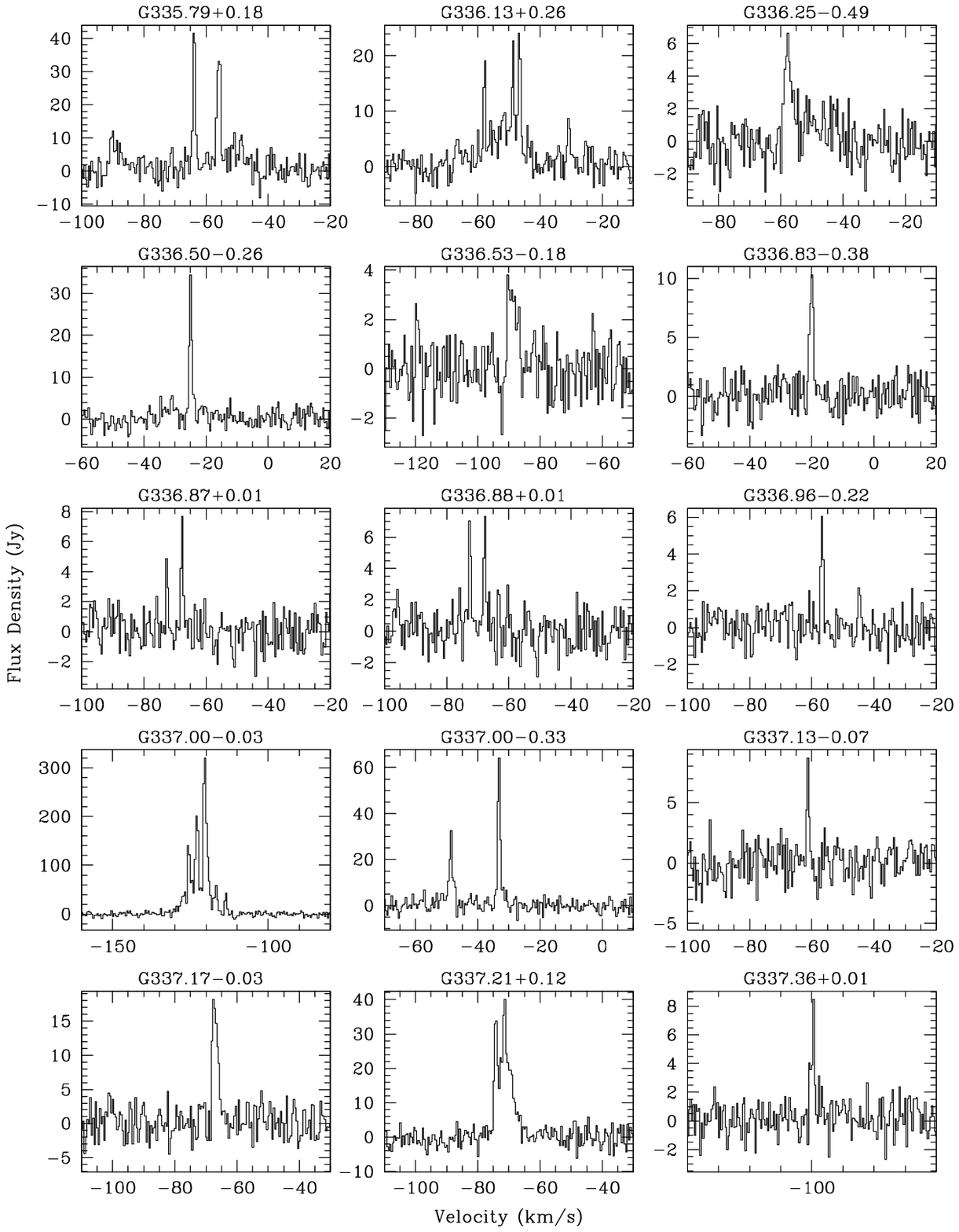}
\contcaption{...}
\end{figure*}

\begin{figure*}
\includegraphics[width=\textwidth]{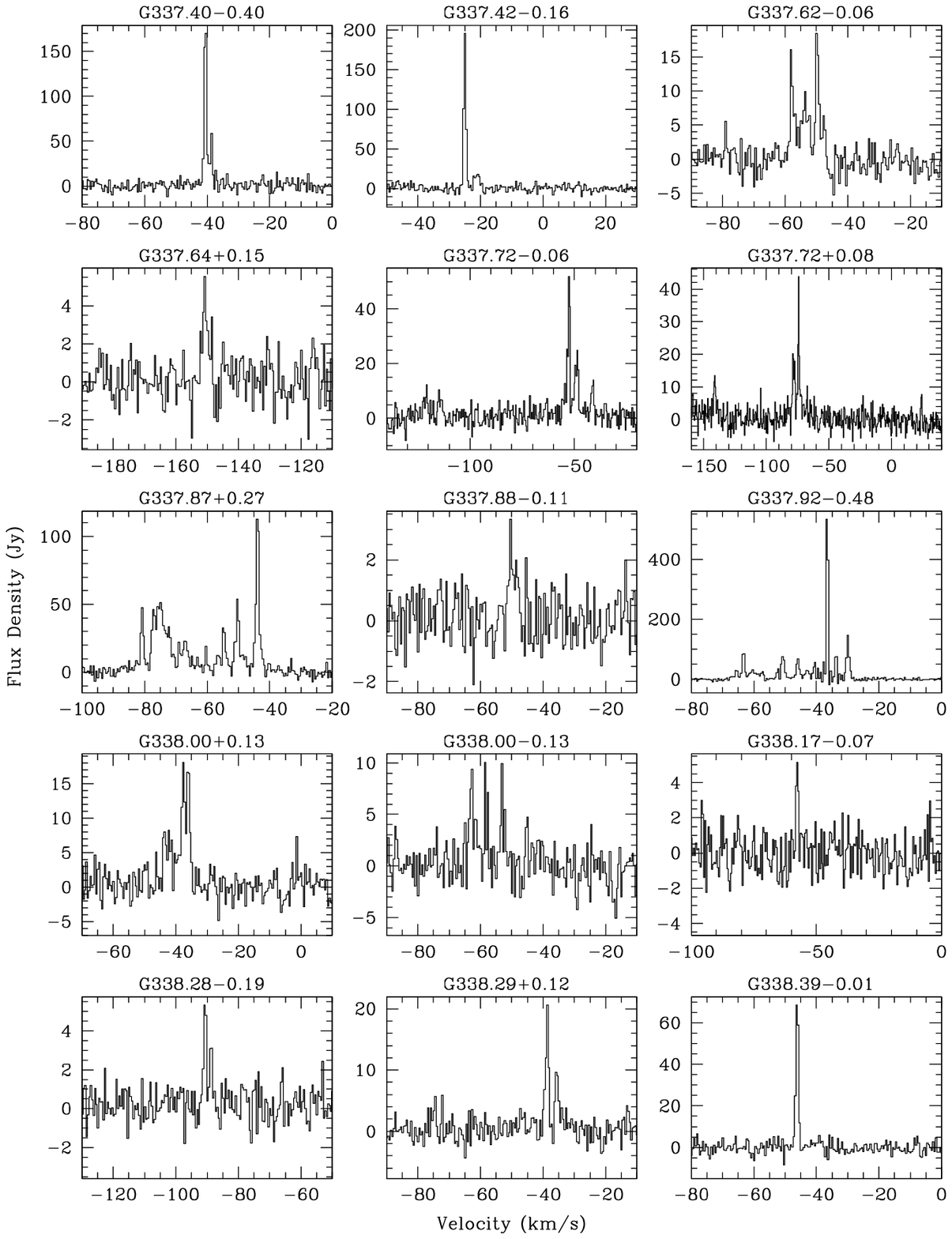}
\contcaption{...}
\end{figure*}

\clearpage
\begin{figure*}
\includegraphics[width=\textwidth]{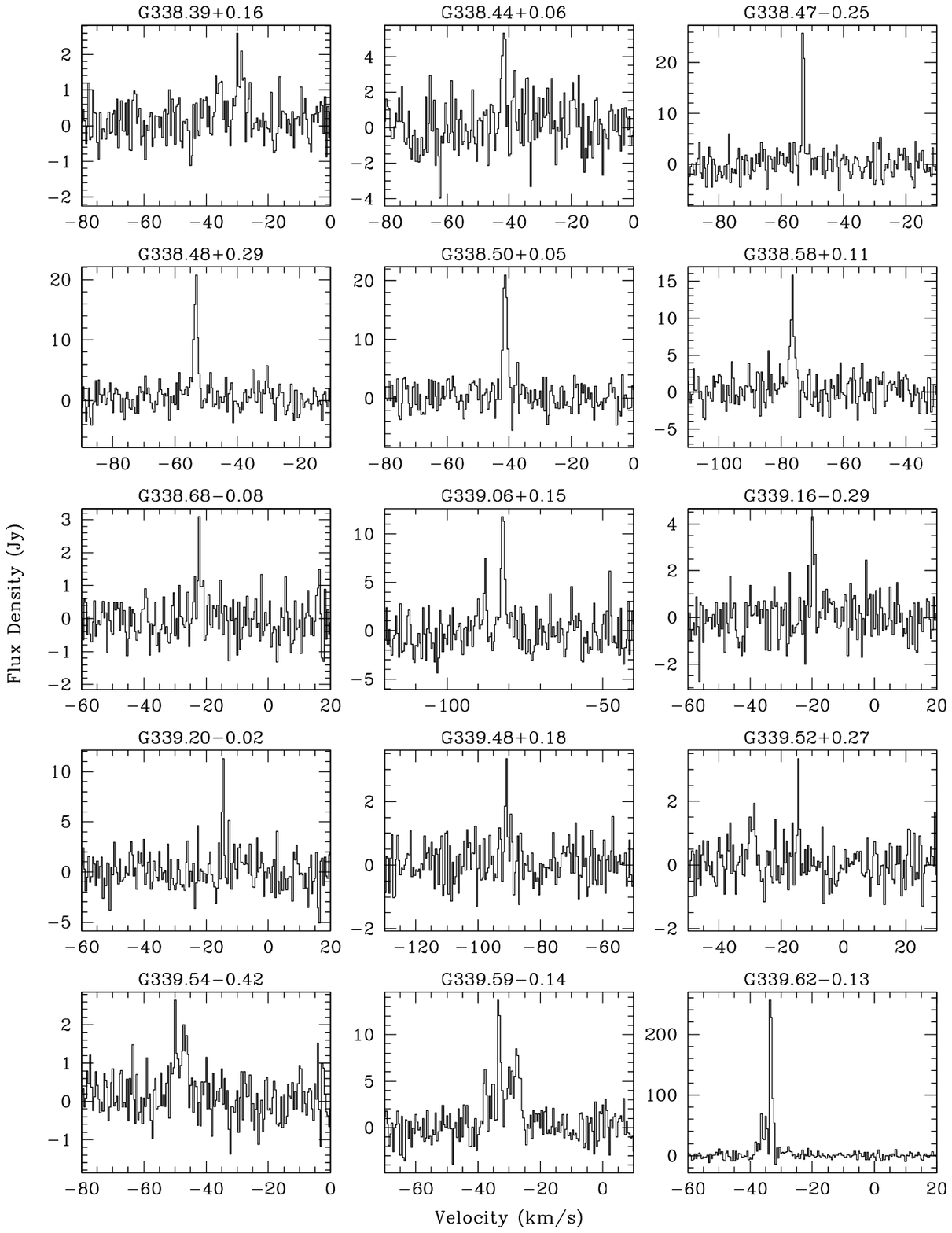}
\contcaption{...}
\end{figure*}

\begin{figure*}
\includegraphics[width=\textwidth]{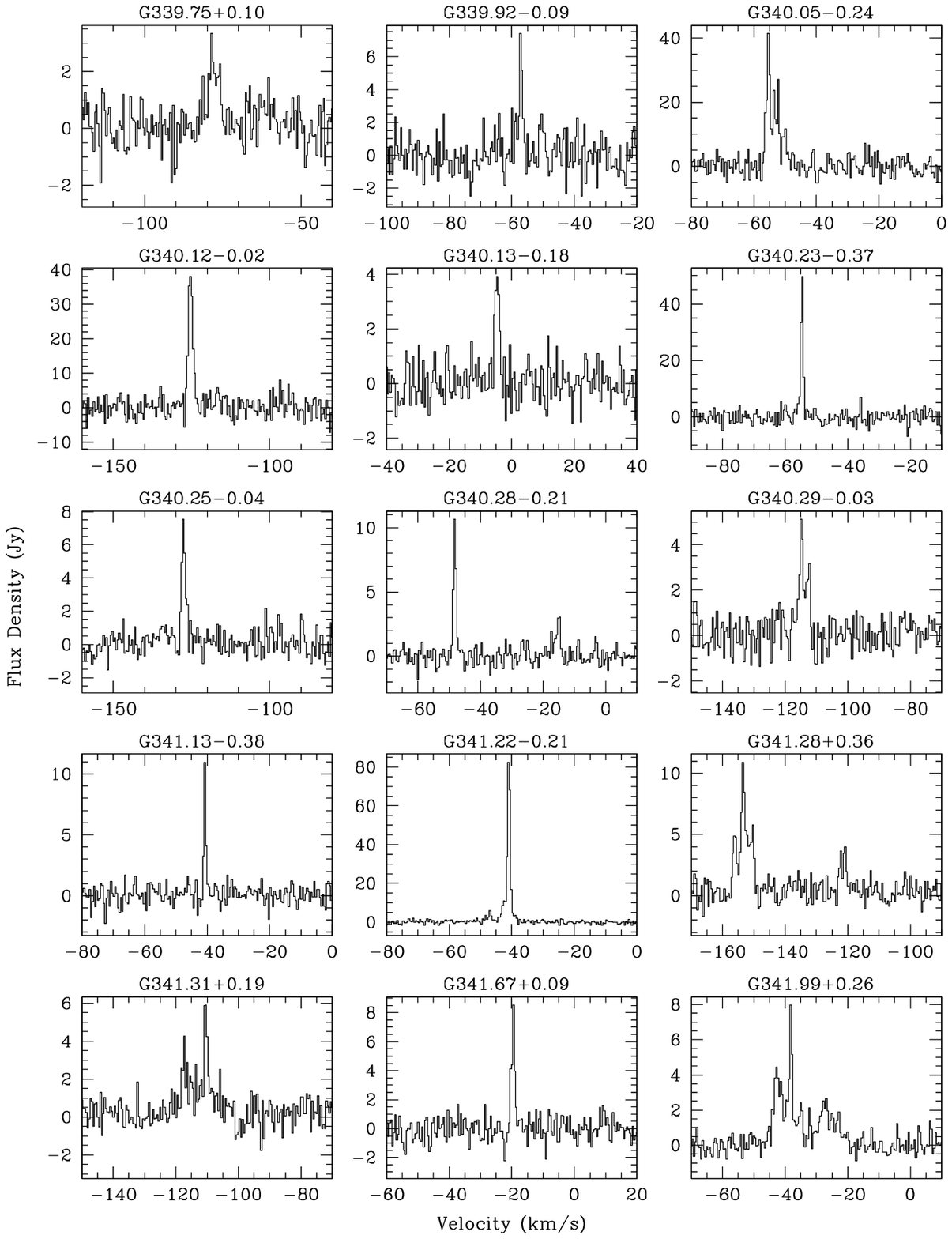}
\contcaption{...}
\end{figure*}

\begin{figure*}
\includegraphics[width=\textwidth]{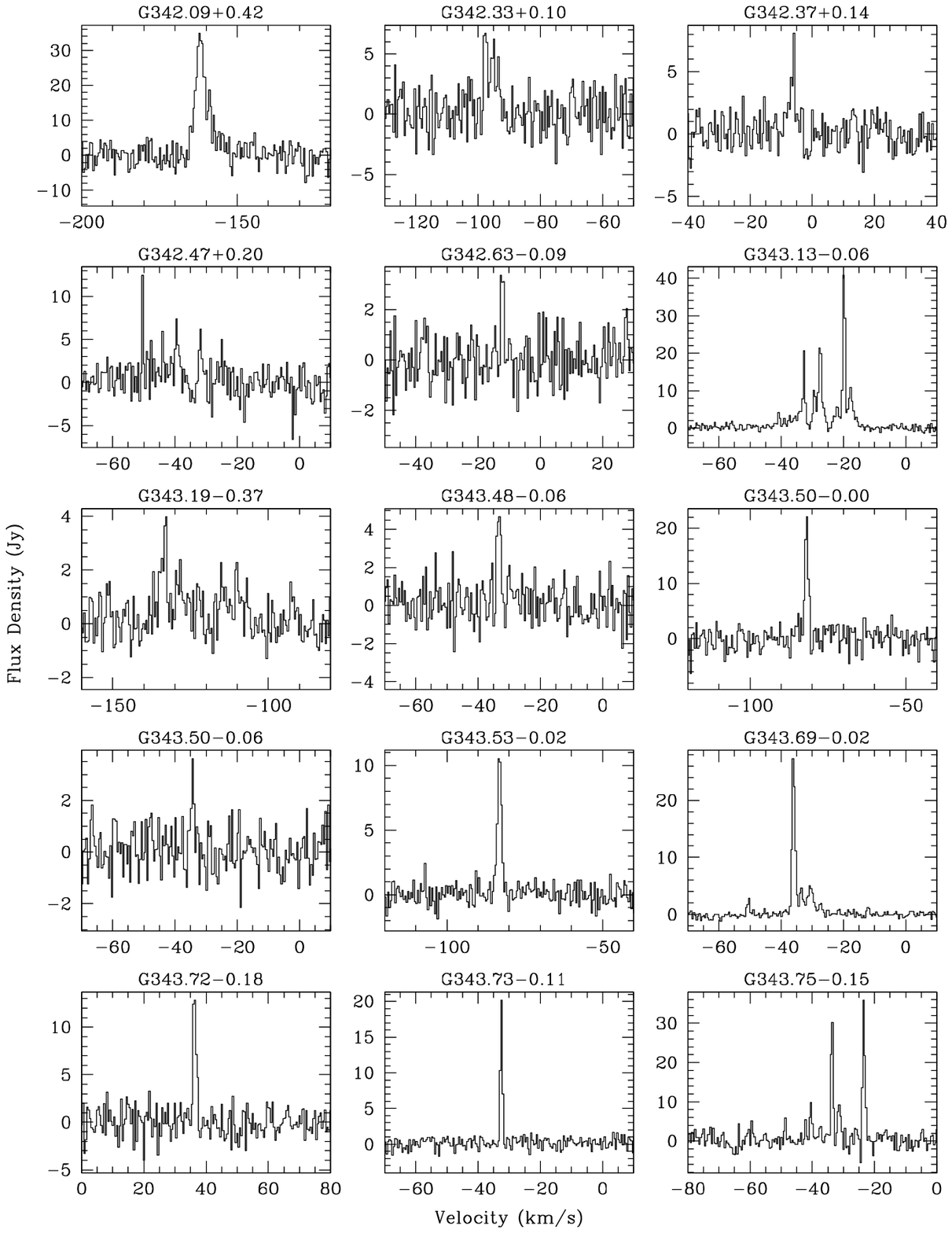}
\contcaption{...}
\end{figure*}

\clearpage
\begin{figure*}
\includegraphics[width=\textwidth]{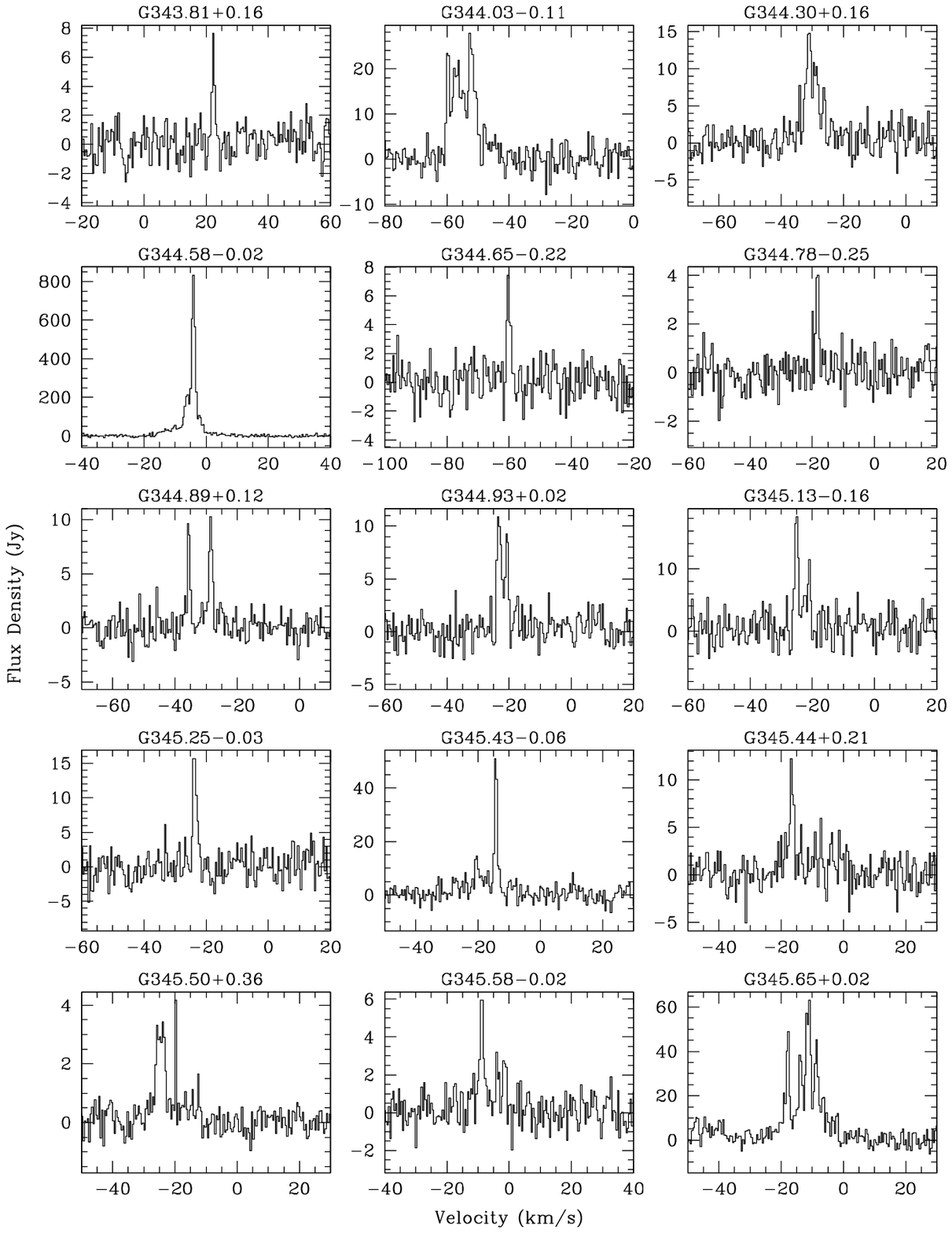}
\contcaption{...}
\end{figure*}

\begin{figure*}
\includegraphics[width=\textwidth]{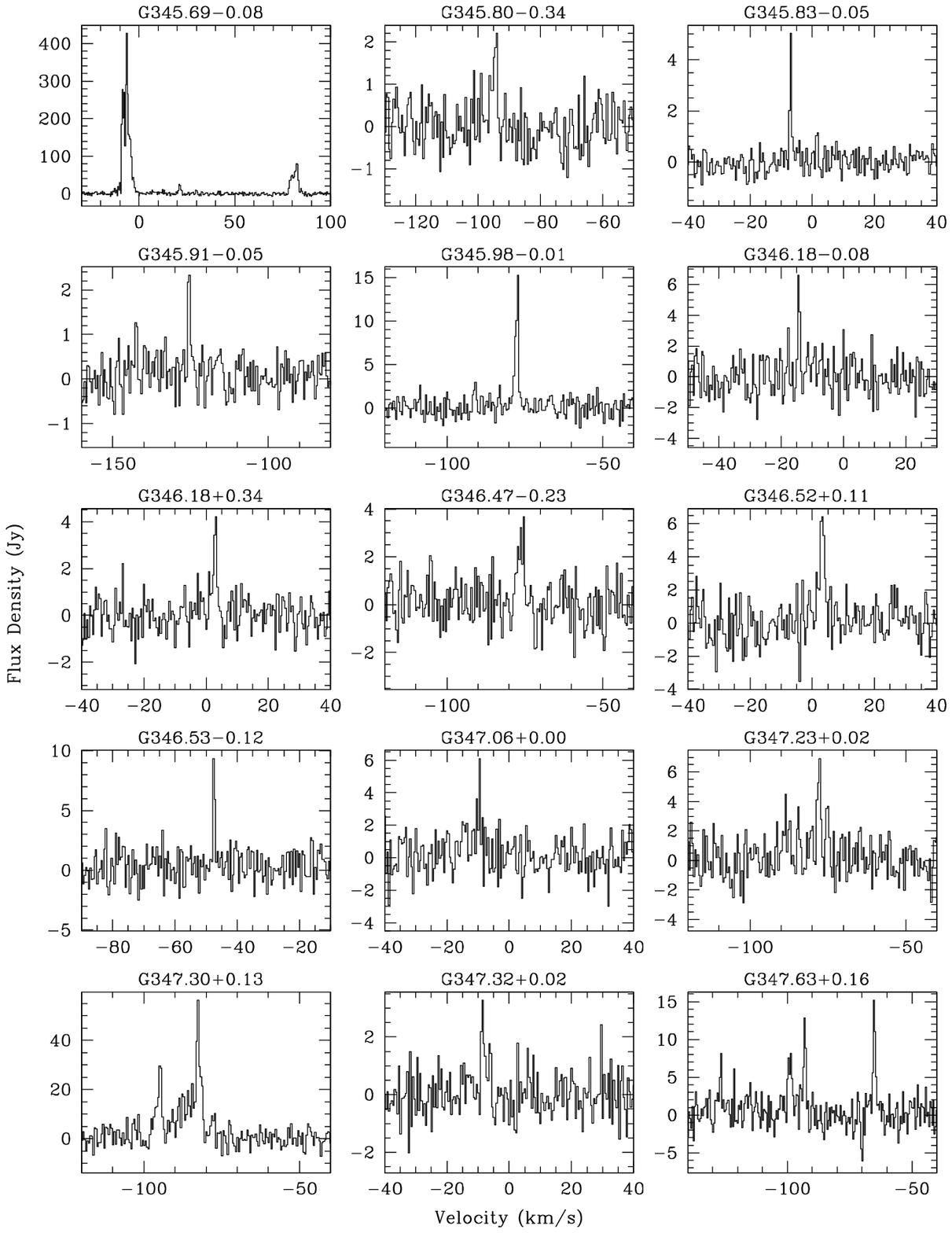}
\contcaption{...}
\end{figure*}

\begin{figure*}
\includegraphics[width=\textwidth]{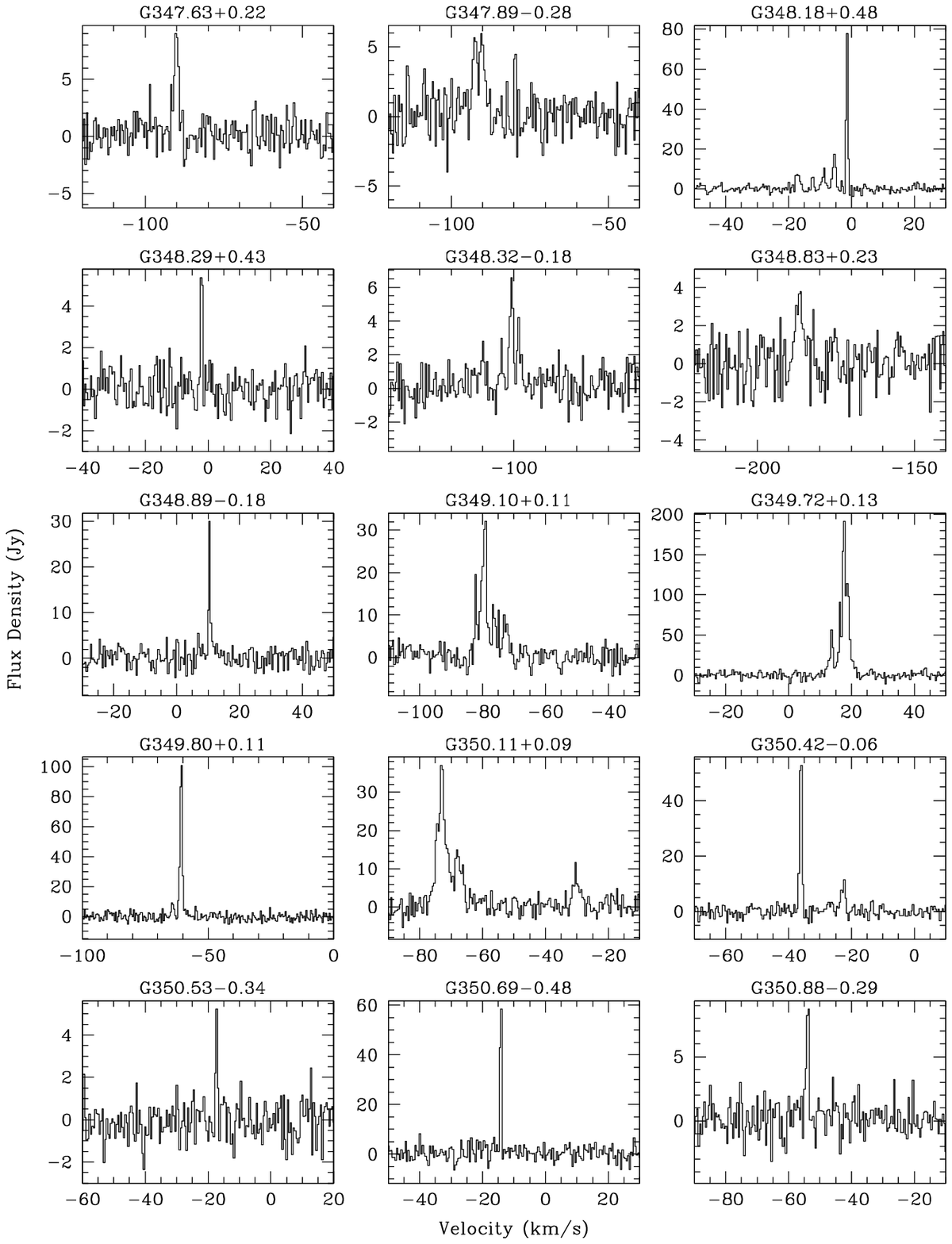}
\contcaption{...}
\end{figure*}

\clearpage
\begin{figure*}
\includegraphics[width=\textwidth]{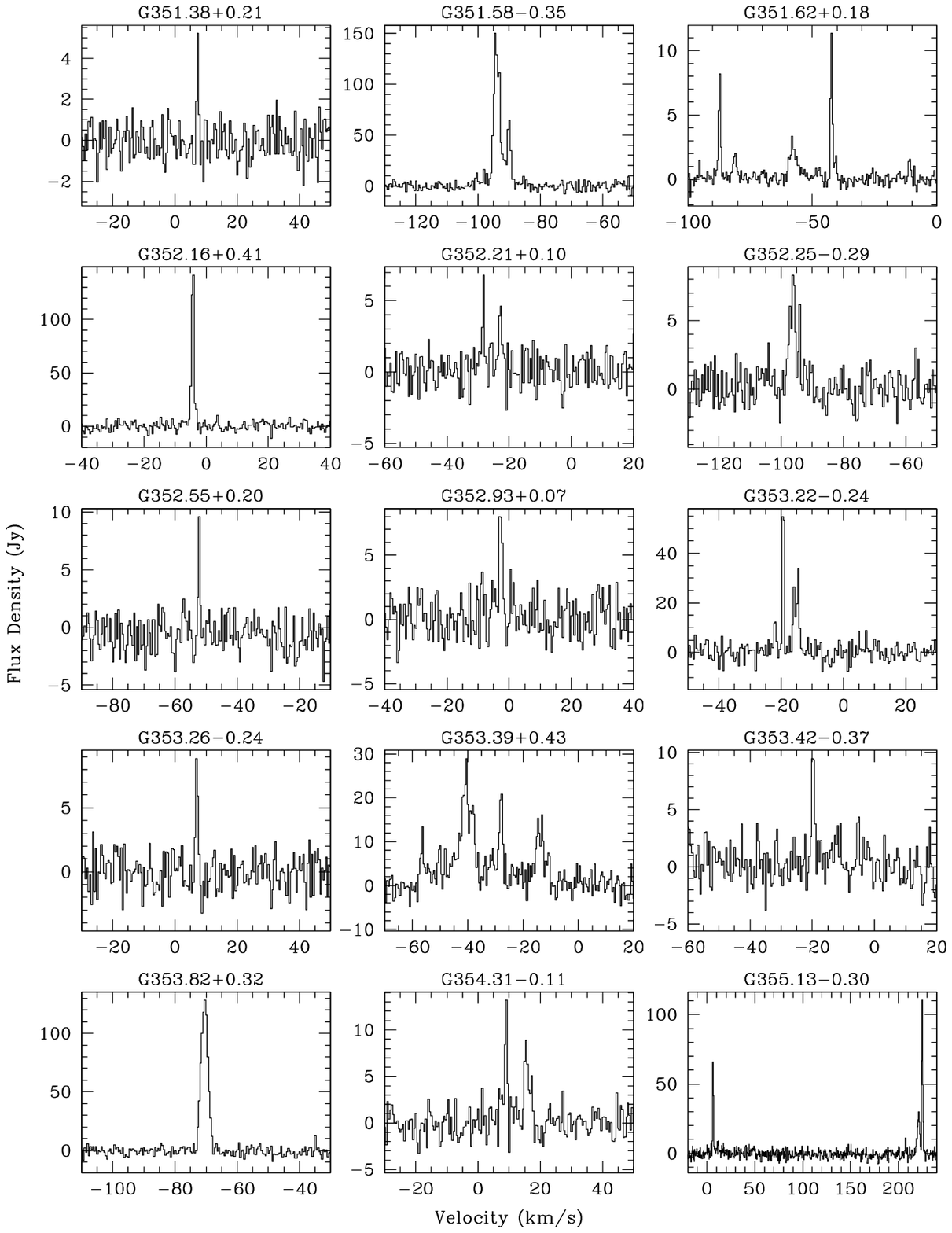}
\contcaption{...}
\end{figure*}

\begin{figure*}
\includegraphics[width=\textwidth]{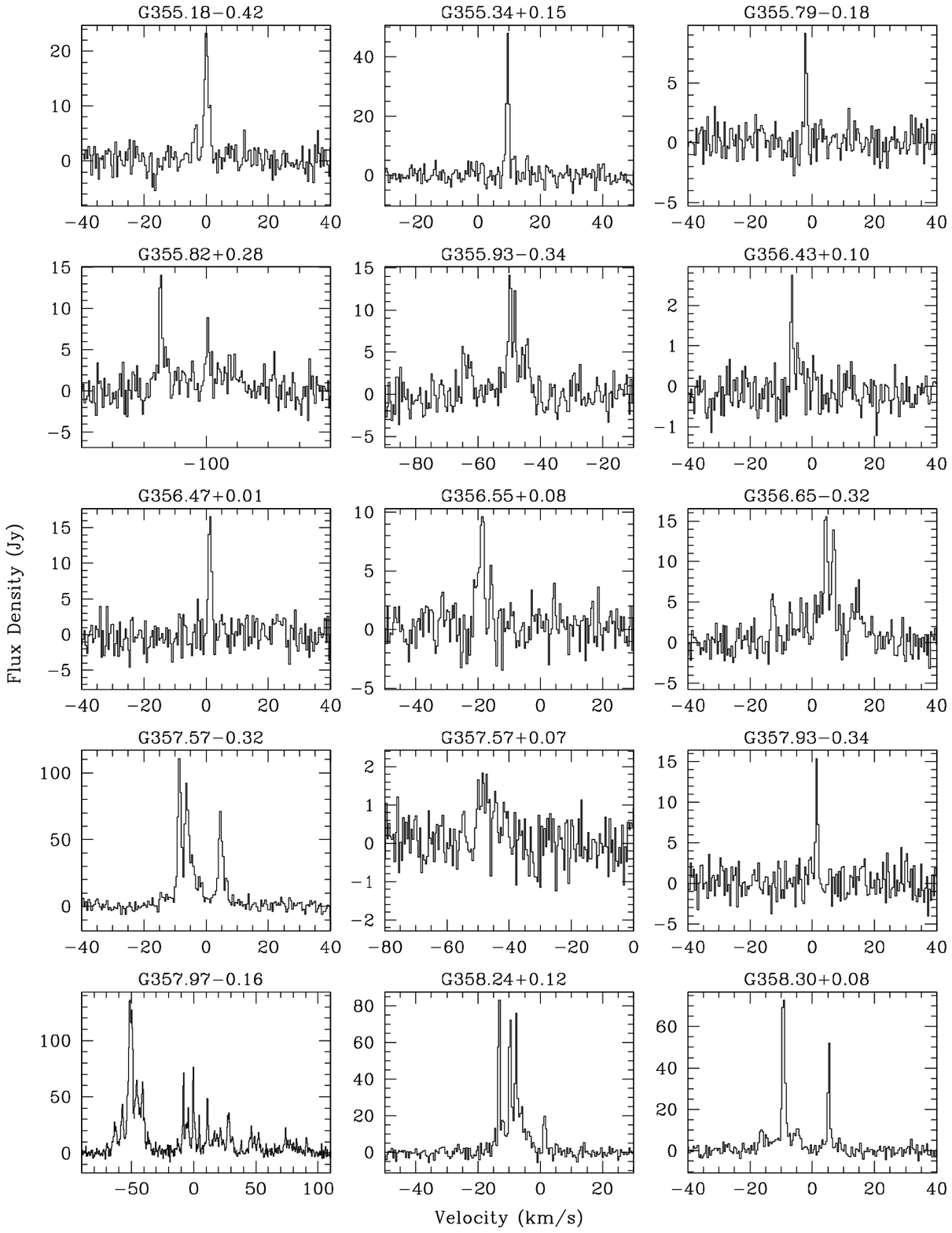}
\contcaption{...}
\end{figure*}

\begin{figure*}
\includegraphics[width=\textwidth]{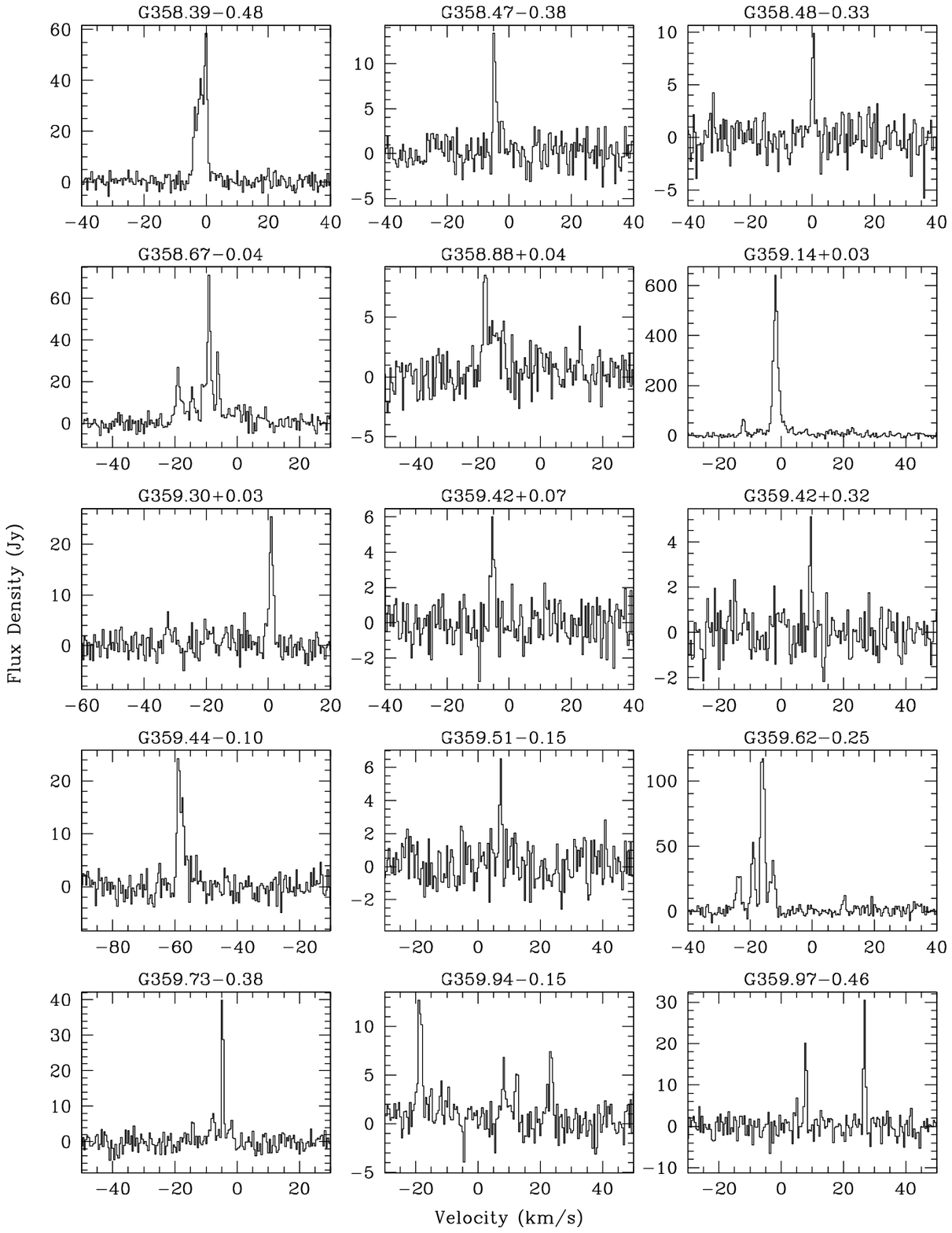}
\contcaption{...}
\end{figure*}

\clearpage

There are 303 \water~maser sites that show more than one peak in the maser
spectrum (we refer to such single peaks in the spectrum as maser spots). Of
these sites with more than one maser spot, the widest velocity distribution
is exhibited by G9.10$-$0.40, over 351.3\,\kms. This maser site was previously
reported by \citet{walsh09}, based on HOPS data, and is a water fountain associated with a
post-AGB star. Subsequent observations by \citet{walsh09} found maser
emission over an even wider range of 398\,\kms.

\begin{figure}
\includegraphics[width=0.5\textwidth]{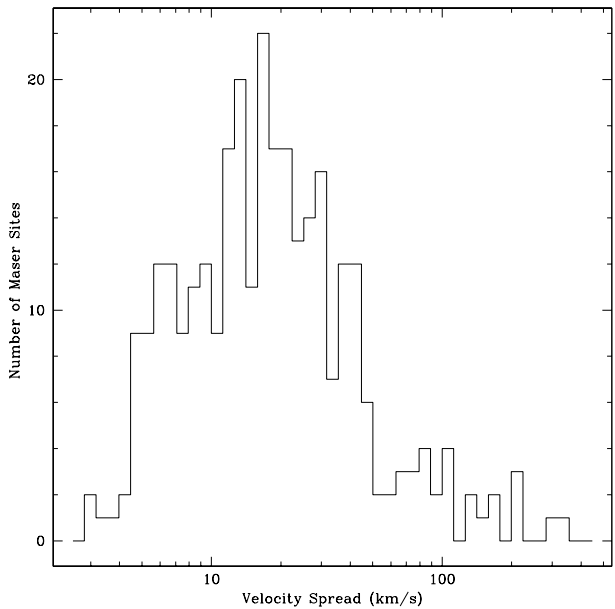}
\caption{Distribution of \water~maser velocity spreads for maser
sites showing more than one maser spot. The majority of maser sites
are confined to a velocity spread of less than 17\,\kms~and 90\%
of maser sites have a velocity spread of less than 50\,\kms.}
\label{vel_spread}
\end{figure}

In total, there are 14 maser
sites with maser spots spread over a velocity range greater than 100\,\kms.
The median velocity spread is 17\,\kms. The distribution of velocity spreads
of maser sites with multiple maser spots can be seen in Figure \ref{vel_spread}.
The Figure shows that the occurrence of maser sites with high velocity
features is relatively rare, but there is a population of about 10\% of
maser sites with maser spots that span velocity ranges greater than 50\,\kms.

When comparing the line FWHM of maser spots, we find that G27.08$+$0.20
exhibits the widest FWHM of 13.3\,\kms. However, because maser sites
usually consist of multiple peaks in the spectrum, it is likely that
any single peak may comprise multiple unresolved
components in the spectrum. This is likely to be the case for G27.08$+$0.20
whose \water~maser emission is very weak and dominated by noise, possibly
masking multiple components. The median line FWHM is 1.2\,\kms and 90\%
of maser spots are no broader than 3.1\,\kms.

In Table \ref{330masers}, the last column indicates whether a maser site is
coincident with the position and shows similar spectral features to the \water~maser
emission from a known evolved star. There are 15 maser sites
so far associated with evolved stars. Due to the 2$^\prime$ beam of the
Mopra observations, we await higher resolution observations to associate all
masers with a particular class of astrophysical object. There are certainly
known evolved stars within 2.2$^\prime$ of \water~masers which we have
not yet associated because we cannot compare our \water~maser spectrum
to a previously published spectrum. We thus expect
there to be more associations with evolved stars to be uncovered, but
with the majority of maser sites associated with star forming regions.
See \S\ref{secglat} for further evidence that leads us to this
expectation.
We plan to conduct follow up observations of the maser sites with the
ATCA, which will give us high positional
accuracy, allowing us to identify the majority of maser sites with known
astrophysical objects.

\section{Discussion}

\subsection{Noise measurements}
We expect the noise level in our spectra to vary by about a factor of two
because observations were made at a range of elevations and weather conditions,
which gave rise to system temperature variations in the range 50 to 120\,K.
In order to determine the sensitivity limits of the water maser data, we measure
the rms noise levels for each quarter-square-degree. This is done by taking a
single channel from the data cube where no emission is expected
(typically a channel at a velocity of +250\,\kms~was used). Then the rms noise
is measured for each quarter-square-degree using the {\sc miriad} task
{\sc imstats}. We also checked the noise levels at different parts of each zoom
band and found that the noise does not vary significantly within each zoom.
This is done for each of the IFs containing \water~maser,
NH$_3$ (1,1), HCCCN (3--2) and H69$\alpha$ emission. These IFs were chosen
because they contain the most important emission lines in the survey and because they
sample close to the full range of frequencies observed in HOPS, from 19.6\,GHz
(H69$\alpha$) to 27.3\,GHz (HCCCN (3--2)).

\begin{figure}
\includegraphics[width=0.5\textwidth]{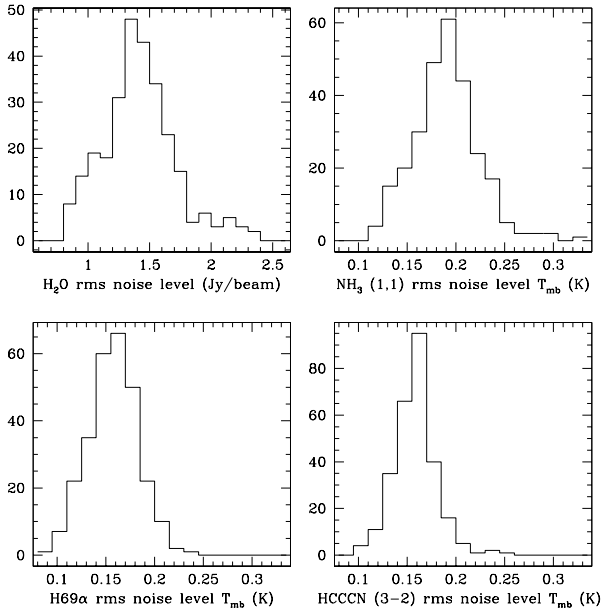}
\caption{Histograms of rms noise levels. {\bf Top-left} H$_2$O maser noise
levels in units of Jy/beam, assuming a conversion factor of 12.4\,Jy/K.
{\bf Top-right} NH$_3$ (1,1) rms noise levels in
units of main beam brightness temperature. {\bf Bottom-left} H69$\alpha$
radio recombination line rms noise levels in units of main beam brightness
temperature. {\bf Bottom-right} HCCCN (3--2) rms noise levels in units of
main beam brightness temperature.}
\label{noisefig}
\end{figure}

Figure \ref{noisefig} shows histograms of the noise levels measured for each
of the IFs. The rms noise levels are typically around 0.15 to 0.2\,K in terms
of main beam brightness temperature. We show the \water~maser rms noise level
in terms of Jy, as this is a more typical intensity scale used for masers
(assuming 12.4\,Jy\,K$^{-1}$).
For \water~masers, we find the rms noise level is below 2\,Jy for 95\% of the
survey and below 1.4\,Jy for 50\% of the survey. The full range
of noise levels varies by a factor of 2.9, but we find that 87\% of the
\water~maser observations are characterised by a rms noise level between 1
and 2\,Jy.

We find that 90\% of the rms noise levels in the IFs containing NH$_3$ (1,1),
H69$\alpha$ and HCCCN (3--2) emission are between 0.12 and 0.24\,K. Thus, we
consider that the rms noise levels in the survey are well characterised by
variation of a factor of two, but note that approximately 10\% of the rms noise
levels in the survey lie outside this range.

\subsection{\water~maser completeness limits}
\label{completeness}
In order to estimate the completeness of the \water~maser observations, we took a
\water~maser data cube, with a representative noise level of 1.5\,Jy (rms), but without
any identified \water~masers. We injected fake signals into the spectrum, using
the {\sc imgen} task in {\sc miriad}, to mimic
the typical characteristics of a real \water~maser: a three-dimensional gaussian
with the size of the beam in the spatial axes and line width the same as the median
\water~maser line width of 1.2\,\kms. The same data reduction routines and visualisation
methods were used to identify the fake masers. In each test we used 100 fake masers
and each test was repeated for four discrete input flux densities. We found
the following successful recovery rate of fake masers in the data cube: at 11.1\,Jy, we
recovered 100\% of the fake masers, at 8.4\,Jy we recovered 98\%, at 6.7\,Jy we recovered
86\% and at 5.5\,Jy we recoved 50\%. Therefore, we consider that for typical observing
conditions, we are complete to about 10\,Jy and 50\% complete to about 5.5\,Jy.

\begin{figure}
\includegraphics[width=0.5\textwidth]{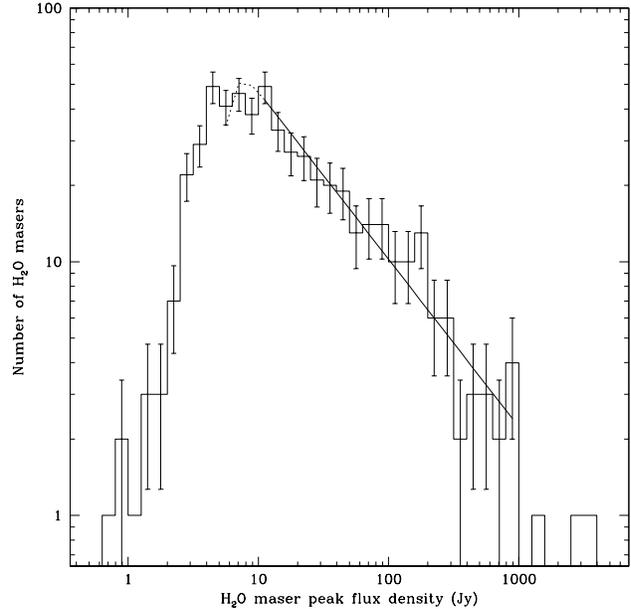}
\caption{Distribution of \water~maser peak flux densities, based on the strongest
maser spot in each maser site. The histogram shows the distribution of detected masers, with
error bars representing $\sqrt{N}$ statistics.
The solid straight line is a power law fit to the distribution above the 100\% completeness
limit of 10\,Jy, with slope -0.66. The dotted line represents the apparent turnover in the
distribution due to the completeness limit. This dotted line is consistent with the
histogram implying that we do not see a significant turnover in the population of
\water~maser flux densities.}
\label{peakfluxes}
\end{figure}

Figure \ref{peakfluxes} shows the distribution of peak flux densities of the strongest
maser spot in each maser site. This figure shows that the number of masers
sharply rises from the strongest masers down to about 5\,Jy, The distribution above
the 100\% completeness limit (10\,Jy) is well matched to
a power law with slope -0.66 down to this point (the solid straight line in Figure
\ref{peakfluxes}). The dotted curve extending from the straight line indicates the expected
numbers of masers, assuming that the real distribution of maser fluxes would follow the
straight line extended to lower flux levels, together with the above completeness figures.
The apparent turnover seen in the dotted line is consistent with the distribution of detected
maser flux densities, shown by the histogram, given error bars that are based on $\sqrt{N}$
statistics. This means that although the distribution shows a peak of masers, this is likely
due to the completeness limit of the survey and not due to an intrinsic turnover in the
maser flux density population.

\subsection{Other maser detections}
\label{othermasers}
Within the HOPS region we made occasional detections of other maser lines
towards 11 regions. These include detections from the
CH$_3$OH (J$_2$ -- J$_1$) E series near 25\,GHz and the previously unknown masing
transition of CH$_3$OH at 23.444\,GHz. The 23.444\,GHz CH$_3$OH detection
has been confirmed as a maser by \citet{voronkov11}. In addition
to this, we detected possible maser emission in the NH$_3$ inversion
transitions (11,9), (8,6) and (3,3).
Details of the NH$_3$ detections are given below.


\subsubsection{NH$_3$ (11,9) and (8,6)}
\begin{figure}
\includegraphics[width=0.5\textwidth]{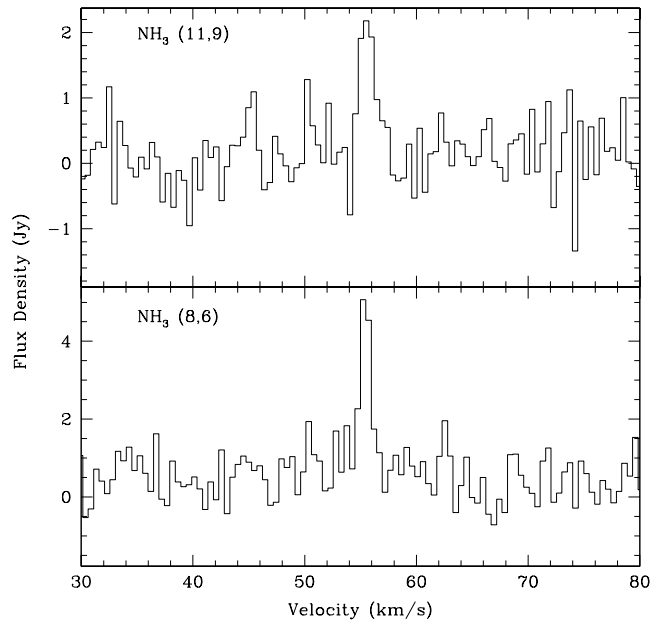}
\caption{NH$_3$ emission spectra towards G19.61-0.23 in the (11,9) inversion
transition ({\bf top}) and (8,6) transition ({\bf bottom}). Emission in the
(11,9) transition has a peak flux density of 2.2\,Jy, line velocity of
55.6\,\kms and line FWHM of 1.5\,\kms. Emission in the (8,6) transition has
a peak flux density of 4.9\,Jy, line velocity of 55.4\,\kms and line FWHM
of 1.1\,\kms.}
\label{nh3masers}
\end{figure}

Emission in these two inversion transitions of NH$_3$ have previously
been reported towards one position: NGC\,6334\,I \citep{walsh07b} and confirmed
as new masing transitions. We detect emission from both transitions in HOPS towards
one new position: G19.61-0.23. Spectra of emission in the two transitions are
shown in Figure \ref{nh3masers}. We also detect \water~maser emission at this
position, as detailed in Table \ref{330masers} and shown in Figure \ref{spectra}.
However, we note that the emission of the NH$_3$ transitions at approximately
55.5\,\kms~is on the edge of the extent of the \water~maser emission and
about 11\,\kms~offset from the strongest \water~maser spot.

We cannot confirm that these two emission lines are due to masers with the current
data, but the fact that the lines are typically narrow (1.5 and 1.1\,\kms~for (11,9)
and (8,6), respectively) and that these transitions have only been 
previously detected as masers,
suggests that a maser origin is more likely than a thermal origin.

G19.61-0.23 is a well known region of high mass star formation in the Galaxy,
containing OH \citep{caswellandhaynes83} and CH$_3$OH masers \citep{caswell95,walsh98},
including the rare 9.9\,GHz CH$_3$OH maser \citep{voronkov10},
cold dust continuum emission \citep{walsh03,thompson06}, molecular line emission
\citep{santangelo10} and multiple ultracompact (UC) \hii~regions \citep{garay98}. \citet{kolpak03}
determine a distance to G19.61-0.23 of 12.3\,kpc, based on H{\scriptsize I} observations.
It is surprising that one of only two known sources of emission in these NH$_3$ lines
is at such a large distance. This suggests
G19.61-0.23 may be an unusual region of maser activity in the Galaxy otherwise
we would expect to detect other maser sites closer to us. \citet{garay98} concluded
that in this region an ionisation front is driven by an expanding \hii region into
a molecular cloud. Together with the presence of the rare 9.9\,GHz CH$_3$OH maser,
which is known to trace strong shocks, it is possible that the NH$_3$ masers are
created in the shock interface between \hii region and molecular cloud.

\subsubsection{NH$_3$ (3,3)}
\begin{figure}
\includegraphics[width=0.5\textwidth]{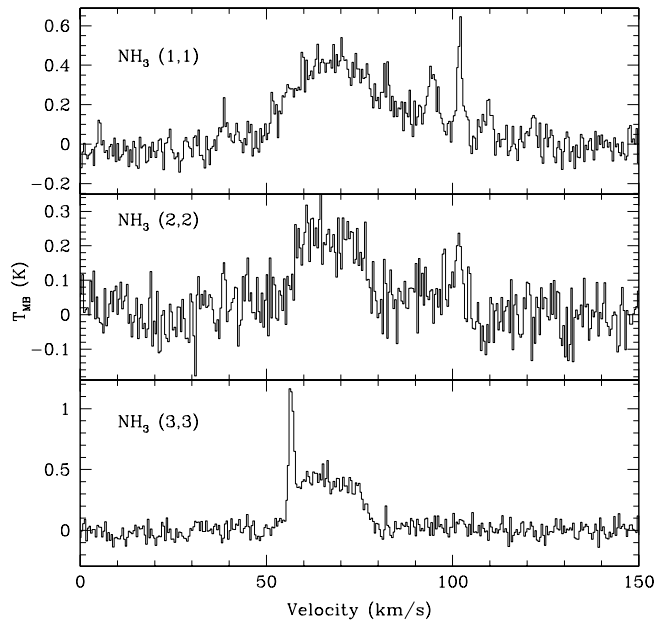}
\caption{NH$_3$ emission spectra towards G23.33-0.30. ({\bf Top}) NH$_3$ (1,1)
emission showing a broad peak centred at approximately 70\,\kms~and a line FWHM
of approximately 40\,\kms. There is a second NH$_3$ (1,1) source
centred at approximately 102\,\kms, with line FWHM of 1.9\,\kms. The second source
also exhibits hyperfine structure. ({\bf Middle}) NH$_3$ (2,2) emission also showing
the two features identified in the (1,1) spectrum at a somewhat weaker level. ({\bf Bottom})
NH$_3$ (3,3) emission showing a broad emission feature, as for the (1,1) and (2,2) spectra,
with peak around 64\,\kms~and line FWHM approximately 20\,\kms. A new narrow-lined feature
is seen in the (3,3) spectrum at 56.7\,\kms, with a line width of 1.2\,\kms. This feature
is suspected to be a maser.}
\label{nh333}
\end{figure}

We find widespread thermal NH$_3$ (3,3) emission throughout the HOPS region. However,
here we report on the possible detection of a maser in this transition. The possible
maser is located at G23.33-0.30. The spectra of NH$_3$ (1,1), (2,2) and (3,3)
emission at the position of G23.33-0.30 are shown in Figure \ref{nh333}.

We suspect that the narrow-lined emission peak at 56.7\,\kms~is a maser because this feature
is not seen in either of the (1,1) or (2,2) spectra. It also shows an unusually narrow line
FWHM of 1.2\,\kms, ie. much narrower than the typical (1,1) and (2,2) thermal linewidths
\citep{longmore07}. We await further observations to confirm, or otherwise, the maser
nature of this emission feature, as it is not possible to conclude given the current data.
The NH$_3$ (3,3) transition has previously been detected as a maser in DR\,21(OH)
\citep{mangum94}, W51 \citep{zhang95}, NGC\,6334\,I \citep{kraemer95} and G5.89-0.39
\citep{hunter08}. However, in all previous detections, the NH$_3$ (3,3) maser has never
been seen stronger than 0.5\,Jy. Our observations of the suspected maser in G23.33-0.30
give a peak flux density of 9.7\,Jy above the NH$_3$ (3,3) thermal emission, which would make it the
brightest NH$_3$ (3,3) maser known by over an order of magnitude.

G23.33-0.30 is associated with an infrared dark cloud \citep{peretto09} and
cold dust continuum source \citep{difrancesco08}. It is also associated with a
Class II CH$_3$OH maser \citep{szymczak00}, however, the CH$_3$OH maser emission
is seen over a velocity range that covers between 63 and 76\,\kms~and does not 
overlap with the suspected NH$_3$ (3,3) maser. Based on the possible associations, the
suspected NH$_3$ (3,3) maser is likely to arise from a high mass star forming
region, although the exact nature of any associations will almost certainly
require higher spatial resolution observations to investigate the spatial
coincidence of the NH$_3$ (3,3) emission and other features in this region.

\subsection{\water~maser Galactic latitude distribution}
\label{secglat}
\begin{figure}
\includegraphics[width=0.5\textwidth]{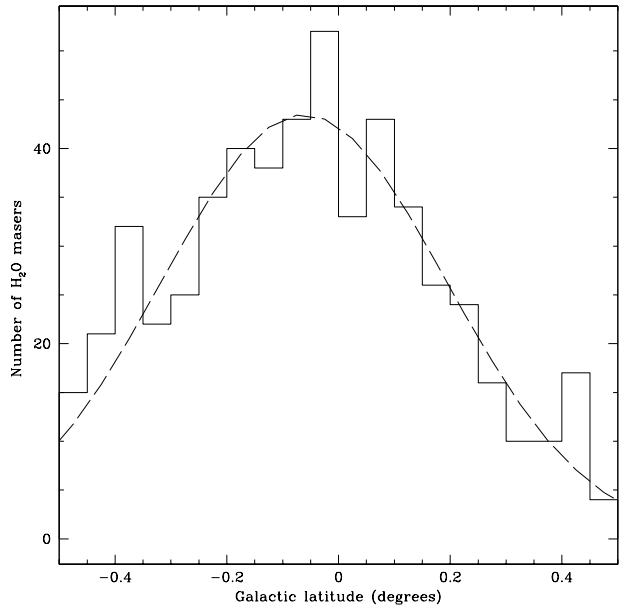}
\caption{Galactic latitude distribution of \water~masers detected in HOPS.
The number of water masers is shown by the histogram and a fitted Gaussian
to the distribution is shown as the dashed line. The fitted Gaussian has a
peak at -0.062$^\circ$ and FWHM of 0.60$^\circ$.}
\label{glat}
\end{figure}

Figure \ref{glat} shows the Galactic latitude distribution for water masers detected
in HOPS.
The dashed line in Figure \ref{glat} shows a fitted Gaussian to the distribution,
which has a peak at $-0.062^\circ$ and FWHM of 0.60$^\circ$. The FWHM corresponds
to a scale height of 0.5$^\circ$, which is slightly smaller than the scale height for
\uchii regions of 0.6$^\circ$ \citep{wood89b} and slightly larger than the scale
height found for Class II CH$_3$OH masers of 0.4$^\circ$ \citep{walsh97}. Both
\uchii~regions and Class II CH$_3$OH masers are reliable tracers of high mass star
formation and their scale height appears to be the smallest known for any class
of Galactic object. Since the \water~masers appear to be as tightly constrained
as the \uchii~regions and the CH$_3$OH masers, we surmise that the majority of detected
\water~masers are likely drawn from the same population as regions of high mass
star formation.

Given the fitted Gaussian to the Galactic latitude distribution, we can estimate
the number of water masers that are detectable, but lie at
Galactic latitudes outside the survey area, between Galactic longitudes of
290$^\circ$ and 30$^\circ$.
This is done by integrating under the entire Gaussian curve and comparing this number
to the 540 detected masers in the area under the curve between
$-0.5^\circ$ and $+0.5^\circ$. We find that this analysis yields about 32 undetected
maser sites, or 6\%. We expect this number of undetected masers is a lower limit. This is
because, as mentioned above, the Gaussian is dominated by maser sites associated
with high mass star formation. But \water~masers are known to also be associated
with evolved stars, which have a larger scale height about the Galactic plane.
Based on a 1612\,MHz OH maser survey towards evolved stars \citep{sevenster97},
we estimate the scale height of evolved stars is 1.5$^\circ$.
Therefore, we expect more detections of water masers at higher Galactic latitudes
from evolved stars. Only 15 maser sites in our survey have been associated with
evolved stars, with possibly some other maser sites yet to be associated. We
estimate that probably no more than 10\% of all detected masers will be associated
with evolved stars, making them a minor contribution to the total number of masers.
Assuming up to 10\% of masers are associated with evolved stars, which have a scale
height of 1.5$^\circ$, we estimate that no more than about 73 detectable masers lie in
this region. 

\subsection{\water~maser Galactic longitude distribution}
\label{secglon}
Note that we have only considered the regions at greater
Galactic latitudes, but within the same Galactic longitude range, as the survey area.
This covers approximately 28\% of all Galactic longitudes. If we assume the
distribution of \water~masers is symmetrical about the Galactic centre in Galactic
longitude, then given there are 339 masers between $l=290^\circ$ and 0$^\circ$,
we estimate there at least 680 masers between $l=290^\circ$ and $l=70^\circ$
and $|b|<0.5^\circ$. Based on the scale heights given above, we estimate
that there are about 800 detectable masers at all Galactic latitudes, within this
Galactic longitude range.
Without knowing the true distribution of \water~masers in the outer Galaxy, we cannot
reliably extrapolate to the number of detectable masers in the entire Galaxy.
However, we expect that the remaining 45\% of the Galaxy will not contain as
many detectable \water~masers as the inner Galaxy, so we place an approximate upper
limit of 1500 in the entire Galaxy. In conclusion, we estimate that there are
between 800 and 1500 \water~masers in the Galaxy that are detectable with a survey
sensitivity comparable to HOPS.

\begin{figure}
\includegraphics[width=0.5\textwidth]{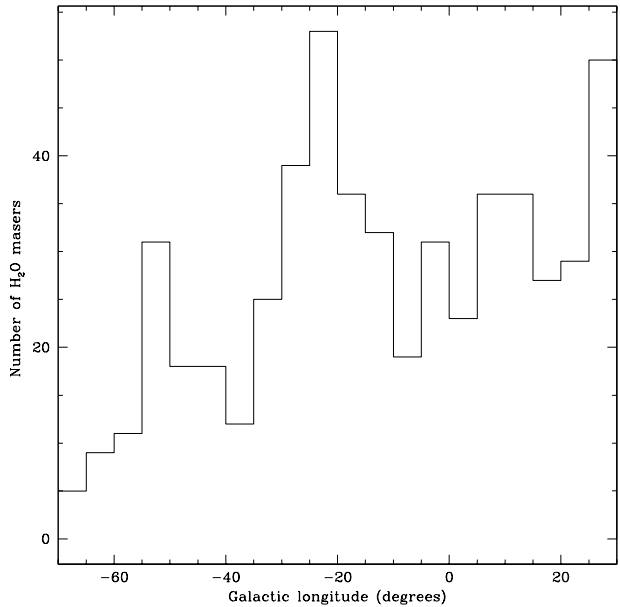}
\caption{Galactic longitude distribution of \water~masers detected in HOPS.
Three peaks are seen at approximately 310$^\circ$, 335$^\circ$ and 25$^\circ$,
which correspond to tangent points of the Galactic spiral arms. There is no prominent
peak towards the Galactic centre.}
\label{glon}
\end{figure}

Figure \ref{glon} shows the distribution of \water~masers in Galactic longitude.
There are three prominent concentrations of \water~masers at approximately
310$^\circ$, 335$^\circ$ and 25$^\circ$. These positions in the Galaxy correspond to tangent
points of Galactic spiral arms. Towards the Galactic centre, we do not find any
peak in the number of \water~masers, compared to other longitudes within 20$^\circ$
of the Galactic centre. This is somewhat surprising as the Galactic centre is very
prominent in many spectral line tracers, such as NH$_3$, reported in Longmore et al.
({\em in preparation}). We will compare the positions of \water~masers and
dense molecular gas in future papers.

\section{Conclusions}
We have completed observations of 100 square degrees of the Galactic plane, using
the Mopra radiotelescope in the 12\,mm band, in a survey we call HOPS (the
{\bf H}$_2${\bf O} Southern Galactic {\bf P}lane {\bf S}urvey). Our observations
concentrated on spectral line emission from the \water~maser at 22.2\,GHz, the
NH$_3$ inversion transitions (1,1), (2,2), (3,3), (6,6) and (9,9), HCCCN (3--2),
and radio recombination lines H69$\alpha$ and H62$\alpha$. We have reported
on the survey design, covering Galactic longitudes from $l=290^\circ$ to
$l=30^\circ$ and Galactic latitudes $|b|<0.5^\circ$.

Within the survey region, we found 540 \water~masers, of which 334 (62\%) are new
detections. We find maser peak flux densities ranging from 0.7 to 3933\,Jy,
with 62 masers (9\%) over 100\,Jy. We find maser spot velocities spread
over ranges up to 351.3\,\kms, with 14 maser sites showing spot velocities
spread over at least 100\,\kms. We estimate that at least 90\% of the detected
masers oucld be associated with high mass star formation, with the remainder being
associated with evolved stars or low mass star formation. This is based on the number
of known associations
of \water~masers and evolved stars in our survey (15) and the Galactic latitude
distribution of the \water~masers, which closely resembles the distribution
of high mass star formation sites.

Our rms noise levels for the \water~maser observations are between 1 and 2\,Jy
for 87\% of the survey, with 95\% of the survey having an rms noise level
below 2\,Jy. We estimate that the survey is 98\% complete down to a flux
limit of about 8.4\,Jy and 50\% complete down to about 5.5\,Jy.
We estimate that there are between 800 and 1500 \water~masers in
the Galaxy that are detectable in a survey with such completeness limits.

We detect possible masers in transitions of CH$_3$OH and NH$_3$. This includes
the detection of NH$_3$ (11,9) and (8,6) emission towards G19.61-0.23. If these
transitions are confirmed as masers, they will be only the second known examples
of these transitions showing maser activity. We also report possible maser
emission in the NH$_3$ (3,3) line towards G23.33-0.30. If confirmed as a maser,
it will be the strongest known maser in this line by at least an order of
magnitude.

In Paper II (Purcell et al. 2011, {\em in preparation}) we present
NH$_3$ (1,1) and (2,2) data, the source finding algorithm and thermal line fitting
routines. Paper III (Longmore et al. 2011, {\em in preparation}) will detail
properties of all other thermal lines detected in HOPS.

\section*{Acknowledgments}
The HOPS team would like to thank the dedicated work of CSIRO Narrabri staff who
supported the observations beyond the call of duty. PAJ acknowledges partial support
from Centro de Astrof\'\i sica FONDAP 15010003 and the GEMINI-CONICYT FUND.
NL acknowledges partial support from the Center of Excellence in Astrophysics
and Associated Technologies (PFB 06) and Centro de Astrof\'{i}sica
FONDAP\,15010003. NL's postdoctoral position at CEA/Irfu was funded by the
Ile-de-France Region.
The University of New South Wales Digital Filter Bank used for the observations (MOPS)
with the Mopra Telescope was provided with support from the Australian Research Council,
CSIRO, The University of New South Wales, Monash University and The University of
Sydney. The Mopra radio telescope is part of the Australia Telescope National Facility
which is funded by the Commonwealth of Australia for operation as a National
Facility managed by CSIRO.

\label{lastpage}

\end{document}